\documentclass[%
 twocolumn,
 amsmath,
 amssymb,
 aps,
 nofootinbib
]{revtex4-1}

\usepackage{simplewick}
\usepackage{footnote}
\usepackage{graphicx}
\usepackage{dcolumn}
\usepackage{bm}
\usepackage{stmaryrd}
\usepackage{subcaption}

\newcommand{\s}{\hspace{2 pt}}
\newcommand{\matrixbb}[4]{\left(\hspace{0 pt}\begin{tabular}{ c c } ${#1}$ & ${#2}$ \\ ${#3}$ & ${#4}$ \end{tabular}\hspace{0 pt}\right)}
\newcommand{\ltr}{\llbracket}
\newcommand{\rtr}{\rrbracket}

\newcommand{\RED}{R_{\text{5D}}}
\newcommand{\LEH}{\mathcal{L}_{\text{EH}}}
\newcommand{\LCC}{\mathcal{L}_{\text{CC}}}
\newcommand{\MPlED}{M_{\text{Pl,5D}}}
\newcommand{\LED}{\mathcal{L}_{\text{5D}}}
\newcommand{\LDDeff}{\mathcal{L}_{\text{4D}}^{(\text{eff})}}

\newcommand{\etaRS}{\eta^{(\text{RS})}_{MN}}
\newcommand{\GRS}{G_{MN}^{(\text{RS})}}
\newcommand{\LRS}{\mathcal{L}_{\text{5D}}^{(\text{RS})}}
\newcommand{\LEHRS}{\mathcal{L}_{\text{EH}}^{(\text{RS})}}
\newcommand{\LCCRS}{\mathcal{L}_{\text{CC}}^{(\text{RS})}}

\newcommand{\GEDOT}{G_{MN}^{(\text{5DOT})}}

\newcommand{\LRSf}[1]{\mathcal{L}_{#1}^{(\text{RS})}}
\newcommand{\LA}[1]{\overline{\mathcal{L}}_{A:#1}}
\newcommand{\LB}[1]{\overline{\mathcal{L}}_{B:#1}}
\newcommand{\LRSfeff}[1]{\mathcal{L}_{#1}^{(\text{RS,eff})}}
\newcommand{\LAeff}[1]{\mathcal{L}_{A:#1}^{(\text{eff})}}
\newcommand{\LBeff}[1]{\mathcal{L}_{B:#1}^{(\text{eff})}}

\newcommand{\kDD}{\kappa_{\text{4D}}}

\newcommand{\an}[1]{a_{#1}}
\newcommand{\bn}[1]{b_{#1}}

\newcommand{\bhhr}[2]{b_{{#1}\hspace{0.5 pt}{#2}\hspace{0.5 pt}r}}
\newcommand{\brhh}[2]{b_{r\hspace{0.5 pt}{#1}\hspace{0.5 pt}{#2}}}
\newcommand{\ahhh}[3]{a_{{#1}{#2}{#3}}}
\newcommand{\bhhh}[3]{b_{{#1}\hspace{1 pt}{#2}\hspace{1 pt}{#3}}}

\usepackage{tikz}
\usetikzlibrary{arrows,decorations.pathmorphing}

\newcommand{\sDklrmn}[1]{\begin{tikzpicture}[scale={#1},photon/.style={decorate,decoration={snake,post length=1mm}}]
        \draw (0,4) -- (0.5,2);
        \draw[photon] (0,4) -- (0.5,2);
        \draw node[anchor=east] at (0,4) {\large $n_{1}$};
        \draw (0,0) -- (0.5,2);
        \draw[photon] (0,0) -- (0.5,2);
        \draw node[anchor=east] at (0,0) {\large $n_{2}$};
        \draw (0.5,2) -- (3.5,2);
        \draw (3.5,2) -- (4,4);
        \draw[photon] (3.5,2) -- (4,4);
        \draw node[anchor=west] at (4,4) {\large $n_{3}$};
        \draw (3.5,2) -- (4,0);
        \draw[photon] (3.5,2) -- (4,0);
        \draw node[anchor=west] at (4,0) {\large $n_{4}$};
      \node[circle,fill=darkgray,draw=black,inner sep=0pt,minimum size=0.1cm] at (0.5,2) {};
      \node[circle,fill=darkgray,draw=black,inner sep=0pt,minimum size=0.1cm] at (3.5,2) {};
      \draw node[anchor=south] at (2,2) {\large $r$};
\end{tikzpicture}}

\newcommand{\tDklrmn}[1]{\begin{tikzpicture}[scale={#1},photon/.style={decorate,decoration={snake,post length=1mm}}]
        \draw (0,4) -- (2,3.5);
        \draw[photon] (0,4) -- (2,3.5);
        \draw node[anchor=east] at (0,4) {\large $n_{1}$};
        \draw (0,0) -- (2,0.5);
        \draw[photon] (0,0) -- (2,0.5);
        \draw node[anchor=east] at (0,0) {\large $n_{2}$};
        \draw (2,0.5) -- (2,3.5);
        \draw (2,3.5) -- (4,4);
        \draw[photon] (2,3.5) -- (4,4);
        \draw node[anchor=west] at (4,4) {\large $n_{3}$};
        \draw (2,0.5) -- (4,0);
        \draw[photon] (2,0.5) -- (4,0);
        \draw node[anchor=west] at (4,0) {\large $n_{4}$};
      \node[circle,fill=darkgray,draw=black,inner sep=0pt,minimum size=0.1cm] at (2,3.5) {};
      \node[circle,fill=darkgray,draw=black,inner sep=0pt,minimum size=0.1cm] at (2,0.5) {};
      \draw node[anchor=east] at (2,2) {\large $r$};
\end{tikzpicture}}

\newcommand{\uDklrmn}[1]{\begin{tikzpicture}[scale={#1},photon/.style={decorate,decoration={snake,post length=1mm}}]
        \draw (0,4) -- (2,3.5);
        \draw[photon] (0,4) -- (2,3.5);
        \draw node[anchor=east] at (0,4) {\large $n_{1}$};
        \draw (0,0) -- (2,0.5);
        \draw[photon] (0,0) -- (2,0.5);
        \draw node[anchor=east] at (0,0) {\large $n_{2}$};
        \draw (2,0.5) -- (2,3.5);
        \draw (2,3.5) -- (4,0);
        \draw[photon] (2,3.5) -- (4,0);
        \draw node[anchor=west] at (4,0) {\large $n_{4}$};
        \draw (2,0.5) -- (4,4);
        \draw[photon] (2,0.5) -- (4,4);
        \draw node[anchor=west] at (4,4) {\large $n_{3}$};
      \node[circle,fill=darkgray,draw=black,inner sep=0pt,minimum size=0.1cm] at (2,3.5) {};
      \node[circle,fill=darkgray,draw=black,inner sep=0pt,minimum size=0.1cm] at (2,0.5) {};
      \draw node[anchor=east] at (2,2) {\large $r$};
\end{tikzpicture}}

\newcommand{\sDkljmn}[1]{\begin{tikzpicture}[scale={#1},photon/.style={decorate,decoration={snake,post length=1mm}}]
        \draw (0,4) -- (0.5,2);
        \draw[photon] (0,4) -- (0.5,2);
        \draw node[anchor=east] at (0,4) {\large $n_{1}$};
        \draw (0,0) -- (0.5,2);
        \draw[photon] (0,0) -- (0.5,2);
        \draw node[anchor=east] at (0,0) {\large $n_{2}$};
        \draw (0.5,2) -- (3.5,2);
        \draw[photon] (0.5,2) -- (3.5,2);
        \draw (3.5,2) -- (4,4);
        \draw[photon] (3.5,2) -- (4,4);
        \draw node[anchor=west] at (4,4) {\large $n_{3}$};
        \draw (3.5,2) -- (4,0);
        \draw[photon] (3.5,2) -- (4,0);
        \draw node[anchor=west] at (4,0) {\large $n_{4}$};
      \node[circle,fill=darkgray,draw=black,inner sep=0pt,minimum size=0.1cm] at (0.5,2) {};
      \node[circle,fill=darkgray,draw=black,inner sep=0pt,minimum size=0.1cm] at (3.5,2) {};
      \draw node[anchor=south] at (2,2) {\large $j$};
\end{tikzpicture}}

\newcommand{\tDkljmn}[1]{\begin{tikzpicture}[scale={#1},photon/.style={decorate,decoration={snake,post length=1mm}}]
        \draw (0,4) -- (2,3.5);
        \draw[photon] (0,4) -- (2,3.5);
        \draw node[anchor=east] at (0,4) {\large $n_{1}$};
        \draw (0,0) -- (2,0.5);
        \draw[photon] (0,0) -- (2,0.5);
        \draw node[anchor=east] at (0,0) {\large $n_{2}$};
        \draw (2,0.5) -- (2,3.5);
        \draw[photon] (2,0.5) -- (2,3.5);
        \draw (2,3.5) -- (4,4);
        \draw[photon] (2,3.5) -- (4,4);
        \draw node[anchor=west] at (4,4) {\large $n_{3}$};
        \draw (2,0.5) -- (4,0);
        \draw[photon] (2,0.5) -- (4,0);
        \draw node[anchor=west] at (4,0) {\large $n_{4}$};
      \node[circle,fill=darkgray,draw=black,inner sep=0pt,minimum size=0.1cm] at (2,3.5) {};
      \node[circle,fill=darkgray,draw=black,inner sep=0pt,minimum size=0.1cm] at (2,0.5) {};
      \draw node[anchor=east] at (2,2) {\large $j$};
\end{tikzpicture}}

\newcommand{\uDkljmn}[1]{\begin{tikzpicture}[scale={#1},photon/.style={decorate,decoration={snake,post length=1mm}}]
        \draw (0,4) -- (2,3.5);
        \draw[photon] (0,4) -- (2,3.5);
        \draw node[anchor=east] at (0,4) {\large $n_{1}$};
        \draw (0,0) -- (2,0.5);
        \draw[photon] (0,0) -- (2,0.5);
        \draw node[anchor=east] at (0,0) {\large $n_{2}$};
        \draw (2,0.5) -- (2,3.5);
        \draw[photon] (2,0.5) -- (2,3.5);
        \draw (2,3.5) -- (4,0);
        \draw[photon] (2,3.5) -- (4,0);
        \draw node[anchor=west] at (4,0) {\large $n_{4}$};
        \draw (2,0.5) -- (4,4);
        \draw[photon] (2,0.5) -- (4,4);
        \draw node[anchor=west] at (4,4) {\large $n_{3}$};
      \node[circle,fill=darkgray,draw=black,inner sep=0pt,minimum size=0.1cm] at (2,3.5) {};
      \node[circle,fill=darkgray,draw=black,inner sep=0pt,minimum size=0.1cm] at (2,0.5) {};
      \draw node[anchor=east] at (2,2) {\large $j$};
\end{tikzpicture}}

\newcommand{\sgDklmn}[1]{\begin{tikzpicture}[scale={#1},photon/.style={decorate,decoration={snake,post length=1mm}}]
        \draw (0,4) -- (2,2);
        \draw[photon] (0,4) -- (2,2);
        \draw node[anchor=east] at (0,4) {\large $n_{1}$};
        \draw (0,0) -- (2,2);
        \draw[photon] (0,0) -- (2,2);
        \draw node[anchor=east] at (0,0) {\large $n_{2}$};
        \draw node[anchor=west] at (4,4) {\large $n_{3}$};
        \draw (2,2) -- (4,0);
        \draw[photon] (2,2) -- (4,0);
        \draw node[anchor=west] at (4,0) {\large $n_{4}$};
        \draw (2,2) -- (4,4);
        \draw[photon] (2,2) -- (4,4);
      \node[circle,fill=darkgray,draw=black,inner sep=0pt,minimum size=0.1cm] at (2,2) {};
      \node[circle,fill=darkgray,draw=black,inner sep=0pt,minimum size=0.1cm] at (2,2) {};
\end{tikzpicture}}

\newcommand{\sDA}[1]{\begin{tikzpicture}[scale={#1},photon/.style={decorate,decoration={snake,post length=1mm}}]
        \draw (0,4) -- (0.5,2);
        \draw[photon] (0,4) -- (0.5,2);
        \draw node[anchor=east] at (0,4) {\large $1$};
        \draw (0,0) -- (0.5,2);
        \draw[photon] (0,0) -- (0.5,2);
        \draw node[anchor=east] at (0,0) {\large $4$};
        \draw (0.5,2) -- (3.5,2);
        \draw[photon] (0.5,2) -- (3.5,2);
        \draw (3.5,2) -- (4,4);
        \draw[photon] (3.5,2) -- (4,4);
        \draw node[anchor=west] at (4,4) {\large $2$};
        \draw (3.5,2) -- (4,0);
        \draw[photon] (3.5,2) -- (4,0);
        \draw node[anchor=west] at (4,0) {\large $3$};
      \node[circle,fill=darkgray,draw=black,inner sep=0pt,minimum size=0.1cm] at (0.5,2) {};
      \node[circle,fill=darkgray,draw=black,inner sep=0pt,minimum size=0.1cm] at (3.5,2) {};
      \draw node[anchor=south] at (2,2) {\large $5$};
\end{tikzpicture}}

\newcommand{\tDA}[1]{\begin{tikzpicture}[scale={#1},photon/.style={decorate,decoration={snake,post length=1mm}}]
        \draw (0,4) -- (2,3.5);
        \draw[photon] (0,4) -- (2,3.5);
        \draw node[anchor=east] at (0,4) {\large $1$};
        \draw (0,0) -- (2,0.5);
        \draw[photon] (0,0) -- (2,0.5);
        \draw node[anchor=east] at (0,0) {\large $4$};
        \draw (2,0.5) -- (2,3.5);
        \draw[photon] (2,0.5) -- (2,3.5);
        \draw (2,3.5) -- (4,4);
        \draw[photon] (2,3.5) -- (4,4);
        \draw node[anchor=west] at (4,4) {\large $2$};
        \draw (2,0.5) -- (4,0);
        \draw[photon] (2,0.5) -- (4,0);
        \draw node[anchor=west] at (4,0) {\large $3$};
      \node[circle,fill=darkgray,draw=black,inner sep=0pt,minimum size=0.1cm] at (2,3.5) {};
      \node[circle,fill=darkgray,draw=black,inner sep=0pt,minimum size=0.1cm] at (2,0.5) {};
      \draw node[anchor=east] at (2,2) {\large $1$};
\end{tikzpicture}}

\newcommand{\uDA}[1]{\begin{tikzpicture}[scale={#1},photon/.style={decorate,decoration={snake,post length=1mm}}]
        \draw (0,4) -- (2,3.5);
        \draw[photon] (0,4) -- (2,3.5);
        \draw node[anchor=east] at (0,4) {\large $1$};
        \draw (0,0) -- (2,0.5);
        \draw[photon] (0,0) -- (2,0.5);
        \draw node[anchor=east] at (0,0) {\large $4$};
        \draw (2,0.5) -- (2,3.5);
        \draw[photon] (2,0.5) -- (2,3.5);
        \draw (2,3.5) -- (4,0);
        \draw[photon] (2,3.5) -- (4,0);
        \draw node[anchor=west] at (4,0) {\large $2$};
        \draw (2,0.5) -- (4,4);
        \draw[photon] (2,0.5) -- (4,4);
        \draw node[anchor=west] at (4,4) {\large $3$};
      \node[circle,fill=darkgray,draw=black,inner sep=0pt,minimum size=0.1cm] at (2,3.5) {};
      \node[circle,fill=darkgray,draw=black,inner sep=0pt,minimum size=0.1cm] at (2,0.5) {};
      \draw node[anchor=east] at (2,2) {\large $2$};
\end{tikzpicture}}

\newcommand{\sgDA}[1]{\begin{tikzpicture}[scale={#1},photon/.style={decorate,decoration={snake,post length=1mm}}]
        \draw (0,4) -- (2,2);
        \draw[photon] (0,4) -- (2,2);
        \draw node[anchor=east] at (0,4) {\large $1$};
        \draw (0,0) -- (2,2);
        \draw[photon] (0,0) -- (2,2);
        \draw node[anchor=east] at (0,0) {\large $4$};
        \draw node[anchor=west] at (4,4) {\large $2$};
        \draw (2,2) -- (4,0);
        \draw[photon] (2,2) -- (4,0);
        \draw node[anchor=west] at (4,0) {\large $3$};
        \draw (2,2) -- (4,4);
        \draw[photon] (2,2) -- (4,4);
      \node[circle,fill=darkgray,draw=black,inner sep=0pt,minimum size=0.1cm] at (2,2) {};
      \node[circle,fill=darkgray,draw=black,inner sep=0pt,minimum size=0.1cm] at (2,2) {};
\end{tikzpicture}}

\newcommand{\propSc}[1]{\begin{tikzpicture}[scale={#1},photon/.style={decorate,decoration={snake,post length=1mm}}]
        \draw (0,0) -- (2,0);
\end{tikzpicture}}

\newcommand{\propGr}[1]{\begin{tikzpicture}[scale={#1},photon/.style={decorate,decoration={snake,post length=1mm}}]
        \draw (0,0) -- (2,0);
        \draw[photon] (0,0) -- (2,0);
        \draw node[anchor=east] at (0,0) {$\mu\nu$};
        \draw node[anchor=west] at (2,0) {$\rho\sigma$};
\end{tikzpicture}}

\newcommand{\verHHH}[1]{\begin{tikzpicture}[scale={#1},photon/.style={decorate,decoration={snake,post length=1mm}}]
        \draw (0,4) -- (2,2);
        \draw[photon] (0,4) -- (2,2);
        \draw node[anchor=east] at (0,4) {\large $n_{1}$};
        \draw (0,0) -- (2,2);
        \draw[photon] (0,0) -- (2,2);
        \draw node[anchor=east] at (0,0) {\large $n_{2}$};
        \draw node[anchor=west] at (4,2) {\large $n_{3}$};
        \draw (2,2) -- (4,2);
        \draw[photon] (2,2) -- (4,2);
      \node[circle,fill=darkgray,draw=black,inner sep=0pt,minimum size=0.1cm] at (2,2) {};
      \node[circle,fill=darkgray,draw=black,inner sep=0pt,minimum size=0.1cm] at (2,2) {};
\end{tikzpicture}}

\newcommand{\verHHR}[1]{\begin{tikzpicture}[scale={#1},photon/.style={decorate,decoration={snake,post length=1mm}}]
        \draw (0,4) -- (2,2);
        \draw[photon] (0,4) -- (2,2);
        \draw node[anchor=east] at (0,4) {\large $n_{1}$};
        \draw (0,0) -- (2,2);
        \draw[photon] (0,0) -- (2,2);
        \draw node[anchor=east] at (0,0) {\large $n_{2}$};
        \draw node[anchor=west] at (4,2) {\large $r$};
        \draw (2,2) -- (4,2);
      \node[circle,fill=darkgray,draw=black,inner sep=0pt,minimum size=0.1cm] at (2,2) {};
      \node[circle,fill=darkgray,draw=black,inner sep=0pt,minimum size=0.1cm] at (2,2) {};
\end{tikzpicture}}

\newcommand{\verHHHH}[1]{\begin{tikzpicture}[scale={#1},photon/.style={decorate,decoration={snake,post length=1mm}}]
        \draw (0,4) -- (2,2);
        \draw[photon] (0,4) -- (2,2);
        \draw node[anchor=east] at (0,4) {\large $n_{1}$};
        \draw (0,0) -- (2,2);
        \draw[photon] (0,0) -- (2,2);
        \draw node[anchor=east] at (0,0) {\large $n_{2}$};
        \draw node[anchor=west] at (4,4) {\large $n_{3}$};
        \draw (2,2) -- (4,0);
        \draw[photon] (2,2) -- (4,0);
        \draw node[anchor=west] at (4,0) {\large $n_{4}$};
        \draw (2,2) -- (4,4);
        \draw[photon] (2,2) -- (4,4);
      \node[circle,fill=darkgray,draw=black,inner sep=0pt,minimum size=0.1cm] at (2,2) {};
      \node[circle,fill=darkgray,draw=black,inner sep=0pt,minimum size=0.1cm] at (2,2) {};
\end{tikzpicture}}

\begin{document}

\title{Massive Spin-2 Scattering Amplitudes in Extra-Dimensional Theories}
\author{R. Sekhar Chivukula$^{a,b}$}
\author{Dennis Foren$^{a,b}$}
\author{Kirtimaan A. Mohan$^{b}$}
\author{Dipan Sengupta$^{a}$}
\author{Elizabeth H. Simmons$^{a,b}$}
\affiliation{$^{a}$ Department of Physics and Astronomy, 9500 Gilman Drive,
 University of California, San Diego }
 \affiliation{$^{b}$ Department of Physics and Astronomy, 567 Wilson Road, Michigan State University, East Lansing}
\date{\today}

\begin{abstract}
In this paper we describe in detail the computation of the scattering amplitudes of massive spin-2 Kaluza-Klein excitations
in a gravitational theory with a single compact extra dimension, whether flat or warped. These scattering amplitudes are characterized by intricate cancellations between different contributions: although individual contributions may grow as fast as ${\cal O}(s^5)$, the full results grow only as ${\cal O}(s)$. We demonstrate that the cancellations persist for all incoming and outgoing particle helicities and examine how truncating the computation to only include a finite number of intermediate states impacts the accuracy of the results. We also carefully assess the range of validity of the low energy effective Kaluza-Klein theory. In particular, for the warped case we demonstrate  directly how an emergent low energy scale controls the size of the scattering amplitude, as conjectured by the AdS/CFT correspondence. 
\end{abstract}

\preprint{MSUHEP-20-003}

\maketitle

\section{\label{sec:level1}Introduction}

Theories of gravity with compact extra dimensions were initially introduced to unify gravity and electromagnetism via the Kaluza-Klein (KK) \cite{Kaluza:1984ws,Klein:1926tv} construction. Constructions involving gravity with ``large"  \cite{Antoniadis:1990ew,ArkaniHamed:1998rs, Appelquist:2000nn} and  ``warped" \cite{Randall:1999ee,Randall:1999vf} extra dimensions gained renewed interest in the last two decades as potential solutions to the Standard Model hierarchy problem, and also within the broader context of string theory. A key feature of all extra-dimensional gravitational theories is the emergence of an infinite tower of massive spin-2 KK resonances in four dimensions. These extra-dimensional models are being probed by the LHC \cite{Aaboud:2017yyg}, where we can search for TeV scale KK excitations as a signature of physics beyond the Standard Model. Extra-dimensional theories that additionally incorporate neutral stable matter motivate certain dark matter models, where particulate dark matter interacts with the Standard Model through a massive spin-2 mediator (see, for example, \cite{Lee:2013bua,Garny:2015sjg,Rueter:2017nbk,Carrillo-Monteverde:2018phy,Goyal:2019vsw,Folgado:2019sgz,Folgado:2019gie}).\footnote{For reviews of extra-dimensional theories (especially in connection to the LHC and phenomenological consequences) and the holographic principle, refer to, for example, \cite{Rattazzi:2003ea,Csaki:2004ay,Gherghetta:2010cj}}

The extra-dimensional gravitational action in these models gives rise to interactions between the massive spin-2 KK resonances. Since the underlying gravitational interactions arise through operators of dimension greater than four, all tree-level scattering amplitudes of the KK modes grow with center-of-mass energy. The compactified theory must therefore be understood as a low-energy effective field theory (EFT), and the energy scale at which these scattering amplitudes would violate partial wave unitarity provides an upper bound on the ``cutoff scale", the energy scale beyond which the EFT fails.

In this paper we describe in greater detail and further
build upon the work initially reported in \cite{Chivukula:2019rij,Chivukula:2019zkt}. In that work we reported that the scattering amplitudes of the helicity-zero modes of the massive spin-2 KK resonances grow no faster than $\mathcal{O}(s)$ due to subtle cancellations between different contributions to these amplitudes.  Here we provide a complete description of the computation of the tree-level scattering amplitudes of massive spin-2 states in compactified theories of five-dimensional gravity, consider the scattering of different combinations of incoming and outgoing particle helicities, address the impact on accuracy when truncating the computation to only include a finite number of intermediate states, and carefully assess the range of validity of the low-energy EFT which arises. 

In the remainder of this introductory section, we review the physics of a theory with a single massive spin-2 particle, summarize our previously reported results for extra-dimensional theories of gravity as well as their connection to the prior literature, and briefly describe the extended results presented here. We then provide an outline of the explicit computations reported in this paper.

\subsection{\label{sec:Ia} Scattering of Single Massive Spin-2 Particle\footnote{Our discussion here follows closely the review in Sec. 8 of \cite{Hinterbichler:2011tt}.}}

Before considering the complexities of a compactified five-dimensional theory and its tower of massive spin-2 KK modes, we set the stage by reviewing the high-energy scattering behavior in a theory of a single self-interacting massive spin-2 particle in four dimensions. 

Physicists have investigated massive spin-2 particles on a four-dimensional spacetime since the initial work of Fierz and Pauli (FP)\cite{Fierz:1939ix}.\footnote{Note that the mass of the graviton is stringently constrained by gravitational wave experiments to be $m<  1.22 \times 10^{-22} {\rm eV}/c^{2}$ \cite{Abbott:2016blz}.} The FP theory is constructed by adding a Lorentz-invariant mass term\footnote{The relative coefficients between the two terms in the mass term are chosen to avoid propagating ghost degrees of freedom. For more details, consult footnote \ref{footnote:noghosts}.} to the Einstein-Hilbert action,
\begin{equation}
S_{G} = \int d^{4}x \left\{  \dfrac{M_{\text{Pl}}^{2}}{2}  \sqrt{-g}\, R + \dfrac{m^{2}}{2}\Big[h^{2}-(h_{\mu\nu})^2\Big]\right\}.
\label{eq:4dgr}
\end{equation}
where $R$ is the Ricci scalar computed from the metric $g_{\mu\nu}$, $g\equiv{\rm det}(g)$, $M_{\text{Pl}} = 1/\sqrt{8\pi G_{N}} \simeq 2.435\times 10^{15}\text{ TeV}$ is the (reduced) Planck mass, and $m$ is the mass of the spin-2 field $h_{\mu\nu}$. We expand the metric around a flat Minkowski background according to $g_{\mu\nu}= \eta_{\mu\nu} +  2h_{\mu\nu}/M_{\text{Pl}}$, use (here and throughout this paper) a mostly-minus flat Lorentz metric $\eta_{\mu\nu}$, and $h\equiv\eta^{\mu\nu}h_{\mu\nu}$. The mass term breaks diffeomorphism invariance, causing $h_{\mu\nu}$ to propagate additional longitudinal polarization modes relative to a massless graviton.\footnote{The helicity-zero mode couples to the trace of the stress energy tensor, and does not decouple in the $m \to 0$ limit, acting instead as a Brans-Dicke scalar \cite{Brans:1961sx}. This feature---known as the vDVZ discontinuity \cite{vanDam:1970vg,Zakharov:1970cc}---seemingly implies light should bend around massive bodies differently than is observed experimentally, and thus eliminates massive gravity as a description of reality; however, further investigations \cite{Vainshtein:1972sx} revealed the perturbative calculation could not be trusted at those experimental distance scales. By incorporating nonlinear effects, the predictions of general relativity are restored.}

The growth of scattering amplitudes with respect to energy in this theory can be studied using the St{\"u}kelberg formalism \cite{Hinterbichler:2011tt,ArkaniHamed:2002sp,Green:1991pa,Siegel:1993sk}.\footnote{Alternatively, one can use the deconstruction formalism  \cite{ArkaniHamed:2002sp,ArkaniHamed:2003vb}.}  Schematically, the St{\"u}kelberg formalism introduces spurious fields through which one formally restores diffeomorphism invariance. The St{\"u}kelberg fields $A_{\mu}$ and $\phi$ are introduced via the replacement
\begin{eqnarray}
h_{\mu\nu} & \to & h_{\mu\nu} + \dfrac{1}{m} \left[\partial_{(\mu}A_{\nu)}\right]+ \dfrac{2}{m^{2}}\left[\partial_\mu\partial_\nu \phi\right]+\ldots~,
\label{eq:stukelberg}
\end{eqnarray}
where $A_{\mu}$ is a vector gauge field with two transverse degrees of freedom, $\phi$ is a real scalar with one degree of freedom, and the ellipses denote additional nonlinear terms which are listed explicitly in \cite{Hinterbichler:2011tt}. 

Crucially, the nonlinear terms \cite{Hinterbichler:2011tt,ArkaniHamed:2002sp} in Eq. (\ref{eq:stukelberg}) are chosen to restore the diffeomorphism invariance of the FP Lagrangian of Eq. (\ref{eq:4dgr}) with respect to the full metric $g_{\mu\nu}=\eta_{\mu\nu}+2h_{\mu\nu}/M_{\rm Pl}$, when one simultaneously does a gauge transformation on  $A_{\mu}$ and a related transformation on $\phi$.  Restoring diffeomorphism invariance in this way \cite{Hinterbichler:2011tt,ArkaniHamed:2002sp}, one finds that the field $A_\mu$ always appears with one derivative in the combination $(\partial A/m)$, $\phi$ with two derivatives $(\partial^2\phi/m^2)$, and that higher order terms in Eq. (\ref{eq:stukelberg}) are suppressed by factors of $M_{\rm Pl}$. The genuine diffeomorphism invariance of the Einstein-Hilbert term in the Lagrangian implies that all interactions for the $A$ and $\phi$ fields come from the mass term in Eq. (\ref{eq:4dgr}).\footnote{The St{\"u}kelberg field redefinition also provides an assurance that the FP mass term in Eq. (\ref{eq:4dgr}) does not generate terms with more than two time derivatives on $\phi$, and thus the theory avoids the Ostrogradsky ghost instabilities which generally plague higher-derivative theories \cite{Boulware:1973my,Woodard:2015zca}.\label{footnote:noghosts}} The FP Lagrangian can be recovered by going to the ``unitary" gauge where the spurious $A_\mu$ and $\phi$ fields are set to zero.

Following \cite{Hinterbichler:2011tt,ArkaniHamed:2002sp}, in gauges other than unitary gauge the $h_{\mu\nu}$, $A_{\mu}$, and $\phi$ can be used to track the helicity-two, helicity-one, and helicity-zero polarization modes respectively of the massive spin-2 field, and their interactions provide an understanding of the ``power-counting" (dependence on energy) of helicity-dependent scattering amplitudes. 
Expanding the FP mass term (while ensuring the necessary non-linear terms from Eq. (\ref{eq:stukelberg}) are correctly included) results in an infinite series of multipoint interactions among $h_{\mu\nu}$ and the  St{\"u}kelberg fields. The prototypical interaction term derived in this way is of the (schematic) form \cite{ArkaniHamed:2002sp}
\begin{align}
    \dfrac{m^{2}}{2}\, \left(\dfrac{2}{M_{\text{Pl}}}\right)^{-2} \left[\dfrac{2}{M_{\text{Pl}}} h\right]^{n_{h}}\left[\dfrac{2}{M_{\text{Pl}}} \dfrac{\partial A}{m}\right]^{n_{A}}\left[\dfrac{2}{M_{\text{Pl}}} \dfrac{\partial^{2} \phi}{m^{2}}\right]^{n_{\phi}}
\end{align}
where $n_{h}$, $n_{A}$, and $n_{\phi}$ count how many instances of $h$, $A$, and $\phi$ are present in this interaction term respectively.

Neglecting powers of $2$, the various factors of graviton mass $m$ and reduced Planck mass $M_{\text{Pl}}$ multiplying the interaction term may be collected together into an interaction scale $\Lambda_{\lambda}$ \cite{ArkaniHamed:2002sp}, like so:
\begin{align}
    \left(\Lambda_{\lambda}\right)^{4 - n_{h} - 2n_{A} - 3n_{\phi}}\, h^{n_{h}} \, (\partial A)^{n_{A}} \, (\partial^{2} \phi)^{n_{\phi}}
\end{align}
where
\begin{align}
    \Lambda_{\lambda} \equiv (m^{\lambda-1}\, M_{\text{Pl}})^{1/\lambda}\label{eq:int}
\end{align}
and $\lambda \equiv (4 - n_{h} - 2n_{A} - 3n_{\phi})/(2 - n_{h} - n_{A} - n_{\phi})$. Assuming $m \ll M_{\text{Pl}}$, a larger $\lambda$ implies the corresponding interaction is suppressed by a lower energy scale $\Lambda_{\lambda}$. 

To study the growth of scattering amplitudes at high energy \cite{ArkaniHamed:2002sp}, we focus on the least-suppressed interaction vertex in the expansion: the one with the largest value of $\lambda$.  For interaction terms, we have $(n_{h} + n_{A} + n_{\phi}) \geq 3$; the largest $\lambda$ arises from  $n_{\phi} = 3$ and $n_{h} = n_{A} = 0$. This corresponds to the cubic-scalar interaction term $(\partial^{2}\phi)^{3}/\Lambda_{5}^{5}$ where $\Lambda_{5} = (m^{4} M_{\text{Pl}})^{1/5}$. We can `build' a 2-to-2 scattering amplitude $\phi\phi\rightarrow \phi\phi$ by `gluing' two instances of this cubic interaction together, such that the corresponding diagram naively grows like $s^{5}/\Lambda_{5}^{10}$ at large incoming center-of-mass energy-squared $s$. In FP gravity, this expectation has been confirmed by direct computation \cite{Aubert:2003je,Cheung:2016yqr}.

Because the preceding discussion did not depend on the specific details of FP gravity, this power-counting argument \cite{ArkaniHamed:2002sp,ArkaniHamed:2003vb,Hinterbichler:2011tt} suggests that the scattering amplitude of helicity-zero modes in {\it any} massive spin-2 theory should typically grow with energy like ${\cal O}(s^{5})$ -- {\it i.e.} that a theory with a single massive spin-2 particle will be a $\Lambda_5$ theory.  By introducing additional polynomial $h_{\mu\nu}$ interaction terms to the FP Lagrangian of Eq. (\ref{eq:4dgr}), cancellations {\it between} diagrams can occur such that the overall scattering amplitude for $\phi\phi \to \phi \phi$ can be reduced to ${\cal O}(s^3)$, resulting in a $\Lambda_{3}$ theory which is valid to higher energies \cite{deRham:2010kj,vanDam:1970vg,ArkaniHamed:2002sp,ArkaniHamed:2003vb,deRham:2010ik,Cheung:2016yqr,Alberte:2019lnd}. However it is is not possible to raise the scale any further in a theory with a single massive spin-2 particle, even after adding arbitrarily many vector or scalar particles \cite{Bonifacio:2018vzv,Bonifacio:2018aon,Bonifacio:2019mgk}.\footnote{For further details on massive gravity (including bigravity theories, which include a massless graviton alongside a massive spin-2 particle) refer to, for example, \cite{Hinterbichler:2011tt,deRham:2014zqa}. All of the theories described, however, are $\Lambda_3$ theories - or worse.}

\subsection{\label{sec:Ib} Scattering of Massive Spin-2 Particles in Compactified 5D Theories}

In a compactified extra-dimensional theory of gravity, the UV behavior of the four-dimensional KK mode scattering amplitudes must be governed by the high-energy behavior of the underlying theory. For a 5D theory in particular, dimensional analysis implies that the five-dimensional graviton scattering amplitudes must grow like $s^{3/2}/M_{\text{Pl,5D}}^{3}$, where $M_{\text{Pl,5D}}$ is the 5D Planck scale.\footnote{The Feynman amplitude for $2 \to 2$ scattering in 5D has units of (mass)$^{-1}$ and, compared to 4D, an additional factor of energy arises in the 5D partial wave expansion \cite{Soldate:1986mk,Chaichian:1987zt}.} However, this implies that, after compactification and decomposing the 5D graviton field into KK modes, the scattering amplitudes of the massive spin-2 modes must grow {\it slower} than ${\cal O}(s^3)$, which was the slowest growth achievable in theories of a single massive spin-2 particle. Moreover, this must be true even though the compactified theory includes terms like $(\partial^{2}\phi)^{3}/\Lambda_{5}^{5}$ for each massive spin-2 field in its St{\"u}kelberg analysis. 

The motivation of our present work is to reconcile the apparent contradiction between the behavior of the underlying extra-dimensional gravitational theory, and the argument in Sec. \ref{sec:Ia} above which would suggest that massive spin-2 modes have scattering amplitudes which grow like ${\cal O}(s^5)$ (or at best ${\cal O}(s^3)$).
Recently, \cite{Chivukula:2019rij} demonstrated that the (elastic) scattering amplitudes for massive spin-2 KK modes in a compactified 5D theory in fact grow only like ${\cal O}(s)$.\footnote{While each individual KK mode scattering amplitudes grow only like ${\cal O}(s)$, as in the case of compactified Yang-Mills theory \cite{SekharChivukula:2001hz} there are coupled channels of the first $N$ KK modes
whose scattering amplitudes grow like $Ns/M^2_{\text{Pl}}$. Identifying the mass of the highest mode to be of order the maximum energy scale of the EFT, one reproduces the expected $s^{3/2}/M^3_{\text{Pl,5D}}$ behavior of the continuum theory.}. This paper amplifies and extends those results. 

More specifically, a tree-level $(n_{1},n_{2})\rightarrow (n_{3},n_{4})$ KK spin-2 scattering process may proceed via any of several diagrams, which we may
organize into the following sets:
\begin{align}
    \mathcal{M}_{\text{c}} & \equiv \raisebox{-0.45\height}{\sgDklmn{0.20}} \\
    \mathcal{M}_{\text{r}} &\equiv \raisebox{-0.45\height}{\sDklrmn{0.20}} +  \raisebox{-0.45\height}{\tDklrmn{0.20}} +  \raisebox{-0.45\height}{\uDklrmn{0.20}}\nonumber\\
    \mathcal{M}_{j} &\equiv \raisebox{-0.45\height}{\sDkljmn{0.20}} +  \raisebox{-0.45\height}{\tDkljmn{0.20}} +  \raisebox{-0.45\height}{\uDkljmn{0.20}}\nonumber
\end{align}
where subscript ``c" denotes the contact diagram,  ``r" denotes the sum of diagrams mediated by the radion (a scalar mode arising from the 5D metric), and ``$j$" denotes the contribution arising from exchange of a spin-2 KK mode $j$. The external $n_{i}$ label the KK numbers of different massive KK mode excitations. The total tree-level matrix element is thus
\begin{align}
    \mathcal{M} \equiv \mathcal{M}_{\text{c}} + \mathcal{M}_{\text{r}} + \sum_{j=0}^{+\infty} \mathcal{M}_j~,
    \label{eq:matrix-element}
\end{align}
The arguments given in Sec. \ref{sec:Ia} imply that for helicity-zero-polarized external states we expect that
\begin{align}
\mathcal{M}_{\text{c}}\hspace{5 pt}\text{and}\hspace{5 pt}\mathcal{M}_{j} &\sim \mathcal{O}(s^5)~,\\
\mathcal{M}_{\text{r}} &\sim \mathcal{O}(s^3)~.
\end{align}
The calculations reported in \cite{Chivukula:2019rij} demonstrate that, although the individual contributions to the helicity-zero spin-2 KK mode scattering amplitudes do indeed grow as fast as ${\cal O}(s^5)$, there are intricate cancellations between different contributions. These cancellations invalidate the naive power-counting analysis given in Sec. \ref{sec:Ia}, which therefore does not apply to a compactified KK theory with multiple massive gravitons.  Indeed, Ref. \cite{Chivukula:2019rij} has  demonstrated the cancellations both in the case of a toroidal (flat) compactification and in the more phenomenologically interesting case of the Randall-Sundrum (RS1)\cite{Randall:1999ee,Randall:1999vf} model with a warped extra dimension. The cancellations are enforced by a set of sum rules \cite{Chivukula:2019zkt} which interrelate the masses and couplings of the various modes.\footnote{Spin-2 KK mode scattering was previously considered by \cite{Schwartz:2003vj}, which used deconstruction to prove that the KK mode scattering amplitudes grew no faster than ${\cal O}(s^3)$ for a flat extra dimension.}\footnote{Following the appearance of \cite{Chivukula:2019rij}, and as \cite{Chivukula:2019zkt} was being completed, the authors of \cite{Bonifacio:2019ioc} independently proved that the scattering amplitudes of helicity-zero modes of massive spin-2 KK modes in extra-dimensional theories grow only like ${\cal O}(s)$ for compactifications on arbitrary Ricci-flat manifolds. Their proof does not encompass the case of RS1, which is the focus of our work. See Appendix \ref{sec:levelE} a discussion of the relationship between our results and those conjectured by \cite{Bonifacio:2019ioc} in more general situations.} 

In this paper we provide a detailed account of the calculations reported in \cite{Chivukula:2019rij,Chivukula:2019zkt}, specifying all conventions and information needed for building upon our work in the future. We also report substantial new results including a study of the behavior of the scattering amplitudes for arbitrary external polarization, the ``truncation" error which results from the (numerically necessary) limitation of summing over a finite number of KK modes in the intermediate states $j$ above, and a study of the emergence of a dynamical low-energy scale \cite{Randall:1999ee,Randall:1999vf} from the behavior of the scattering amplitudes in RS1.

\subsection{\label{sec:Ic} Guide to the Paper}

Here is an outline of the material presented in the main text and appendices of the paper.

In Sec. \ref{sec:level2} we describe the 5D RS1 model, specify our conventions for the metric, describe the field content, and outline the procedure used for the (5D) weak field expansion. We provide the details of the weak field expansion itself, including specifying the form of the interactions among up to four 5D fields, within Appendix \ref{sec:levelA}. 

In Sec. \ref{sec:level3} we carry out the KK mode expansion, thereby obtaining the 4D particle content of the model, and discuss the form of the interactions among the 4D fields. A general analysis of the properties of the extra-dimensional wavefunctions is given in Appendix \ref{sec:levelB}, and the more detailed description of the 4D interactions is given in Appendix \ref{sec:levelC}.

In Sec. \ref{sec:level4} we begin our analysis of the scattering amplitudes of the massive spin-2 KK modes. Sec. \ref{sec:level4a} gives details of our kinematic and helicity conventions. As described above, the full tree-level scattering amplitudes will (in general) require summing over the exchange of all intermediate states, and we will find that the cancellations needed to reduce the growth of scattering amplitudes from ${\cal O}(s^5)$ to ${\cal O}(s)$ will only completely occur once all states are included. In this section we therefore introduce two ``partial" forms of the scattering amplitudes which will facilitate our discussion of the cancellations [a] truncated matrix elements, which include only exchange of KK modes below some mode number, and [b] the expansion of the matrix elements in powers of energy. In Sec. \ref{sec:level4b} we analyze the case of KK mode scattering in the case in which the curvature of the internal manifold vanishes: the 5D Orbifolded Torus model. 

Sec. \ref{sec:level5} describes in detail the computation of the elastic scattering amplitudes of massive spin-2 KK modes in the RS1 model, for arbitrary values of the curvature of the internal space. For all nonzero curvatures, every KK mode in the infinite tower contributes to each scattering process. We discuss, elaborate upon, and apply the sum rules introduced in \cite{Chivukula:2019zkt}. A new analytic proof for a relation arising from the $s^3$ and $s^2$ sum rules is discussed in Appendix \ref{sec:levelD}, and the relationships of our couplings and sum rules to those conjectured in \cite{Bonifacio:2019ioc} are given in Appendix \ref{sec:levelE}. Finally, Sec. \ref{sec:nonlongitudinal} analyzes the (milder) high-energy behavior of the scattering of nonlongitudinal helicity modes (helicities other than zero) of the massive spin-2 KK modes.

Sec. \ref{sec:level6} presents a detailed numerical analysis of the scattering in the RS1 model. In Sec. \ref{sec:level6a} we demonstrate that the cancellations demonstrated for elastic scattering occur for inelastic scattering channels as well, with the cancellations becoming exact as the number of included intermediate KK modes increases. In Sec. \ref{sec:level6b} we examine the truncation error arising from keeping only a finite number of intermediate KK mode states. We then return, in Sec. \ref{sec:level6c} to the question of the validity of the KK mode EFT. In particular, using the results derived in Appendix \ref{sec:largekrc} for large values of the AdS curvature, we demonstrate directly from the scattering amplitudes that the cutoff scale is proportional to the RS1 emergent scale \cite{ArkaniHamed:2000ds,Rattazzi:2000hs}
\begin{align}
    \Lambda_\pi = M_{\text{Pl}}\, e^{-k\pi r_c}~,
\end{align}
which is related to the location of the IR (TeV) brane \cite{Randall:1999ee,Randall:1999vf}.

Finally, Sec. \ref{sec:level7} contains our conclusions.

\section{\label{sec:level2}The 5D Randall-Sundrum Model and its Weak Field Expansion}

In this section, we describe the 5D RS1 model, specify our conventions for the metric, describe the field content, and outline the procedure used for the (5D) weak field expansion. Appendix \ref{sec:levelA} provides the details of the weak field expansion itself, including specifying the form of the interactions among up to four 5D fields.

\subsection{\label{sec:level2a}General Considerations and Notation}
Our investigation concerns the RS1 model without matter \cite{Randall:1999ee,Randall:1999vf}, in which gravity permeates a five-dimensional (5D) bulk that is bounded by two four-dimensional (4D) branes at $y = 0$ and $y=\pi r_{c}$. The length $r_{c}$ is known as the compactification radius of the extra dimension. This 5D spacetime is parameterized by coordinates $x^{M}\equiv (x^{\mu},y)$, where the $x^{\mu}$ act like our usual 4D spacetime coordinates and $y$ is an extra-dimensional spatial coordinate. By using an orbifold symmetry that associates every 5D point $(x,y)$ with a point $(x,-y)$ and restricting the field content to include only fields even under orbifold parity, the coordinate $y$ can be extended to cover the interval $[-\pi r_{c},+\pi r_{c}]$ and thereby parameterizes a circle of radius $r_{c}$. In this orbifolded set-up, the branes are located at the orbifold fixed points of the extended spacetime. Oftentimes we will use factors of $r_{c}$ to replace the dimensionful variables with dimensionless equivalents, such as replacing $y$ with $\varphi \equiv y/r_{c} \in [-\pi,\pi]$ when it is convenient to do so. 

In general, we will denote a 4D Lorentz index with a lowercase Greek letter such as $\mu =0,1,2,3$, whereas a 5D index will be denoted by an uppercase Latin letter such as $M=0,1,2,3,5$. The 4D flat metric $\eta_{\mu\nu} = \text{Diag}(+1,-1,-1,-1)$ is used to raise/lower 4D indices, e.g. $x_\mu \equiv \eta_{\mu\nu} x^{\nu}$.

The 5D RS1 metric is of the following form:
\begin{align}
    G_{MN} = \matrixbb{w(x,y)g_{\mu\nu}}{0}{0}{-v(x,y)^2}~. \label{GMN general}
\end{align}
This is expressed in coordinates $x^M \equiv (x^\mu,y)$ such that the corresponding invariant interval $ds^2$ equals
\begin{align}
    ds^2 = (G_{MN}) dx^M dx^N = (wg_{\mu\nu}) dx^\mu dx^\nu - (v^2)dy^2~, \label{GMN ds2}
\end{align}
allowing for warping of the transverse four-dimensional space. Meanwhile, the inverse metric equals
\begin{align}
    \tilde{G}^{MN} = \matrixbb{\tilde{g}^{\mu\nu}/w(x,y)}{0}{0}{-1/v(x,y)^2}~,
\end{align}
where we denote the inverse with a tilde (e.g. $\tilde{G} \equiv G^{-1}$ and $\tilde{g} \equiv g^{-1}$). Several quantities related to the spacetime geometry are directly calculable from $G_{MN}$. For instance, the Christoffel symbols, Ricci curvature, and scalar curvature equal
\begin{align}
    \Gamma^{P}_{MN} &= \dfrac{1}{2} \tilde{G}^{PQ}\left(\partial_M G_{NQ} + \partial_N G_{MQ} - \partial_Q G_{MN}\right)~,\nonumber\\
    R_{MN} &= \partial_N \Gamma^{P}_{MP} - \partial_P \Gamma^{P}_{MN} + \Gamma^{P}_{NQ} \Gamma^{Q}_{MP} - \Gamma^{P}_{PQ} \Gamma^{Q}_{MN}~,\nonumber\\
    \RED &= \tilde{G}^{MN} R_{MN}~,
\end{align}
respectively. When going from the metric to the scalar curvature, exactly two derivatives are applied in every term, a fact that proves important when we organize the eventual 4D effective theory.

Integrals over the 5D spacetime are weighted by the invariant volume element $\sqrt{\det G}\hspace{3 pt}d^4x\hspace{3 pt}dy$, which factors into a 4D piece and an extra-dimensional piece:
\begin{align}
    \sqrt{\det G}\hspace{5 pt}d^4x\hspace{3 pt}dy =\left[w^2 \sqrt{-\det g}\hspace{5 pt} d^4x\right]\cdot \left(v\hspace{5 pt} dy\right)
\end{align}
The quantity in square brackets is the 4D projection of the 5D invariant volume element and thereby acts as an effective 4D volume element on a 4D sheet at constant $y$.

The pure gravity RS1 Lagrangian consists of two pieces. The first piece is the Einstein-Hilbert Lagrangian $\LEH$, which is defined as
\begin{align}
    \LEH \equiv \dfrac{2}{\kappa^2} \sqrt{\det G}\hspace{3 pt}\RED = \dfrac{2}{\kappa^2}w^2 v \sqrt{-\det g}\hspace{3 pt}\RED
\end{align}
where $\kappa$ has units of (Energy)$^{-3/2}$. This implies that the 5D Planck mass $\MPlED$ and 5D quantity $\kappa$ are related according to $\kappa^2\MPlED^3=4$. The second piece is the cosmological constant Lagrangian $\LCC$, which  can be written as
\begin{align}
    \LCC &\equiv \dfrac{12}{\kappa^2}kr_c\bigg\{2\sqrt{\det G}\hspace{1 pt} (\partial_\varphi|\varphi|)^2 - w^2\sqrt{-\det g}\hspace{1 pt} (\partial_\varphi^2 |\varphi|)\bigg\}~.
\label{eq:CCLag}
\end{align}
$\LCC$ generates two types of terms: terms proportional to $(\partial_\varphi|\varphi|)^2$ provide a 5D cosmological constant in the bulk whereas terms proportional to $(\partial_\varphi^2 |\varphi|)$ generate tension on the branes (a prime indicates differentiation with respect to $y$, e.g. $f^\prime = \partial_y f$). The coefficients of these terms are chosen so as to guarantee a solution of Einstein's equations that is 4D Poincar\'{e} invariant; namely, the vacuum solution they imply equals
\begin{align}
\etaRS \equiv \matrixbb{e^{-2k|y|} \eta_{\mu\nu}}{0}{0}{-1} \label{RS vacuum}
\end{align}
as expressed in coordinates $x^{M} = (x^{\mu}, y)$, where $k$ is the nonnegative warping parameter and has units of $(\text{Energy})^{+1}$. 

Combining $\LEH$ and $\LCC$ yields $\LED$, the Lagrangian of the matter-free 5D theory:
\begin{align}
    \LED = \LEH + \LCC~. \label{L5Dx}
\end{align}
The 4D effective theory is then defined from the action:
\begin{align}
    S = \int d^4x\hspace{2 pt}\bigg[dy\hspace{5 pt}\LED\bigg] \equiv \int d^4x\hspace{5 pt}\LDDeff \label{L4Deff Def}
\end{align}
i.e. the Lagrangian $\LDDeff$ is obtained by integrating $\LED$ across the extra dimension. The form of  $\LCC$ specifically prevents a nonzero 4D cosmological constant in the effective theory described by $\LDDeff$.

This gravitational Lagrangian will be expanded in the weak field approximation as a perturbation series in fields in order to obtain particle interactions and calculate matrix elements. Upon expanding Eq. \eqref{L5Dx}, each term will contain either two spatial derivatives $\partial_{\mu}$ or two extra-dimensional derivatives $\partial_{y}$. However, certain terms in the expansion of Eq. \eqref{L5Dx} will contain instances of $\partial^{2}_{y}$ which obscure the coupling structure of the 4D theory. We can ensure no two extra-dimensional derivatives ever act on the same field in the expansion by adding a total derivative to Eq. \eqref{L5Dx}. Specifically, we can eliminate all instances of $\partial^{2}_{y}$ in the expanded Lagrangian without changing the physics by adding the total derivative\footnote{The orbifold boundary conditions we employ will require all normal derivatives of the metric to vanish on the branes, and hence this term is purely for convenience and does not change the physics.}
\begin{align}
    \Delta\mathcal{L} = \dfrac{2}{\kappa^2}\partial_{y}\left[\dfrac{w^2}{\sqrt{v}} \sqrt{-\det g}\left(\ltr \tilde{g} g^\prime \rtr + \dfrac{(\partial_{y} w)}{w}\right)\right]
\end{align}
where a prime indicates differentiation with respect to $y$ and twice-squared bracket notation indicates a cyclic contraction of Lorentz indices, e.g. $\ltr \hat{h}^\prime \hat{h}^\prime \rtr \equiv (\partial_{y} h^{\mu\nu})(\partial_{y} h_{\nu\mu})$. Therefore, in practice we use
\begin{align}
    \LED = \LEH + \LCC + \Delta \mathcal{L}~. \label{L5Dy}
\end{align}
Of course, in order to weak field expand this Lagrangian, we must first establish the relevant fields.

\subsection{\label{sec:level2b}The Weak Field Expansion}

Now that we have a generic path from the 5D metric $G$ to the 4D effective Lagrangian $\LDDeff$, we may discuss the field content of the RS1 theory.  The gravitational particle content is obtained by perturbing the vacuum with field-dependent functions. To ensure correct units and assist the Lagrangian's eventual weak field expansion, we will introduce the fields alongside an explicit factor of $\kappa$. 
We choose to utilize the Einstein frame parameterization \cite{Appelquist:1983vs,Charmousis:1999rg,Wehus:2002se,Rattazzi:2003ea}, which eliminates mixing between the scalar and tensor modes -- and ultimately yields a canonically-normalized 4D effective Lagrangian. In this parameterization, $w$ and $v$ in Eq. (\ref{GMN general}) may be written as
\begin{align}
    w(x,y) &= e^{-2(k|y|+\hat{u})} \hspace{30 pt}v(x,y)= 1+2\hat{u}
    \label{eq:weakfieldi}
\end{align}
where $\hat{u}(x,y)$, as we will soon see, is related to the 5D radion field.  Furthermore, we identify $g_{\mu\nu}$ as weakly perturbed from the flat value $\eta_{\mu\nu}$, e.g.
\begin{align}
    \eta_{\mu\nu} \hspace{15 pt}\mapsto\hspace{15 pt} g_{\mu\nu} \equiv \eta_{\mu\nu} + \kappa \hat{h}_{\mu\nu}~,  \label{EtaPerturbRS}
\end{align}
where the symmetric tensor field $\hat{h}_{\mu\nu}(x,y)$ contains the spin-2 modes. The metric is then
\begin{align}
    \GRS = \matrixbb{e^{-2(k|y|+\hat{u})}(\eta_{\mu\nu} + \kappa \hat{h}_{\mu\nu})}{0}{0}{-(1+2\hat{u})^2}~. \label{GMNRS}
\end{align}

The 5D radion, $\hat{r}(x,y)$ is related to $\hat{u}(x,y)$ via
\begin{align}
    \hat{u}(x,y) \equiv \dfrac{\kappa \hspace{2 pt} \hat{r}(x,y)}{2\sqrt{6}}\hspace{2 pt}e^{+k(2|y|-\pi r_c)}~.
\end{align}
Unlike $\hat{h}_{\mu\nu}$, the 5D radion field can be made $y$-independent via a gauge transformation \cite{Callin:2004zm}, and so we choose $\hat{r}(x,y)=\hat{r}(x)$. 

In some 5D models, the off-diagonal elements $G^{(\text{RS})}_{5\mu}$ and $G^{(\text{RS})}_{\mu5}$ give rise to an orbifold-odd graviphoton excitation which can also be made $y$-independent via gauge symmetries \cite{Callin:2004zm}; however, the RS1 scenario possesses an orbifold symmetry which removes this degree of freedom and ensures $G^{(\text{RS})}_{\mu5} = G^{(\text{RS})}_{5\mu} = 0$. Meanwhile, the graviton and radion fields must be even functions of $y$ to ensure the interval $ds^2$ described by $\GRS$ is invariant under the orbifold transformation. Both of these 5D fields have units of $(\text{Energy})^{+3/2}$.

As outlined in the previous subsection, the metric $\GRS$ determines a Lagrangian $\LRS \equiv \LEHRS + \LCCRS + \Delta \mathcal{L}^{(\text{RS})}$. We calculate $\LRS$ as a perturbation series in $\kappa$ and thereby obtain its weak field expansion (WFE). In particular, because we are ultimately concerned with 2-to-2 tree-level scattering of massive spin-2 states, we require several of the three- and four-particle interactions present in the $\mathcal{O}(\kappa^2)$ WFE $\LRS$. The details of this procedure and its results are summarized in Appendix \ref{AppendixWFE}.

\section{\label{sec:level3}The 4D Effective Theory}

In this section, we carry out the KK mode expansion, thereby obtaining the 4D particle content of the model, and discuss the form of the interactions among the 4D fields. A general analysis of the properties of the extra-dimensional wavefunctions is given in Appendix \ref{sec:levelB}, and the more detailed description of the 4D interactions is given in Appendix \ref{sec:levelC}.

\subsection{\label{sec:level3a}4D Particle Content}
The 4D particle content is determined by employing the KK decomposition ansatz \cite{Kaluza:1984ws,Klein:1926tv,Goldberger:1999wh}:
\begin{align}
    \hat{h}_{\mu\nu}(x,y) &= \dfrac{1}{\sqrt{\pi r_c}} \sum_{n=0}^{+\infty} \hat{h}^{(n)}_{\mu\nu}(x)\psi_n(\varphi)~, \nonumber\\
    \hat{r}(x) &= \dfrac{1}{\sqrt{\pi r_c}}
   \hat{r}^{(0)}(x) \psi_0~,\label{KKreduce}
\end{align}
where we recall that $\varphi=y/r_c$.
The operators $\hat{h}^{(n)}_{\mu\nu}$ and $\hat{r}^{(0)}$ are 4D spin-2 and spin-0 fields respectively, while each $\psi_{n}$ is a wavefunction which solves the following Sturm-Liouville equation
\begin{align}
    \partial_{\varphi}\left[\varepsilon^{-4} (\partial_{\varphi} \psi_{n})\right] = - \mu_{n}^{2} \varepsilon^{-2} \psi_{n} \label{SLeqx}
\end{align}
subject to the boundary condition $(\partial_{\varphi} \psi_{n})=0$ at $\varphi = 0$ and $\pi$, where $\varepsilon \equiv e^{k |y|}=e^{kr_{c}|\varphi|}$ \cite{Goldberger:1999wh}. Up to normalization, there exists a unique solution $\psi_{n}$ per eigenvalue $\mu_{n}$, each of which we index with a discrete KK number $n \in \{0,1,2,\cdots\}$ such that $\mu_{0}=0 < \mu_{1} < \mu_{2} < \cdots$. Given a KK number $n$, the quantity $\mu_{n}$ and wavefunction $\psi_{n}(\varphi)$ are entirely determined by the value of the unitless nonnegative combination $kr_{c}$. Additional details about this ansatz (including explicit expressions for the wavefunctions and their derivation in a slightly more general circumstance) comprise Appendix B. For now, we note that with proper normalization the $\psi_{n}$ satisfy two convenient orthonormality conditions:
\begin{align}
    \dfrac{1}{\pi}\int_{-\pi}^{+\pi} d\varphi\hspace{5 pt} \varepsilon^{-2}\psi_{m}\psi_{n} &= \delta_{m,n}~,\label{onAx}\\
    \dfrac{1}{\pi}\int_{-\pi}^{+\pi} d\varphi\hspace{5 pt} \varepsilon^{-4}(\partial_\varphi\psi_m)(\partial_\varphi\psi_n) &= \mu_{n}^{2}\delta_{m,n}~.\label{onBx}
\end{align}
Furthermore, the $\{\psi_{n}\}$ form a complete set, such that the following completeness relation holds:
\begin{align}
    \delta(\varphi_{2}-\varphi_{1}) = \sum_{j=0}^{+\infty} \dfrac{1}{\pi} \varepsilon^{-2} \psi_{j}(\varphi_{1})\s \psi_{j}(\varphi_{2})~.
    \label{eq:completeness}
\end{align}

The KK number $n=0$ corresponds to $\mu_{n} = 0$, for which Eq. $\eqref{SLeqx}$ admits a flat wavefunction solution $\psi_0$ corresponding to the massless 4D graviton. Upon normalization via Eq. \eqref{onAx}, this wavefunction is
\begin{align}
    \psi_0 = \sqrt{\dfrac{\pi k r_c}{1-e^{-2\pi kr_c}}}
\end{align}
up to a phase that we set to $+1$ by convention. This is the wavefunction that Eq. $\eqref{KKreduce}$ associates with the fields $\hat{h}^{(0)}$ and $\hat{r}^{(0)}$. The lack of higher modes in the KK decomposition of $\hat{r}$ reflects its $y$-independence. In this sense, choosing to associate $\psi_{0}$ with $\hat{r}^{(0)}$ in Eq. $\eqref{KKreduce}$ is merely done for convenience.

Before employing KK decomposition to compute the interactions of the 4D states, we apply the ansatz to the simpler quadratic terms. In particular, the 5D quadratic graviton Lagrangian equals (from Appendix \ref{AppendixWFE})
\begin{align}
    \LRSf{hh} = \hspace{3 pt} \varepsilon^{-2}\s\LA{hh} + \varepsilon^{-4}\s\LB{hh}~, \label{L5DRS hh}
\end{align}
where
\begin{align}
     \LA{hh} =& - \hat{h}_{\mu\nu}(\partial^\mu \partial^\nu \hat{h}) + \hat{h}_{\mu\nu} (\partial^\mu \partial_\rho \hat{h}^{\rho\nu})\nonumber\\
     &- \dfrac{1}{2} \hat{h}_{\mu\nu}(\square \hat{h}^{\mu\nu}) + \dfrac{1}{2} \hat{h}(\square \hat{h})~,\label{5DhhA}\\
    \LB{hh} =& - \dfrac{1}{2} \ltr \hat{h}^\prime \hat{h}^\prime \rtr + \dfrac{1}{2} \ltr \hat{h}^\prime \rtr^2~,\label{5DhhB}
\end{align}
where we recall that a prime indicates differentiation with respect to $y$ and a twice-squared bracket indicates a cyclic contraction of Lorentz indices.

Similarly, the quadratic 5D radion Lagrangian equals,
\begin{align}
    \LRSf{rr} = e^{-2 \pi kr_{c}}\s\varepsilon^{+2}\s \LA{rr}~,\label{Lrr RS}
\end{align}
where
\begin{align}
    \LA{rr} = \dfrac{1}{2}(\partial_\mu\hat{r})(\partial^\mu \hat{r})~.\label{LrrA RS}
\end{align}
To obtain the 4D effective equivalents of the above 5D expressions, we must integrate over the extra dimension and employ the KK decomposition ansatz.

First, the graviton: the first term in \eqref{L5DRS hh} becomes
\begin{align}
    \LAeff{hh} &\equiv \int_{-\pi r_c}^{+\pi r_c}dy\hspace{5 pt} \varepsilon^{-2}\s\LA{hh}\\
    &\hspace{-25 pt}=\int_{-\pi r_c}^{+\pi r_c} dy\hspace{5 pt} \varepsilon^{-2}\left[-\hat{h}_{\mu\nu} (\partial^\mu \partial^\nu \hat{h})\right.\nonumber\\
    &\left. + \hat{h}_{\mu\nu} (\partial^\mu \partial_\rho \hat{h}^{\rho\nu})- \dfrac{1}{2} \hat{h}_{\mu\nu} (\square \hat{h}^{\mu\nu}) + \dfrac{1}{2} \hat{h} (\square\hat{h})\right]\nonumber\\
    &\hspace{-25 pt}= \sum_{m,n=0}^{+\infty} \left[-\hat{h}^{(m)}_{\mu\nu} (\partial^\mu \partial^\nu \hat{h}^{(n)}) + \hat{h}^{(m)}_{\mu\nu} (\partial^\mu \partial_\rho \hat{h}^{(n)\rho\nu})\right.\nonumber\\
    &\left.- \dfrac{1}{2} \hat{h}^{(m)}_{\mu\nu} (\square \hat{h}^{(n)\mu\nu}) + \dfrac{1}{2} \hat{h}^{(m)} (\square\hat{h}^{(n)})\right]\nonumber\\
    &\times \dfrac{1}{\pi}\int_{-\pi}^{+\pi} d\varphi\hspace{5 pt} \varepsilon^{-2}\psi_{m} \psi_{n} ~,\nonumber
\end{align}
whereas its second term becomes
\begin{align}
    \LBeff{hh} &\equiv \int_{-\pi r_c}^{+\pi r_c} dy\hspace{5 pt}\varepsilon^{-4}\s\LB{hh}\\
    &\hspace{-25 pt}= \int_{-\pi r_c}^{+\pi r_c} dy\hspace{5 pt} \varepsilon^{-4}\left[- \dfrac{1}{2}\ltr \hat{h}^\prime \hat{h}^\prime\rtr + \dfrac{1}{2} \ltr \hat{h}^\prime\rtr^2\right]\nonumber\\
    &\hspace{-25 pt}= \sum_{m,n=0}^{+\infty}\left[- \dfrac{1}{2}\ltr \hat{h}^{(m)}\hat{h}^{(n)}\rtr +\dfrac{1}{2} \ltr \hat{h}^{(m)}\rtr \ltr \hat{h}^{(n)} \rtr \right]\nonumber\\
    &\times\dfrac{1}{\pi r_c^2} \int_{-\pi}^{+\pi} d\varphi\hspace{5 pt} \varepsilon^{-4}(\partial_\varphi \psi_m)(\partial_\varphi \psi_n)~.\nonumber
\end{align}
These are simplified via the orthonormality relations Eqs. \eqref{onAx} and \eqref{onBx}, such that the 4D effective Lagrangian resulting from $\LRSf{hh}$ equals
\begin{align}
    \LRSfeff{hh} &= \LAeff{hh}+ \LBeff{hh}\\
    &\hspace{-25 pt} = \mathcal{L}_{\text{Kin}}^{(S=2)}(\hat{h}^{(0)})+ \sum_{n=1}^{+\infty} \mathcal{L}_{\text{FP}}(m_n,\hat{h}^{(n)})~,\nonumber
\end{align}
wherein $m_{n}\equiv \mu_{n}/r_{c}$. Therefore, KK decomposition of the 5D field $\hat{h}_{\mu\nu}$ results in the following 4D particle content: a single massless spin-2 mode $\hat{h}^{(0)}$, and countably many massive spin-2 modes $\hat{h}^{(n)}$ with $n\in\{1,2,\cdots\}$ (each having a corresponding Fierz-Pauli mass term). The zero mode $\hat{h}^{(0)}$ is consistent with the usual 4D graviton, and will be identified as such. The 4D graviton has dimensionful coupling constant $\kDD = 2/M_{\text{Pl}} = \psi_{0} \kappa/\sqrt{\pi r_c}$ where $M_{\text{Pl}}$ is the reduced 4D Planck mass. In terms of the reduced 4D Planck mass, the full 4D Planck mass equals $\sqrt{8\pi}M_{\text{Pl}}$.

Meanwhile, the 4D effective equivalent of $\mathcal{L}^{(\text{RS})}_{rr}$ from Eq. \eqref{Lrr RS} equals:
\begin{align}
    \LRSfeff{rr} &= \int_{-\pi r_c}^{+\pi r_c} dy\hspace{5 pt} \LRSf{rr}\\
    &\hspace{-25 pt}= \int_{-\pi r_c}^{+\pi r_c} dy\hspace{5 pt} e^{-2\pi kr_{c}}\s\varepsilon^{+2}\left[\dfrac{1}{2}  (\partial_\mu \hat{r})(\partial^\mu\hat{r})\right]\nonumber \\ &\hspace{-25 pt}=\dfrac{1}{2}(\partial_\mu \hat{r}^{(0)})(\partial^\mu \hat{r}^{(0)}) \cdot \dfrac{{\psi_{0}}^2}{\pi r_c} \int_{-\pi r_c}^{+\pi r_c} dy\hspace{5 pt} e^{+2k(|y|-\pi r_c)} \nonumber \\
    &\hspace{-25 pt}= \mathcal{L}^{S=0}_{\text{Kin}}(\hat{r}^{(0)})~.\nonumber
\end{align}
Therefore, KK decomposing the 5D radion yields only a single massless spin-0 mode $\hat{r}^{(0)}$. Like its 5D progenitor, this 4D state is called the radion. Note the exponential factor in Eq. \eqref{Lrr RS} is inconsistent with the orthonormality equation \eqref{onAx}, so we had to calculate the integral explicitly. Thankfully, the $y$-independent radion must possess a flat extra-dimensional wavefunction and so the exponential factor can at most affect its normalization. This would not be the case if the radion's $y$-dependence could not be gauged away.

The RS1 model has three independent parameters according to the above construction: the extra-dimensional radius $r_{c}$, the warping parameter $k$, and the 5D coupling strength $\kappa$. However, we use a more convenient set of independent parameters in practice: the unitless extra-dimensional combination $kr_{c}$, the mass $m_{1}$ of the first massive KK mode $\hat{h}^{(1)}$, and the reduced 4D Planck mass $M_{\text{Pl}}$. These sets are related according to the following relations:
\begin{align}
    m_{1} &\equiv\dfrac{1}{r_{c}}\mu_{1}(kr_{c})\hspace{10 pt}\text{via Eq. \eqref{SLeqx}}~,\\
    M_{\text{Pl}} &\equiv \dfrac{2}{\kappa\sqrt{k}} \sqrt{1 - e^{-2kr_{c}\pi}}~.
\end{align}
In our analysis, we will consider $M_{\text{Pl}}$ and $m_1$ fixed, and vary $k r_c$. 
When explicit values are used, we will choose $kr_{c} \in [0,10]$, $m_{1} = 1\text{ TeV}$, and $M_{\text{Pl}}= 2.435\times 10^{15}\text{ TeV}$.

\subsection{\label{sec:level3b}Beyond Quadratic Order}
Deriving the quadratic terms proceeded so cleanly in part because all wavefunctions with a nonzero KK number occur in pairs and are thus subject to orthonormality relations. Such simplifications are seldom possible when dealing with a product of three or more 5D graviton fields, and instead the integrals must be dealt with explicitly. Consequently, the RS1 model possesses many nonzero triple couplings and calculating a matrix element for $2$-to-$2$ scattering of massive KK modes typically requires a sum over infinitely many diagrams, each of which is mediated by a different massive KK mode and contains various products of these overlap integrals.

Keeping this in mind, consider all terms in the weak field expanded Lagrangian $\LRS$ that have exactly $H$ spin-2 fields and no radion fields. After KK decomposition, terms with two 4D derivatives (designated as A-Type) are proportional to overlap integrals
\begin{align}
    \an{\vec{n}} \equiv \dfrac{1}{\pi} \int_{-\pi}^{+\pi}d\varphi\hspace{5 pt} \varepsilon^{-2} \prod_{i=1}^{H} \psi_{n_i}~,
    \label{eq:define-a}
\end{align}
and those containing two extra-dimensional derivatives (designated as B-Type) are proportional to integrals
\begin{align}
    \bn{\vec{n}} \equiv \dfrac{1}{\pi} \int_{-\pi}^{+\pi}d\varphi\hspace{5 pt} \varepsilon^{-4} (\partial_\varphi \psi_{n_1})(\partial_\varphi \psi_{n_2})\prod_{i=3}^{H} \psi_{n_i}~,
    \label{eq:define-b}
\end{align}
where $\vec{n} = (n_1\cdots n_H)$ are the KK numbers of the relevant spin-2 fields.\footnote{Every term in $\mathcal{L}^{(\text{RS})}_{\text{5D}}$ contains exactly two derivatives. Because even-spin fields carry an even number of Lorentz indices and the Lagrangian is a Lorentz scalar, those two derivatives must either both be 4D derivatives or both be extra-dimensional derivatives, no matter how many spin-2 or spin-0 fields are present. Therefore, A-Type and B-Type couplings exhaust the possible wavefunction integrals encountered in the RS1 model.} These integrals are unitless and entirely determined by the value of $kr_{c}$. Note that $a_{\vec{n}}$ is fully symmetric in all KK numbers, whereas $b_{\vec{n}}$ is symmetric in the first pair and remaining KK numbers separately. Pictorially, we indicate the vertices associated with these couplings as small filled circles attached to the appropriate number of particle lines,
\begin{align}
    \raisebox{-0.45\height}{\verHHH{0.20}} \hspace{10 pt}&\supset\hspace{10 pt} a_{n_{1}n_{2}n_{3}}\hspace{10 pt} b_{\mathcal{P}[n_{1},n_{2},n_{3}]}\\
    \raisebox{-0.45\height}{\verHHHH{0.20}} \hspace{10 pt}&\supset\hspace{10 pt} a_{n_{1}n_{2}n_{3}n_{4}}\hspace{10 pt} b_{\mathcal{P}[n_{1},n_{2},n_{3},n_{4}]}
\end{align}
where overlapping straight and wavy lines indicate a spin-2 particle, and $\mathcal{P}$ indicates that all permutations of its arguments should be considered. If we set $n_{3}=0$ in the triple spin-2 coupling, the corresponding wavefunction $\psi_{0}$ is flat; either $\psi_{0}$ is differentiated (in which case the integral vanishes) or it can be factored out of the $y$-integral thereby allowing us to invoke the wavefunction orthogonality relations on the remaining wavefunction pair. In this way, the triple spin-2 couplings imply that the massless 4D graviton couples diagonally to the other spin-2 states, as required by 4D general covariance:
\begin{align}
    \ahhh{n_1}{n_2}{0} &= \psi_0\s \delta_{n_1,n_2}~,\\
    \bhhh{n_1}{n_2}{0} &= \mu_{n_1}^{2} \psi_{0} \s \delta_{n_1,n_2}~,\nonumber\\
    \bhhh{0}{n_1}{n_2} &= 0~.\nonumber
\end{align}
The Sturm-Liouville problem that defines the wavefunctions $\{\psi_{n}\}$ also relates various A-Type and B-Type couplings to each other; we will explore this further in Sec. \ref{sec:level5}.

When calculating matrix elements of massive KK mode scattering, we must also consider radion-mediated diagrams. These involve coupling a radion to a pair of spin-2 states, which requires the integral
\begin{align}
    \bhhr{n_{1}}{n_{2}} \equiv \dfrac{\psi_{0}}{\pi}e^{-kr_c\pi}\int_{-\pi}^{+\pi} d\varphi\hspace{5 pt}\varepsilon^{-2}(\partial_\varphi\psi_{n_1})(\partial_\varphi\psi_{n_2})~.
\end{align}
This is defined analogously to the pure spin-2 couplings in the sense that we indicate the role of the radion wavefunction within the coupling (e.g. differentiated vs. undifferentiated) through the placement of a pseudo-KK index ``r". The RS1 model lacks an analogous A-Type radion coupling and the $\brhh{n_1}{n_2}$ coupling vanishes for the same reason that the $\bhhh{0}{n_{1}}{n_{2}}$ coupling vanished. Note that the exponential factor in the integrand of $\bhhr{n_1}{n_2}$ prevents use of the orthonormality relations; therefore, the radion typically couples nondiagonally to massive spin-2 modes. Pictorially,
\begin{align}
    \raisebox{-0.45\height}{\verHHR{0.20}}\hspace{10 pt} &\supset\hspace{10 pt} b_{n_{1}n_{2}r}
\end{align}
where unadorned straight lines indicate a radion.

Appendix \ref{AppendixKKResults} describes how the detailed vertices between 4D particles are derived from the 5D theory and summarizes the relevant interactions. These interactions form the building blocks of our matrix elements, which we turn to next.

\section{\label{sec:level4} Elastic Scattering in the 5D Orbifolded Torus Model}

In this section, we begin our analysis of the scattering amplitudes of the massive spin-2 KK modes. Section \ref{sec:level4a} gives details of our kinematic and helicity conventions. As described above, the full tree-level scattering amplitudes will (in general) require summing over the exchange of all intermediate states, and we will find that the cancellations needed to reduce the growth of scattering amplitudes from ${\cal O}(s^5)$ to ${\cal O}(s)$ will only completely occur once all states are included. In this section we therefore introduce two ``partial" forms of the scattering amplitudes which will facilitate our discussion of the cancellations [a] truncated matrix elements, which include only exchange of KK modes below some mode number, and [b] the expansion of the matrix elements in powers of energy. In Sec. \ref{sec:level4b} we analyze the case of KK mode scattering in the case in which the curvature of the internal manifold vanishes: the 5D Orbifolded Torus model. 

\subsection{\label{sec:level4a}Preliminaries}
The preceding sections (and related appendices) described how to determine the vertices relevant to tree-level $2$-to-$2$ scattering of massive spin-2 helicity eigenstates in the center-of-momentum frame. This section calculates and analyzes those matrix elements. For scattering of nonzero KK modes $(n_{1},n_{2})\rightarrow(n_{3},n_{4})$ with helicities $(\lambda_{1},\lambda_{2})\rightarrow(\lambda_{3},\lambda_{4})$, we choose coordinates such that the initial particle pair have 4-momenta satisfying
\begin{align}
    p^\mu_{1} &= (E_{1}, + |\vec{p}_{i}| \hat{z})\hspace{25 pt}p_{1}^2 = m_{n_{1}}^2\\
    p^\mu_{2} &= (E_{2}, - |\vec{p}_{i}| \hat{z})\hspace{25 pt}p_{2}^2 = m_{n_{2}}^2
\end{align}
and the final particle pair have 4-momenta satisfying
\begin{align}
    p^\mu_{3} &= (E_{3}, + \vec{p}_f)\hspace{25 pt}p_{3}^2 = m_{n_{3}}^2\\
    p^\mu_{4} &= (E_{4}, - \vec{p}_f)\hspace{25 pt}p_{4}^2 = m_{n_{4}}^2
\end{align}
where $\vec{p}_f \equiv |\vec{p}_f| (\sin\theta\cos\phi,\sin\theta\sin\phi,\cos\theta)$. That is, the initial pair approach along the $z$-axis and the final pair separate along the line described by the angles $(\theta,\phi)$. The helicity-$\lambda$ spin-2 polarization tensor $\epsilon_{\lambda}^{\mu\nu}(p)$ for a particle with 4-momentum $p$ is defined according to
\begin{align}
    \epsilon_{\pm 2}^{\mu\nu} &= \epsilon_{\pm1}^{\mu} \epsilon_{\pm 1}^{\nu}~,\\
    \epsilon_{\pm 1}^{\mu\nu} &= \dfrac{1}{\sqrt{2}}\left[\epsilon_{\pm 1}^{\mu}\epsilon_{0}^{\nu} + \epsilon_{0}^{\mu} \epsilon_{\pm 1}^{\nu}\right]~\label{ep21}\\
    \epsilon_{0}^{\mu\nu} &= \dfrac{1}{\sqrt{6}}\bigg[\epsilon^{\mu}_{+1} \epsilon^{\nu}_{-1} + \epsilon^{\mu}_{-1}\epsilon^{\nu}_{+1} + 2\epsilon^{\mu}_{0}\epsilon^{\nu}_{0}\bigg]~,
\end{align}
where $\epsilon_{s}^\mu$ are the (particle-direction dependent) spin-1 polarization vectors
\begin{align}
    \epsilon_{\pm 1}^{\mu} &= \pm\dfrac{e^{\pm i\phi}}{\sqrt{2}}\bigg(0,- c_{\theta}c_{\phi} \pm i s_{\phi}, -c_{\theta}s_{\phi} \mp ic_{\phi},s_{\theta}\bigg)~,\\
    \epsilon_{0}^{\mu} &= \dfrac{E}{m}\bigg(\sqrt{1-\dfrac{m^2}{E^2}}, \hat{p}\bigg)~,\label{ep10}
\end{align}
$(c_{x},s_{x})\equiv(\cos x,\sin x)$, and $\hat{p}$ is a unit vector in the direction of the momentum \cite{Han:1998sg}. We use the Jacob-Wick second particle convention, which adds a phase $(-1)^{\lambda}$ to $\epsilon^{\mu\nu}_{\lambda}$ when the polarization tensor describes $h^{(n_{2})}$ or $h^{(n_{4})}$  \cite{Jacob:1959at}. Due to rotational invariance, we may set the azimuthal angle $\phi$ to $0$ without loss of generality. Meanwhile, the propagators for virtual spin-0 and spin-2 particles of mass $M$ and 4-momentum $P$ are, respectively,
\begin{align}
    \raisebox{2 pt}{\propSc{0.5}}\hspace{18 pt} &= \dfrac{i}{P^2-M^2}\\
    \raisebox{-0.35\height}{\propGr{0.5}} &= \dfrac{i B^{\mu\nu,\rho\sigma}}{P^2-M^2}
\end{align}
where we use the spin-2 propagator convention \cite{Han:1998sg}
\begin{align}
    &B^{\mu\nu,\rho\sigma} \equiv \dfrac{1}{2}\bigg[\overline{B}^{\mu\rho}\overline{B}^{\nu\sigma} + \overline{B}^{\nu\rho}\overline{B}^{\mu\sigma} - \dfrac{1}{3}(2+\delta_{0,M}) \overline{B}^{\mu\nu}\overline{B}^{\rho\sigma}\bigg]\nonumber\\
    &\left.\overline{B}^{\alpha\beta}\right|_{M=0} = \eta^{\alpha\beta}\hspace{25 pt}\left.\overline{B}^{\alpha\beta}\right|_{M\neq 0} \equiv \eta^{\alpha\beta} - \dfrac{P^\alpha P^\beta}{M^2}
\end{align}
and $\eta^{\mu\nu} = \text{Diag}(+1,-1,-1,-1)$ is the flat 4D metric. The massless spin-2 propagator is derived in the de Donder gauge by adding a gauge-fixing term $\mathcal{L}_{gf}=-( \partial^{\mu}\hat{h}^{(0)}_{\mu\nu} - \tfrac{1}{2} \partial_{\nu} \ltr \hat{h}^{(0)} \rtr )^{2}$ to the Lagrangian. The Mandelstam variable $s \equiv (p_{1}+p_{2})^2 = (E_{1}+E_{2})^2$ provides a convenient frame-invariant measure of collision energy. The minimum value of $s$ that is kinematically allowed equals $s_{\text{min}}\equiv \max[(m_{n_{1}}+m_{n_{2}})^{2},(m_{n_{3}}+m_{n_{4}})^{2}]$. When dealing with explicit full matrix elements, we will replace $s\in[s_{\text{min}},+\infty)$ with the unitless $\mathfrak{s} \in [0,+\infty)$ which is defined according to $s \equiv s_{\text{min}}(1+\mathfrak{s})$.

As discussed in Sec. \ref{sec:Ib}, any tree-level massive spin-2 scattering amplitude can be written as
\begin{align}
    \mathcal{M} \equiv \mathcal{M}_{\text{c}} + \mathcal{M}_{\text{r}} + \sum_{j=0}^{+\infty} \mathcal{M}_j~,
    \label{eq:Mtot-def}
\end{align}
where we separate the contributions arising from contact interactions, radion exchange, and a sum over the exchanged intermediate KK states $j$ (and where ``0" labels the massless graviton). In practice, this sum cannot be completed in entirety and must instead be truncated. Therefore, we also define the truncated matrix element
\begin{align}
    \mathcal{M}^{[N]} \equiv \mathcal{M}_{\text{c}} + \mathcal{M}_{\text{r}} + \sum_{j=0}^{N} \mathcal{M}_j~,
    \label{eq:M-N-def}
\end{align}
which includes the contact diagram, the radion-mediated diagrams, and all KK mode-mediated diagrams with intermediate KK number less than or equal to $N$.

We are concerned with the high-energy behavior of these matrix elements, and will therefore examine the high-energy behavior of each of the contributions discussed.
Because the polarization tensors $\epsilon^{\mu\nu}_{\pm1}$ introduce odd powers of energy, $\sqrt{s}$ is a more appropriate expansion parameter for generic helicity combinations. Thus, we series expand the diagrams and total matrix element in $\sqrt{s}$ and label the coefficients like so:
\begin{align}
    \mathcal{M}(s,\theta) \equiv \sum_{\sigma \in \tfrac{1}{2}\mathbb{Z}} \overline{\mathcal{M}}^{(\sigma)}(\theta)\cdot s^{\sigma}
\end{align}
and define $\mathcal{M}^{(\sigma)}\equiv \overline{\mathcal{M}}^{(\sigma)}\cdot s^{\sigma}$. In what follows, we demonstrate that $\mathcal{M}^{(\sigma)}$ vanishes for $\sigma > 1$ regardless of helicity combination and we present the residual linear term in $s$ for helicity-zero elastic scattering. However, before we tackle the generic RS1 theory, let us start by analyzing a simpler case: $\LRS$ in the limit of no warping.

\subsection{\label{sec:level4b}The 5D Orbifolded Torus}

Before investigating scattering amplitudes in the general RS1 model, we consider a special case in which the internal space is flat. Taking the limit of the RS1 metric \eqref{GMNRS} as $k r_c$ vanishes, while simultaneously maintaining a nonzero finite first mass $m_1$ (or, equivalently, a nonzero finite $r_c$), yields the 5D Orbifolded Torus (5DOT) model. The 5DOT metric lacks explicit dependence on $y$,
\begin{align}
    \GEDOT = \matrixbb{e^{\tfrac{-\kappa\hat{r}}{\sqrt{6}}}(\eta_{\mu\nu} + \kappa \hat{h}_{\mu\nu})}{0}{0}{-\left(1+\tfrac{\kappa\hat{r}}{\sqrt{6}}\right)^2}~, \label{GMN5DOT}
\end{align}
and as $kr_{c}\rightarrow 0$ the massive wavefunctions go from exponentially-distorted Bessel functions to simple cosines:
\begin{align}
    \psi_n = \begin{cases}
        \psi_0 = \tfrac{1}{\sqrt{2}}\\
        \psi_n = -\cos(n|\varphi|)\hspace{15 pt} 0<n\in\mathbb{Z}
    \end{cases}\label{RSwfxn}
\end{align}
with masses given by $\mu_{n} = m_{n} r_{c} = n$ and 5D gravitational coupling $\kappa = \sqrt{2\pi r_c}\hspace{1 pt}\kDD = \sqrt{8\pi r_{c}}/M_{\text{Pl}}$. In the absence of warp factors, the radion now couples diagonally and spin-2 interactions display discrete KK momentum conservation. Explicitly, an $H$-point vertex $\hat{h}^{(n_1)}\cdots \hat{h}^{(n_H)}$ in the 4D effective 5DOT model has vanishing coupling if there exists no choice of $c_i \in \{-1,+1\}$ such that $c_1 n_1 +\cdots + c_H n_H =0$. For example, the three-point couplings $\ahhh{n_{1}}{n_{2}}{n_{3}}$ and $\bhhh{n_{1}}{n_{2}}{n_{3}}$ are nonzero only when $n_{1}=|n_{2}\pm n_{3}|$. Therefore, unlike when $kr_c$ is nonzero, the 5DOT matrix element $\mathcal{M}^{(\text{5DOT})}$ for a process $(n_{1},n_{2})\rightarrow (n_{3},n_{4})$ consists of only finitely many nonzero diagrams.

For $(n,n)\rightarrow (n,n)$, the 5DOT matrix element arises from four types of diagrams:
\begin{align}
    \mathcal{M}^{(\text{5DOT})}_{(n,n)\rightarrow (n,n)} = \mathcal{M}_{\text{c}} + \mathcal{M}_{\text{r}} +  \mathcal{M}_{0} + \mathcal{M}_{2n}~.
\end{align}
Using Eqs. (\ref{eq:define-a}) and (\ref{eq:define-b}) and the toroidal wavefunctions, we find:
\begin{align}
    n^{2} a_{nnnn} \hspace{5 pt} &= \hspace{5 pt} 3b_{nnnn} \hspace{5 pt} = \hspace{5 pt} \dfrac{3}{4}n^{2}~,\nonumber \\
    n^{2} a_{nn0} \hspace{5 pt} &= \hspace{5 pt} b_{nn0} \hspace{5 pt} = \hspace{5 pt} b_{nnr} \hspace{5 pt} =\hspace{5 pt}  \dfrac{1}{\sqrt{2}}n^{2}~,\label{eq:orbifold-couplings}\\
    n^{2} a_{nn(2n)} \hspace{5 pt} &= \hspace{5 pt} - b_{nn(2n)} \hspace{5 pt} = \hspace{5 pt} \dfrac{1}{2} b_{(2n)nn} \hspace{5 pt} = \hspace{5 pt} -\dfrac{1}{2}n^{2}~, \nonumber 
\end{align}
where here again the subscript ``0" refers to the massless 4D graviton. We focus first on the scattering of helicity-zero states, which have the most divergent high-energy behavior (we return to consider other helicity combinations in Sec. \ref{sec:nonlongitudinal}). Reference \cite{Chivukula:2019rij} lists $\mathcal{M}_{\text{c}}^{(\sigma)}$, $\mathcal{M}_{\text{r}}^{(\sigma)}$, $\mathcal{M}_{0}^{(\sigma)}$, and $\mathcal{M}_{2n}^{(\sigma)}$ for $\sigma\geq 1$, and demonstrates how cancellations occur among them such that $ \overline{\mathcal{M}}^{(\sigma)} = 0$  for $\sigma > 1$ and the leading contribution in incoming energy is
\begin{align}
    \overline{\mathcal{M}}^{(1)} = \dfrac{3\kappa^2}{256\pi r_c} \left[7+\cos(2\theta)\right]^2\csc^{2}\theta\,.\label{M15DOTnnnn}
\end{align}
We report here the results of the full calculation, including subleading terms. 

The complete (tree-level) matrix element for the elastic helicity-zero 5DOT process equals
\begin{align}
    \mathcal{M}^{(\text{5DOT})} = \dfrac{\kappa^{2} n^{2}\left[P_{0} + P_{2} c_{2\theta} + P_{4} c_{4\theta} + P_{6}c_{6\theta} \right] \csc^{2}\theta}{256\pi r_{c}^{3} \mathfrak{s}(\mathfrak{s}+1)(\mathfrak{s}^{2} + 8 \mathfrak{s} + 8 - \mathfrak{s}^{2} c_{2\theta})}~,\label{M5DOTtotal}
\end{align}
where
\begin{align}
    P_{0} &= 510\s \mathfrak{s}^{5} + 3962\s\mathfrak{s}^{4} + 8256\s \mathfrak{s}^{3} + 7344\s \mathfrak{s}^{2}\nonumber\\
    &\hspace{15 pt} + 3216\s \mathfrak{s} + 704~,\\
    P_{2} &= -429 \s\mathfrak{s}^{5} + 393 \s\mathfrak{s}^{4} + 3936 \s\mathfrak{s}^{3} + 5584 \s\mathfrak{s}^{2}\nonumber\\
    &\hspace{15 pt} +3272 \s\mathfrak{s} + 768~,\\
    P_{4} &= -78 \s\mathfrak{s}^{5} - 234 \s\mathfrak{s}^{4} + 192 \s\mathfrak{s}^{3} + 1552 \s\mathfrak{s}^{2}\nonumber\\
    &\hspace{15 pt}+ 1776 \s\mathfrak{s} + 576~,\\
    P_{6} &= -3 \s\mathfrak{s}^{5} - 25 \s\mathfrak{s}^{4} - 96 \s\mathfrak{s}^{3} - 144\s\mathfrak{s}^{2} - 72 \s\mathfrak{s}~,
\end{align}
and $\mathfrak{s}$ is defined such that $s \equiv s_{\text{min}} (1 + \mathfrak{s})$ where in this case $s_{\text{min}} = 4 m_{n}^{2} = 4n^{2}/r_{c}^{2}$.

For a generic helicity-zero 5DOT process $(n_{1},n_{2})\rightarrow (n_{3},n_{4})$, the leading high-energy contribution to the matrix element equals
\begin{align}
    \overline{\mathcal{M}}^{(1)} = \dfrac{\kappa^2}{256\pi r_c}x_{n_{1}n_{2}n_{3}n_{4}} \left[7+\cos(2\theta)\right]^2\csc^{2}\theta~, \label{M15DOT}
\end{align}
where $x$ is fully symmetric in its indices, and satisfies
\begin{align}
    x_{aaaa} = 3,\hspace{15 pt}x_{aabb} = 2,\hspace{5 pt}\text{ otherwise}\hspace{5 pt}
    x_{abcd} = 1~,\nonumber
\end{align}
when discrete KK momentum is conserved (and, of course, vanishes when the process does not conserve KK momentum). 

The multiplicative $\csc^{2}\theta$ factor in Eq. \eqref{M5DOTtotal} is indicative of $t$- and $u$-channel divergences from the exchange of the massless graviton and radion, which introduces divergences at $\theta = 0,\pi$. Such IR divergences prevent us from directly using a partial wave analysis to determine the strong coupling scale of this theory. In order to characterize the strong-coupling scale of this theory, we must instead investigate a nonelastic scattering channel for which  KK momentum conservation implies that no massless states can contribute, $\mathcal{M}_{0} = \mathcal{M}_{r} = 0$. (In this case, the $\csc^{2}\theta$ factor present in Eq. $\eqref{M15DOT}$ is an artifact of the high-energy expansion and is absent from the full matrix element.) 

Consider for example the helicity-zero 5DOT process $(1,4)\rightarrow (2,3)$.  The total matrix element is computed from four diagrams
\begin{align}
    \raisebox{-0.45\height}{\sgDA{0.20}}+\raisebox{-0.45\height}{\sDA{0.20}}+\raisebox{-0.45\height}{\tDA{0.20}}+\raisebox{-0.45\height}{\uDA{0.20}}
\end{align}
which together yield, after explicit computation,
\begin{align}
    \mathcal{M}= \dfrac{ \kappa^{2} \mathfrak{s} }{12800 \pi r_{c}^{3} (\mathfrak{s} + 1)^{2} Q_{+} Q_{-}} \sum_{i=0}^{4} Q_{i} c_{i\theta}~,
\end{align}
where
\begin{align}
    Q_{\pm} &= 25(\mathfrak{s} + 1) \pm \left[3 + \sqrt{(25\s\mathfrak{s} + 16)(25\s\mathfrak{s}+24)} \cos\theta\right]~,\\
    Q_{0} &= 15\left(2578125\s \mathfrak{s}^{4} + 9437500\s\mathfrak{s}^{3} + 12990000\s \mathfrak{s}^{2}\right.\nonumber\\
    &\hspace{15 pt}\left.+ 7971000\s \mathfrak{s} + 1840564\right)~,\\
    Q_{1} &= 72 \sqrt{(25\s\mathfrak{s}+16)(25\s\mathfrak{s}+24)}(50\s\mathfrak{s}+43)(50\s\mathfrak{s}+47)~,\\
    Q_{2} &= 4\left(2734375\s \mathfrak{s}^{4} + 11562500\s\mathfrak{s}^{3} + 18047500\s \mathfrak{s}^{2}\right.\nonumber\\
    &\hspace{15 pt}\left.+ 12340500\s \mathfrak{s} + 3121692\right)~,\\
    Q_{3} &= 24 \sqrt{(25\s\mathfrak{s}+16)(25\s\mathfrak{s}+24)}(50\s\mathfrak{s}+51)(50\mathfrak{s}+59)~,\\
    Q_{4} &= 390625\s\mathfrak{s}^{4} + 2187500\s\mathfrak{s}^{3} + 4360000\s\mathfrak{s}^{2}\nonumber\\
    &\hspace{15 pt} +3729000\s\mathfrak{s} + 1165956~,
\end{align}
and $s_{\text{min}} = 25/r_{c}^{2}$. As expected, unlike the elastic 5DOT matrix element (\ref{M5DOTtotal}), the $(1,4)\rightarrow(2,3)$ 5DOT matrix element is finite at $\theta =0,\pi$.

Given a 2-to-2 scattering process with helicities $(\lambda_{1},\lambda_{2})$ $\rightarrow$ $(\lambda_{3},\lambda_{4})$, the corresponding partial wave amplitudes $a^{J}$ are defined as \cite{Jacob:1959at}
\begin{align}
a^J = \dfrac{1}{32 \pi^2} \int d\Omega\hspace{10 pt} D_{\lambda_i\lambda_f}^J(\theta,\phi) \mathcal{M}(s,\theta,\phi)~,
\end{align}
where $\lambda_{i} = \lambda_{1} - \lambda_{2}$ and $\lambda_{f} = \lambda_{3} - \lambda_{4}$, $d\Omega = d(\cos\theta)\s d\phi$, and the Wigner D functions $D^{J}_{\lambda_{a}\lambda_{b}}$ are normalized according to
\begin{align}
    \int d\Omega \hspace{10 pt} D^J_{\lambda_{a}\lambda_{b}}(\theta,\phi) \cdot D^{J^\prime *}_{\lambda_{a}^{\prime} \lambda_{b}}(\theta,\phi) = \dfrac{4\pi}{2J+1}\cdot  \delta_{JJ^{\prime}}\cdot \delta_{\lambda_{a}\lambda^{\prime}_{a}}~.
\end{align}
Each partial wave amplitude is constrained by unitarity to satisfy
\begin{align}
    \sqrt{1-\dfrac{s_{\text{min}}}{s}}\hspace{5 pt} \mathfrak{R}[a^{J}] \leq \frac{1}{2}~,
\end{align}
where $\mathfrak{R}[a^{J}]$ denotes the real part of $a^{J}$. The leading partial wave amplitude of the $(1,4)\rightarrow(2,3)$ helicity-zero  5DOT matrix element corresponds to $J=0$, and has leading term
\begin{align}
    a^{0} \simeq \dfrac{s}{8\pi M_{\text{Pl}}^{2}}\s \ln\left(\dfrac{s}{s_{\text{min}}}\right)~.
\end{align}
Hence, this matrix element violates unitarity when ${\rm Re}\, a^0 \simeq 1/2$, or equivalently when the value of $E\equiv \sqrt{s}$ is near or greater than $ \Lambda_{\text{strong}}^{(\text{5DOT})} \equiv \sqrt{4\pi} M_{\text{Pl}}$. Because $M_{\text{Pl}}$ labels the reduced Planck mass, $\Lambda_{\text{strong}}^{(\text{5DOT})}$ is roughly the conventional Planck mass.  We will use this inelastic calculation as a benchmark for estimating the strong-coupling scale associated with other processes. 

We now consider the behavior of scattering amplitudes in the RS1 model.

\section{\label{sec:level5} Elastic Scattering in the Randall-Sundrum Model}

This section discusses the computation of the elastic scattering amplitudes of massive spin-2 KK modes in the RS1 model, for arbitrary values of the curvature of the internal space. For any nonzero curvature, every KK mode in the infinite tower contributes to each scattering process and the cancellation from $\mathcal{O}(s^5)$ to $\mathcal{O}(s)$ energy growth only occurs when all of these states are included. We first review and elaborate on the derivation of the sum rules introduced in \cite{Chivukula:2019zkt}. A new analytic proof for a relation arising from the $s^3$ and $s^2$ sum rules is discussed in Appendix \ref{sec:levelD}, and the relationships of our couplings and sum rules to those conjectured  in \cite{Bonifacio:2019ioc} are given in Appendix \ref{sec:levelE}. In the subsequent subsections, we apply the sum rules to determine the leading high-energy behavior of the amplitudes for two-body scattering of helicity-zero modes.  Finally, Sec. \ref{sec:nonlongitudinal} analyses the (milder) high-energy behavior of the scattering of nonlongitudinal helicity modes of the massive spin-2 KK states.

\subsection{Coupling Identities}

Let us now consider the elastic helicity-zero RS1 process $(n,n)\rightarrow (n,n)$. We will approach it by identifying sum rules that enforce cancellations among different contributions to the scattering amplitude at a given order in $s$.  This subsection rederives and elaborates on several results from Ref. \cite{Chivukula:2019zkt}; we apply the coupling relations in the subsequent subsections.

The wavefunctions $\psi_{n}$ solve the Sturm-Liouville problem defined by Eq. \eqref{SLeqx} when subject to the boundary condition $(\partial_{\varphi}\psi_{n})=0$ at $\varphi =0,\pi$ and satisfy the orthonormality relations Eqs. \eqref{onAx} and \eqref{onBx}. In particular, $\psi_{0} = \sqrt{\pi kr_{c}/(1-e^{-2\pi kr_{c}})}$. By integrating by parts and utilizing the Sturm-Liouville equation, we derive the following generic relation:
\begin{align}
    &\mu_{i}^{2} \int d\varphi\hspace{5 pt}\varepsilon^{-2}(\psi_{i}^{N} \cdot \mathbb{X}) = - \int d\varphi\hspace{5 pt}\left[-\mu_{i}^{2} \varepsilon^{-2} \psi_{i}\right]\cdot \psi_{i}^{N-1} \mathbb{X}\nonumber\\
    &\hspace{25 pt}= - \int d\varphi\hspace{5 pt}\partial_{\varphi}\left[\varepsilon^{-4} (\partial_{\varphi} \psi_{i})\right]\cdot \psi_{i}^{N-1}\mathbb{X}\nonumber\\
    &\hspace{25 pt}= \int d\varphi\hspace{5 pt}\varepsilon^{-4} (\partial_{\varphi} \psi_{i})\cdot \partial_{\varphi}\left[\psi_{i}^{N-1}\mathbb{X}\right]\nonumber\\
    &\hspace{25 pt}= (N-1)\int d\varphi\hspace{5 pt}\varepsilon^{-4}(\partial_{\varphi}\psi_{i})^{2}\cdot\psi_{i}^{N-2}\mathbb{X}\nonumber\\
    &\hspace{40 pt}+ \int d\varphi\hspace{5 pt} \varepsilon^{-4}(\partial_{\varphi}\psi_{i})(\partial_{\varphi}\mathbb{X})\psi_{i}^{N-1}~,
\end{align}
where $\mathbb{X}$ is a generic function of $\varphi$. Through appropriate choices of the function $\mathbb{X}$, the number $N$ of instances of $\psi_{i}$, and the KK index $i$, we obtain the following relations:
\begin{align}
    (\mathbb{X},N,i) = (\psi_{j},2,n) \implies\mu_{n}^{2} a_{nnj} &= b_{nnj} + b_{jnn}~,\\
    (\mathbb{X},N,i) = (\psi_{n}^{2},1,j) \implies\mu_{j}^{2} a_{nnj} &= 2 b_{jnn}~,\\
    (\mathbb{X},N,i) = (1,4,n) \implies\mu_{n}^{2} a_{nnnn} &= 3 b_{nnnn}~,
\end{align}
which allow us to rewrite all B-Type couplings of the form $b_{\vec{n}}$ in terms of A-Type couplings:
\begin{align}
    b_{nnj} &= \left(\mu_{n}^{2} - \tfrac{1}{2} \mu_{j}^{2}\right)a_{nnj}~,\\
    b_{jnn} &= \tfrac{1}{2} \mu_{j}^{2}a_{nnj}~,\\
    b_{nnnn} &= \tfrac{1}{3} \mu_{n}^{2} a_{nnnn}~.
\end{align}

Furthermore, the completeness relation Eq. (\ref{eq:completeness}) implies the generic relation
\begin{align}
    &\sum_{j} \left[\dfrac{1}{\pi}\int d\varphi_{1}\hspace{5 pt}\psi_{j}\cdot \mathbb{X}\right]\cdot \left[\dfrac{1}{\pi}\int d\varphi_{2}\hspace{5 pt}\psi_{j}\cdot \mathbb{Y}\right]\nonumber\\
    &\hspace{25 pt}=\dfrac{1}{\pi} \int d\varphi\hspace{5 pt} \varepsilon^{+2}\s \mathbb{X}\cdot \mathbb{Y}~,
\end{align}
where $\mathbb{X}$ and $\mathbb{Y}$ are generic functions of $\varphi$, from which one may derive, for instance,
\begin{align}
    \sum_{j} a_{nnj}^{2} &= a_{nnnn}~,\label{SumRuleO5}\\
    \sum_{j} b_{nnj}a_{nnj} &= b_{nnnn} = \dfrac{1}{3}\mu_{n}^{2}a_{nnnn}~,
\end{align}
and
\begin{align}
    \sum_{j} \mu_{j}^{2} a_{nnj}^{2} &= \sum_{j} \left[2 b_{jnn}\right] a_{nnj}\nonumber\\
    &= 2\sum_{j} \left[\mu_{n}^{2}a_{nnj} - b_{nnj}\right] a_{nnj}\nonumber\\
    &= \dfrac{4}{3} \mu_{n}^{2} a_{nnnn}~.\label{SumRuleO4}
\end{align}
We can continue adding instances of $\mu_{j}^{2}$ to the sum and repeat this procedure with $\sum_{j}\mu_{j}^{4}a_{nnj}^{2}$ and  $\sum_{j}\mu_{j}^{6}a_{nnj}^{2}$. The details of these manipulations are summarized in Appendix \ref{AppendixSumRules}; the principal result is
\begin{align}
    \sum_{j=0}^{+\infty}\left[\mu_{j}^{2} - 5 \mu_{n}^{2} \right] \mu_{j}^{4} a_{nnj}^{2} &= -\dfrac{16}{3}\mu_{n}^{6} a_{nnnn}~.\label{SumRuleOsAx}
\end{align}

Note that the equations in this section relate couplings and spectra, which are determined entirely by the Sturm-Liouville problem and therefore depend only on the value of $kr_{c}$ (i.e. not on $m_{1}$ or $M_{\text{Pl}}$).  Because of their origin, these equations relating 4D masses and couplings must ultimately be expressions of the original 5D diffeomorphism invariance.  

\subsection{Cancellations at $\mathcal{O}(s^{5})$ in RS1}
We will now go through the contributions to the elastic helicity-zero $(n,n)\rightarrow (n,n)$ scattering process in the RS1 model order by order in powers of $s$, and apply the sum rules derived in the previous section. 

As described in Sec. \ref{sec:Ib}, the contact diagram and spin-2-mediated diagrams individually diverge like $\mathcal{O}(s^{5})$. After converting all $b_{\vec{n}}$ couplings into $a_{\vec{n}}$ couplings, their contributions to the elastic helicity-zero RS1 matrix element equal
\begin{align}
    \overline{\mathcal{M}}_{c}^{(5)} &=-\dfrac{\kappa^{2}\s a_{nnnn}}{2304\s\pi r_{c}\s m_{n}^{8}}\left[7+\cos(2\theta)\right]\sin^{2}\theta~,\\
    \overline{\mathcal{M}}_{j}^{(5)} &= \dfrac{\kappa^{2} \s a_{nnj}^{2}}{2304\s\pi r_{c}\s m_{n}^{8}}\left[7+\cos(2\theta)\right]\sin^{2}\theta~,
\end{align}
such that they sum to
\begin{align}
    \overline{\mathcal{M}}^{(5)}&= \dfrac{\kappa^{2}\left[7+\cos(2\theta)\right]\sin^{2}\theta}{2304\s \pi r_{c} \s m_{n}^{8}}\left\{ \sum_{j=0}^{+\infty} a_{nnj}^{2} - a_{nnnn}\right\}~. \label{ELM5RS1}
\end{align}
This vanishes via Eq. \eqref{SumRuleO5}, which we will, henceforth, refer to as the $\mathcal{O}(s^{5})$ sum rule.

\subsection{Cancellations at $\mathcal{O}(s^{4})$ in RS1}
The $\mathcal{O}(s^{4})$ contributions to the elastic helicity-zero RS1 matrix element equal
\begin{align}
    \overline{\mathcal{M}}_{c}^{(4)} &=\dfrac{\kappa^{2}\s a_{nnnn}}{6912\s\pi r_{c}\s m_{n}^{6}}\left[63-196\cos(2\theta)+5\cos(4\theta)\right]~,\\
    \overline{\mathcal{M}}_{j}^{(4)} &= -\dfrac{\kappa^{2}\s a_{nnj}^{2}}{9216\s\pi r_{c}\s m_{n}^{6}}\bigg\{\left[7+\cos(2\theta)\right]^{2}\dfrac{m_{j}^{2}}{m_{n}^{2}}\nonumber\\
    &\hspace{10 pt}+2\left[9-140\cos(2\theta)+3\cos(4\theta)\right]\bigg\}~.
\end{align}
Using the $\mathcal{O}(s^{5})$ sum rule, $\overline{\mathcal{M}}^{(4)}$ equals
\begin{align}
    \overline{\mathcal{M}}^{(4)}&= \dfrac{\kappa^{2}\left[7+\cos(2\theta)\right]^{2}}{9216\s \pi r_{c} \s m_{n}^{6}}\left\{ \dfrac{4}{3}a_{nnnn} - \sum_{j} \dfrac{m_{j}^{2}}{m_{n}^{2}} a_{nnj}^{2}\right\}~. \label{ELM4RS1}
\end{align}
This vanishes via Eq. \eqref{SumRuleO4}, which we shall refer to as the $\mathcal{O}(s^{4})$ sum rule.

\subsection{Cancellations at $\mathcal{O}(s^{3})$ in RS1}
Once the $\mathcal{O}(s^{5})$ and $\mathcal{O}(s^{4})$ contributions are cancelled, the radion-mediated diagrams, which diverge like $\mathcal{O}(s^{3})$, become relevant to the leading behavior of the elastic helicity-zero RS1 matrix element. Furthermore, because of differences between the massless and massive spin-2 propagators, $\overline{\mathcal{M}}_{0}$ and $\overline{\mathcal{M}}_{j>0}$ differ from one another at this order (and lower). The full set of relevant contributions is therefore
\begin{align}
    \overline{\mathcal{M}}^{(3)}_{c} &= \dfrac{\kappa^{2}\s a_{nnnn}}{3456\s \pi r_{c}\s m_{n}^{4}} \left[-185+692\cos(2\theta)+5\cos(4\theta)\right]~,\\
    \overline{\mathcal{M}}^{(3)}_{r} &= -\dfrac{\kappa^{2}}{32 \s \pi r_{c} \s m_{n}^{4}}\left[\dfrac{b_{nnr}^{2}}{(m_{n}r_{c})^{4}}\right] \sin^{2}\theta~,\\
    \overline{\mathcal{M}}^{(3)}_{0} &= \dfrac{\kappa^{2}\s a_{nn0}^{2}}{1152\s \pi r_{c}\s m_{n}^{4}} \left[15-270\cos(2\theta)-\cos(4\theta)\right]~,\\
    \overline{\mathcal{M}}^{(3)}_{j>0} &= \dfrac{\kappa^{2}\s a_{nnj}^{2}}{2304\s \pi r_{c}\s m_{n}^{4}} \bigg\{5\left[1-\cos(2\theta)\right] \dfrac{m_{j}^{4}}{m_{n}^{4}}\nonumber\\
    &\hspace{10 pt}+\left[69+60\cos(2\theta)-\cos(4\theta)\right]\dfrac{m_{j}^{2}}{m_{n}^{2}}\nonumber\\
    &\hspace{10 pt}+2\left[13-268\cos(2\theta)-\cos(4\theta)\right]\bigg\}~,
\end{align}
After applying the $\mathcal{O}(s^{5})$ and $\mathcal{O}(s^{4})$ sum rules, $\overline{\mathcal{M}}^{(3)}$ equals
\begin{align}
    \overline{\mathcal{M}}^{(3)} &= \dfrac{5\s\kappa^{2}\s \sin^{2}\theta}{1152 \s \pi r_{c} \s m_{n}^{4}}\bigg\{ \sum_{j} \dfrac{m_{j}^{4}}{m_{n}^{4}} a_{nnj}^{2} - \dfrac{16}{15} a_{nnnn}\nonumber\\
    &\hspace{10 pt}-\dfrac{4}{5}\left[ \dfrac{9\s b_{nnr}^{2}}{(m_{n}r_{c})^{4}} - a_{nn0}^{2}\right]\bigg\}~. \label{ELM3RS1}
\end{align}
These contributions cancel if the following $\mathcal{O}(s^{3})$ sum rule holds true:
\begin{align}
    \sum_{j=0}^{+\infty} \mu_{j}^{4} a_{nnj}^{2} = \dfrac{16}{15}\mu_{n}^{4} a_{nnnn} + \dfrac{4}{5}\bigg[9b_{nnr}^{2} - \mu_{n}^{4} a_{nn0}^{2}\bigg]\label{SumRuleOs3} 
\end{align}
We do not yet have an analytic proof of this sum rule; however we have verified that the right-hand side numerically approaches the left-hand side as the maximum intermediate KK number $N_{\text{max}}$ is increased to $100$ for a wide range of values of $kr_{c}$, including $kr_{c} \in \{10^{-3},10^{-2},10^{-1},1,2,\ldots,10\}$.\footnote{The cancellations implied by this sum rule can be seen in the vanishing of ${\mathcal R}^{[N](3)}$ Fig. \ref{fig:cancel1111} as $N$ increases.}

\subsection{Cancellations at $\mathcal{O}(s^{2})$ in RS1}
The contributions to the elastic helicity-zero matrix element at $\mathcal{O}(s^{2})$ equal
\begin{align}
    \overline{\mathcal{M}}_{c}^{(2)} &= -\dfrac{\kappa^{2}\s a_{nnnn}}{54\s \pi r_{c} \s m_{n}^{2}}\left[5 +47 \cos(2\theta)\right]~,\\
    \overline{\mathcal{M}}_{r}^{(2)} &= \dfrac{\kappa^{2}}{48\s \pi r_{c} \s m_{n}^{2}}\left[\dfrac{b_{nnr}^{2}}{(m_{n}r_{c})^{4}}\right]\left[7 + \cos(2\theta)\right]~,\\
    \overline{\mathcal{M}}_{0}^{(2)} &= \dfrac{\kappa^{2}\s a_{nn0}^{2}}{576 \s \pi r_{c} \s m_{n}^{2}} \left[175 + 624 \cos(2\theta) + \cos(4\theta)\right]~,\\
    \overline{\mathcal{M}}_{j>0}^{(2)} &= \dfrac{\kappa^{2}\s a_{nnj}^{2}}{6912 \s \pi r_{c}\s m_{n}^{2}} \bigg\{4\left[7+\cos(2\theta)\right]\left[5 - 2\dfrac{m_{j}^{2}}{m_{n}^{2}}\right] \dfrac{m_{j}^{4}}{m_{n}^{4}}\nonumber\\
    &\hspace{10 pt}-\left[1291+1132\cos(2\theta)+9\cos(4\theta)\right] \dfrac{m_{j}^{2}}{m_{n}^{2}}\nonumber\\
    &\hspace{10 pt}+4\left[553+1876\cos(2\theta)+3\cos(4\theta)\right]\bigg\}~.
\end{align}
By applying the $\mathcal{O}(s^{5})$ and $\mathcal{O}(s^{4})$ sum rules (but {\it not} the $\mathcal{O}(s^{3})$ sum rule), the total $\mathcal{O}(s^{2})$ contribution equals
\begin{align}
    \overline{\mathcal{M}}^{(2)} &= \dfrac{\kappa^{2}\s \left[7+\cos(2\theta)\right]}{864\s \pi r_{c} \s m_{n}^{2}}\bigg\{\sum_{j}\left[\dfrac{m_{j}^{2}}{m_{n}^{2}} - \dfrac{5}{2}\right]\dfrac{m_{j}^{4}}{m_{n}^{4}} a_{nnj}^{2}\nonumber\\
    &\hspace{10 pt}+\dfrac{8}{3}a_{nnnn} -2\left[ \dfrac{9\s b_{nnr}^{2}}{(m_{n}r_{c})^{4}} - a_{nn0}^{2}\right]\bigg\}~, \label{ELM2RS1}
\end{align}
which vanishes if the following $\mathcal{O}(s^{2})$ sum rule holds:
\begin{align}
    \sum_{j=0}^{+\infty} \left[ \mu_{j}^{2} - \dfrac{5}{2}\mu_{n}^{2}\right]\mu_{j}^{4} a_{nnj}^{2} &= -\dfrac{8}{3}\mu_{n}^{6} a_{nnnn} \nonumber\\
    &\hspace{10 pt}+ 2\mu_{n}^{2}\bigg[9b_{nnr}^{2} - \mu_{n}^{4} a_{nn0}^{2}\bigg]~.\label{SumRuleOs2}
\end{align}
Again, we do not yet have a proof for this sum rule, despite strong numerical evidence that it is correct (see Sec. \ref{sec:level6}). However, combining the $\mathcal{O}(s^{3})$ and $\mathcal{O}(s^{2})$ sum rules (Eqs. \eqref{SumRuleOs3} and \eqref{SumRuleOs2}), yields an equivalent set 
\begin{align}
    \sum_{j=0}^{+\infty}\left[\mu_{j}^{2} - 5 \mu_{n}^{2} \right] \mu_{j}^{4} a_{nnj}^{2} &= -\dfrac{16}{3}\mu_{n}^{6} a_{nnnn}~,\label{SumRuleOsA}\\
    \dfrac{5}{4} \sum_{j=0}^{+\infty} \mu_{j}^{4} a_{nnj}^{2}-\dfrac{4}{3}\mu_{n}^{4} a_{nnnn} &= 9 b_{nnr}^{2} - \mu_{n}^{4}a_{nn0}^{2}~.\label{SumRuleOsB}
\end{align}
Eq. \eqref{SumRuleOsA} is precisely Eq. \eqref{SumRuleOsAx}, for which we give an analytic proof in Appendix \ref{AppendixSumRules}. Therefore, if the $\mathcal{O}(s^{3})$ sum rule holds true, then the $\mathcal{O}(s^{2})$ must also hold true, and vice versa.

Finally, we note that the sum rules we have derived in RS1 in Eqs. (\ref{SumRuleO5}), (\ref{SumRuleO4}), (\ref{SumRuleOs3}), and (\ref{SumRuleOs2}),
are consistent with those inferred by the authors of \cite{Bonifacio:2019ioc} who assumed that cancellations in the spin-0 scattering amplitude of massive spin-2 modes in KK theories must occur to result in amplitudes which grow like $\mathcal{O}(s)$. A description of the correspondence of our results with theirs is given in Appendix \ref{sec:levelE}.

\subsection{The Residual $\mathcal{O}(s)$ Amplitude in RS1}

After applying all the sum rules above\footnote{The elastic 5D Orbifolded Torus couplings (\ref{eq:orbifold-couplings}) directly satisfy all of these sum rules.} (including Eq. \eqref{SumRuleOsB}, which lacks an analytic proof), the leading contribution to the elastic helicity-zero matrix element is found to be $\mathcal{O}(s)$. The relevant contributions, sorted by diagram type, equal
\begin{align}
    \overline{\mathcal{M}}_{c}^{(1)} &= \dfrac{\kappa^{2}\s a_{nnnn}}{1728\s \pi r_{c}}\left[1505 + 3108 \cos(2\theta) -5 \cos(4\theta)\right]~,\\
    \overline{\mathcal{M}}_{r}^{(1)} &= -\dfrac{\kappa^{2}}{24\s \pi r_{c}}\left[\dfrac{b_{nnr}^{2}}{(m_{n}r_{c})^{4}}\right]\left[9 + 7\cos(2\theta)\right]~,\\
    \overline{\mathcal{M}}_{0}^{(1)} &= \dfrac{\kappa^{2}\s a_{nn0}^{2}\s \csc^{2}\theta}{2304 \s \pi r_{c}} \left[748 + 427 \cos(2\theta)\right.\nonumber\\
    &\hspace{10 pt}\left.+ 1132\cos(4\theta)-3\cos(6\theta)\right]~,\\
    \overline{\mathcal{M}}_{j>0}^{(1)} &= \dfrac{\kappa^{2}\s a_{nnj}^{2} \s \csc^{2}\theta}{6912 \s \pi r_{c}} \bigg\{3\left[7+\cos(2\theta)\right]^{2}\dfrac{m_{j}^{8}}{m_{n}^{8}}\nonumber\\
    &\hspace{-10 pt}-4\left[241+148\cos(2\theta)-5\cos(4\theta)\right]\dfrac{m_{j}^{6}}{m_{n}^{6}}\nonumber\\
    &\hspace{-10 pt}+4\left[787+604\cos(2\theta)-47\cos(4\theta)\right]\dfrac{m_{j}^{4}}{m_{n}^{4}}\nonumber\\
    &\hspace{-10 pt}-\left[3854+5267\cos(2\theta)+98\cos(4\theta)-3\cos(6\theta)\right]\dfrac{m_{j}^{2}}{m_{n}^{2}}\nonumber\\
    &\hspace{-10 pt}+\left[2156+1313\cos(2\theta)+3452\cos(4\theta)-9\cos(6\theta)\right]\bigg\}~.
\end{align}
Combining them, according to Eq. \eqref{eq:Mtot-def}, yields
\begin{align}
    \overline{\mathcal{M}}^{(1)} &= \dfrac{\kappa^{2}\s \left[7+\cos(2\theta)\right]^{2}\csc^{2}\theta}{2304\s \pi r_{c}}\bigg\{\sum_{j}\dfrac{m_{j}^{8}}{m_{n}^{8}} a_{nnj}^{2}\nonumber\\
    &\hspace{10 pt}+\dfrac{28}{15} a_{nnnn} - \dfrac{48}{5}\left[ \dfrac{9\s b_{nnr}^{2}}{(m_{n}r_{c})^{4}} - a_{nn0}^{2}\right]\bigg\}~.\label{ELM1RS1}
\end{align}
This is generically nonzero, and thus represents the true leading high-energy behavior of the elastic helicity-zero RS1 matrix element.

\subsection{\label{sec:nonlongitudinal} Nonlongitudinal Scattering}

The sum rules of the previous subsections were derived by considering what cancellations were necessary to ensure the elastic helicity-zero RS1 matrix element grew no faster than $\mathcal{O}(s)$, a constraint which in turn comes from considering the extra-dimensional physics. This bound on high-energy growth must hold for scattering of all helicities. 

Indeed, upon studying the nonlongitudinal scattering amplitudes, we find that the sum rules derived in the helicity-zero case are sufficient to ensure {\it all} elastic RS1 matrix elements grow at most like $\mathcal{O}(s)$.

\begin{figure*}
\includegraphics[scale=0.15]{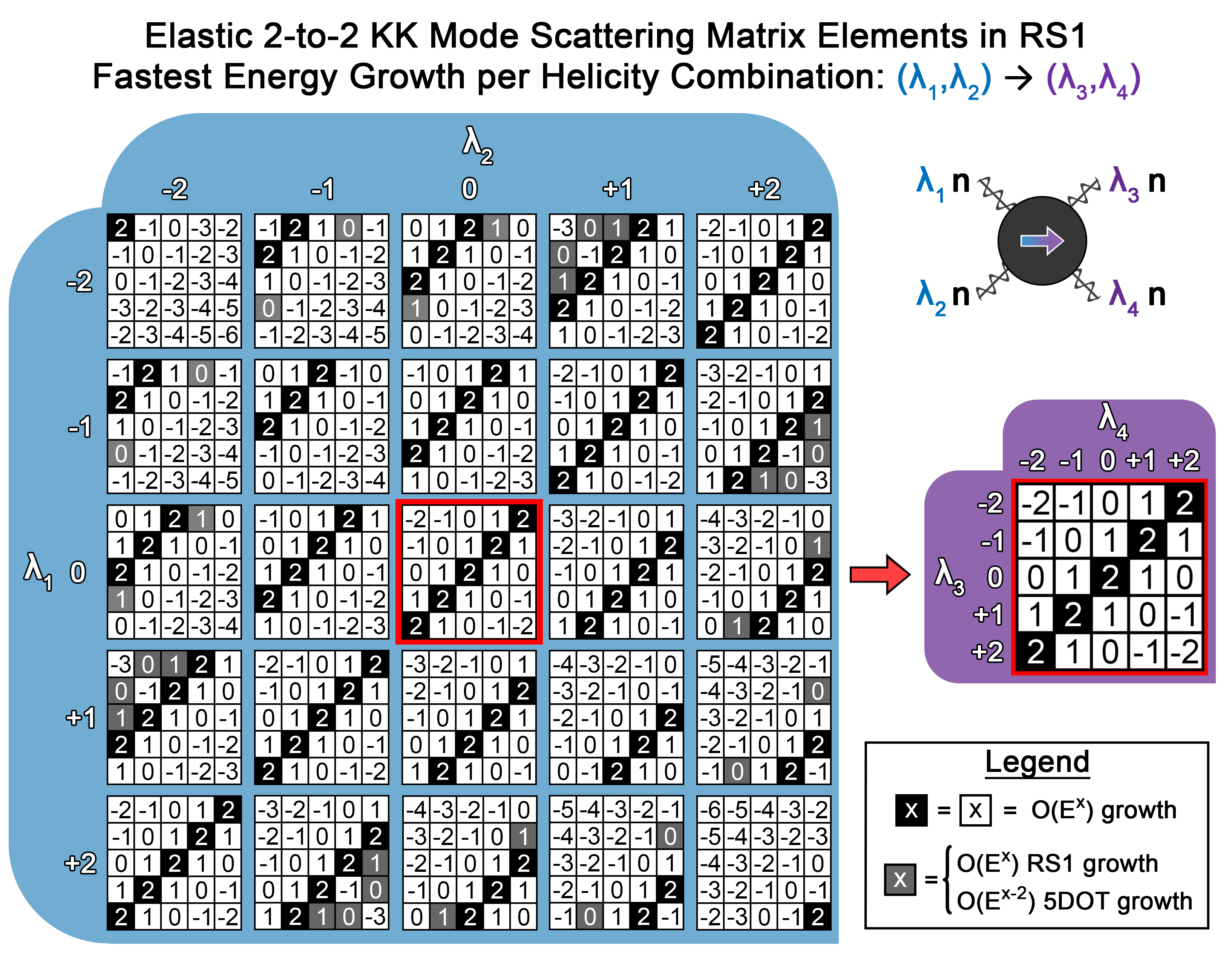}
 \caption{This table gives the leading order (in energy) growth of elastic $(n,n) \to (n,n)$ scattering for different incoming ($\lambda_{1,2}$) and outgoing ($\lambda_{3,4}$) helicity combinations in RS1. In the cases listed in grey, the leading order behavior is softer in the orbifolded torus limit (by two powers of center-of-mass energy).}
\label{fig:tableoftables}
\end{figure*}

Figure \ref{fig:tableoftables} lists the leading high-energy behavior of the elastic RS1 matrix element for each helicity combination after  the sum rules have been applied. These results are expressed in terms of the leading exponent of incoming energy $E\equiv \sqrt{s}$. For example, the elastic helicity-zero matrix element diverges like $\mathcal{O}(s)=\mathcal{O}(E^{2})$ and so its growth is recorded as ``2" in the table. As expected, no elastic RS1 matrix element grows faster than $\mathcal{O}(E^{2})$.

Some matrix elements grow more slowly with energy in the 5DOT model than they do in the more general RS1 model; they are indicated by the gray boxes in Fig. \ref{fig:tableoftables}. For these instances, the leading $\mathcal{M}^{(\sigma)}$ contribution in RS1 is always proportional to the same combination of couplings
\begin{align}
    \left[3 a_{nn0}^{2} + 16 a_{nnnn}\right] \mu_{n}^{4} - 27\s b_{nnr}^{2}~,
\end{align}
which vanishes exactly when $kr_{c}$ vanishes. Regardless of the specific helicity combination considered, no full matrix element vanishes.

\section{\label{sec:level6} Numerical Study of Scattering Amplitudes in the Randall-Sundrum Model}

This section presents a detailed numerical analysis of the scattering in the RS1 model. In Sec. \ref{sec:level6a} we demonstrate that the cancellations demonstrated for elastic scattering occur for inelastic scattering channels as well, with the cancellations becoming exact as the number of included intermediate KK modes increases. In Sec. \ref{sec:level6b} we examine the truncation error arising from keeping only a finite number of intermediate KK mode states. We then return, in Sec. \ref{sec:level6c} to the question of the validity of the KK mode EFT. In particular, we demonstrate directly from the scattering amplitudes that the cutoff scale is proportional to the RS1 emergent scale \cite{ArkaniHamed:2000ds,Rattazzi:2000hs}
\begin{align}
    \Lambda_\pi = M_{\text{Pl}}\, e^{-k\pi r_c}~,
\end{align}
which is related to the location of the IR (TeV) brane \cite{Randall:1999ee,Randall:1999vf}.

\subsection{\label{sec:level6a}Numerical Analysis of Cancellations in Inelastic Scattering Amplitudes}

We have demonstrated that the elastic scattering amplitudes in the Randall-Sundrum model grow only as ${\cal O}(s)$ at high energies, and have analytically derived the sum rules which enforce these cancellations. Physically, we expect similar cancellations and sum rules apply for arbitrary inelastic scattering amplitudes as well. However, we have found no analytic derivation of this property.\footnote{This is to be contrasted with the situation for KK compactifications on Ricci-flat manifolds, where an analytic demonstration of the needed cancellations has been found \cite{Bonifacio:2019ioc}.}

\begin{figure*}
\includegraphics[width=\linewidth]{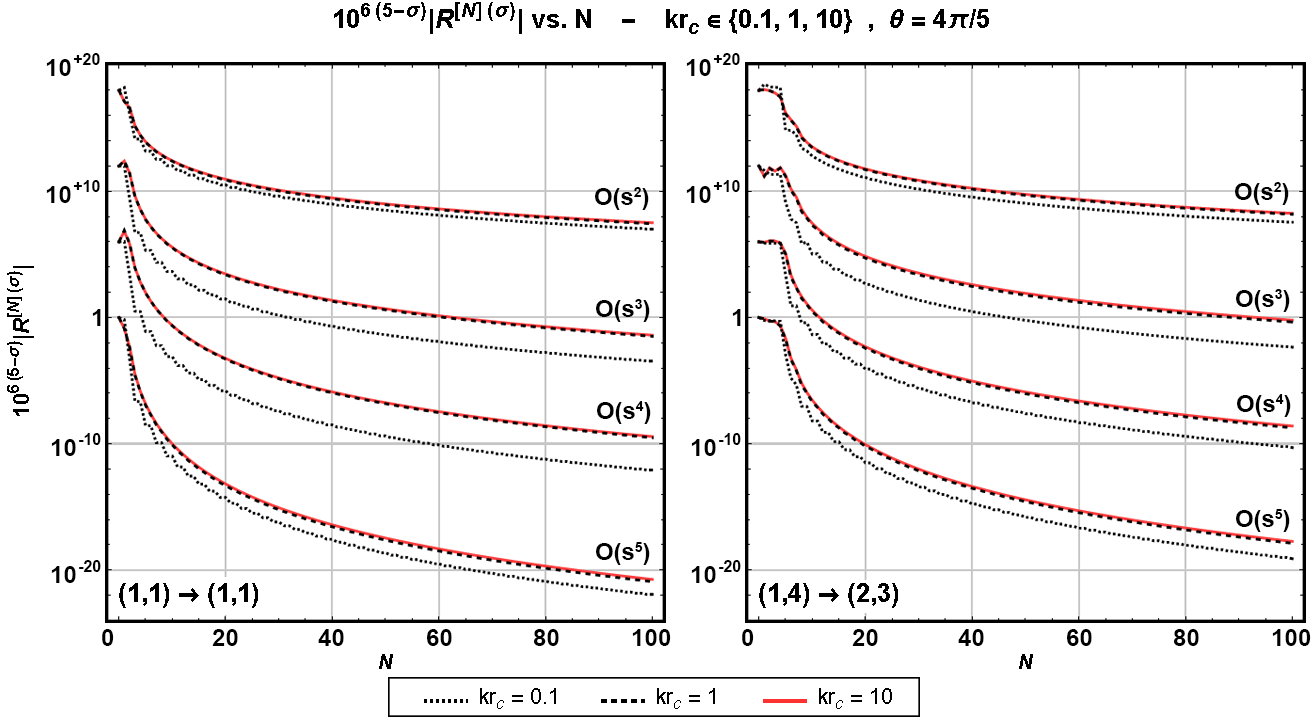}
\caption{These plots show the size of the residual truncation error for the helicity-zero scattering of KK modes $(1,1) \to (1,1)$ (left) and $(1,4) \to (2,3)$ (right) as a function of the number of KK intermediate states included ($N$) relative to not including any massive KK states, $\mathcal{R}^{[N](\sigma)}(kr_{c},\theta)$ from Eq. (\ref{eq:defR}). The curves are shown for $kr_c=0.1,\, 1,\, 10$ for $\theta=4\pi/5$. In all cases, the truncation error falls rapidly with addition of more intermediate states. To visually separate the different curves, the value of the ratio at $N=0$ has been artificially normalized to $(1,10^6,10^{12},10^{18})$ for $\sigma=5,4,3,2$ respectively.}
\label{fig:cancel1111}
\end{figure*}

Instead, we demonstrate here numerical checks with which we observe behavior consistent with the expected cancellations. To do so, we must first rewrite our expressions so we may vary $kr_c$ while keeping $M_{\text{Pl}}$ and $m_1$ fixed. We do so by noting that we may rewrite the common matrix element prefactor as
\begin{align}
    \dfrac{\kappa^{2}}{\pi r_{c}} = \dfrac{\kappa_{\text{4D}}^{2}}{{\psi_{0}}^{2}} = \dfrac{1}{\pi kr_{c}}\left[1- e^{-2kr_{c}\pi}\right]\dfrac{4}{M_{\text{Pl}}^{2}}~,
\end{align}
and that $r_{c} = \mu_{1}/m_{1}$, such that $\mathcal{M}^{(\sigma)}$ can be factorized for any process (and any helicity combination) into three unitless pieces, each of which depends on a different independent parameter:
\begin{align}
    \mathcal{M}^{(\sigma)} \equiv   \left[\mathcal{K}^{(\sigma)}(kr_{c},\theta)\right] \cdot \left[\dfrac{s}{M_{\text{Pl}}^{2}}\right] \cdot \left[\dfrac{\sqrt{s}}{m_{1}}\right]^{2(\sigma-1)}~. \label{MmsE}
\end{align}
This defines the dimensionless quantity $\mathcal{K}^{(\sigma)}$ characterizing the residual growth of order $(\sqrt{s})^{2\sigma}$ in any scattering amplitude . We can apply this decomposition to the truncated matrix element contribution $\mathcal{M}^{[N](\sigma)}$, as defined in Eq. (\ref{eq:M-N-def}) as well. By comparing $\mathcal{M}^{[N](\sigma)}$ to $\mathcal{M}^{[0](\sigma)}$ and increasing $N$ when $\sigma > 1$, we can measure how cancellations are improved by including more KK states in the calculation and do so in a way that depends only on $kr_{c}$ and $\theta$. Therefore, we define
\begin{align}
    \mathcal{R}^{[N](\sigma)}(kr_{c},\theta) \equiv \dfrac{\mathcal{M}^{[N](\sigma)}}{\mathcal{M}^{[0](\sigma)}} = \dfrac{\mathcal{K}^{[N](\sigma)}}{\mathcal{K}^{[0](\sigma)}}~,
    \label{eq:defR}
\end{align}
which vanishes as $N\rightarrow +\infty$ if and only if $\mathcal{M}^{[N](\sigma)}$ vanishes as $N\rightarrow +\infty$. Because $\mathcal{R}^{[N](\sigma)}$ depends continuously on $\theta$, we expect that so long as we choose a $\theta$-value such that $\mathcal{K}^{[N](\sigma)}\neq 0$, its exact value is unimportant to confirming cancellations. Figure \ref{fig:cancel1111} plots $10^{6(5-\sigma)}\mathcal{R}^{[N_{\text{max}}](\sigma)}$ for the helicity-zero processes $(1,1)\rightarrow(1,1)$ and $(1,4)\rightarrow(2,3)$ as functions of $N_{\text{max}}\rightarrow 100$  for $kr_{c}\in \{10^{-1},1,10\}$ and $\theta=4\pi/5$. The factor of $10^{6(5-\sigma)}$ only serves to vertically separate the curves for the reader's visual convenience; without this factor, the curves would all begin at $\mathcal{R}^{[0](\sigma)}=1$ and thus would substantially overlap.

We find that, both for the case of elastic scattering $(1,1) \to (1,1)$ where we have an analytic demonstration of the cancellations and for the inelastic case $(1,4) \to (2,3)$ where we do not, $\mathcal{M}^{[N](\sigma)} \to 0$ as $N\to \infty$. Furthermore, we find that the rate of convergence is similar in the two cases. In addition, and perhaps more surprisingly, the rate of convergence is relatively independent of the value of $kr_c$ for values between $1/10$ and $10$.

\subsection{\label{sec:level6b}Truncation Error}

In the RS1 model, the exact tree-level matrix element for any scattering amplitude requires summing over the entire tower of KK states. In practice, of course, any specific calculation will only include a finite number of intermediate states $N$. In this subsection we investigate the size of the ``truncation error" of such a calculation. For simplicity, in this section we will focus on the helicity-zero elastic scattering amplitude $(1,1) \to (1,1)$ and investigate the size of the truncation error for different values of $kr_c$ and center-of-mass scattering energy. 

For $\sigma>1$, consider the ratio
\begin{align}
    {\mathcal F}^{[N](\sigma)}(kr_{c},s)\equiv \max_{\theta \in [0,\pi]}\left|\dfrac{\mathcal{M}^{[N](\sigma)}(kr_{c},s,\theta)}{\mathcal{M}(kr_{c},s,\theta)}\right|~, 
\end{align}
which measures the size of each truncated matrix element contribution relative to the full amplitude.\footnote{In practice, we approximate the ``full" amplitude by  ${\mathcal{M}^{[N=100]}(kr_{c},s,\theta)}$, which we have checked provides ample sufficient numerical accuracy for the quantities reported here.} For sufficiently large $N$ and $\sigma > 1$ we have confirmed numerically that  the ratio $|\mathcal{M}^{[N](\sigma)}/\mathcal{M}^{[N}|$ reaches a global maximum at $\theta = \pi/2$ for $\sigma > 1$. Therefore
\begin{align}
    {\mathcal F}^{[N](\sigma)}(kr_{c},s) = \left|\dfrac{\mathcal{M}^{[N](\sigma)}(kr_{c},s,\theta)}{\mathcal{M}(kr_{c},s,\theta)}\right|_{\theta = \pi/2}~.
    \label{eq:defF1}
\end{align}

Unlike $\mathcal{M}^{(\sigma)}$ for $\sigma >1$, $\mathcal{M}^{(1)}$ diverges at $\theta \in \{0,\pi\}$ because of a $\csc^{2}\theta$ factor, as indicated in Eq. \eqref{ELM1RS1}, which arises from the $t$- and $u$-channel exchange of light states.\footnote{Formally, the sum over intermediate KK modes in Eq. (\ref{ELM1RS1}) extends over all masses, but the couplings $a_{11n}$ vanish as $n$ grows and suppress the contributions from heavy states.} The total elastic RS1 amplitude $\mathcal{M}$, on the other hand, only has such IR divergences due to the exchange of the massless graviton and radion. For this reason, and as confirmed by the numerical evaluation of $\mathcal{M}^{[N](1)}/\mathcal{M}^{[N]}$,  the divergences at $\theta\in\{0,\pi\}$ of $\mathcal{M}^{[N](1)}$  are actually slightly more severe than the corresponding divergences of $\mathcal{M}^{[N]}$, and so the ratio $\mathcal{M}^{[N](1)}/\mathcal{M}^{[N]}$ grows large in the vicinity of $\theta \in \{0,\pi\}$. However, this unphysical divergence is confined to nearly forward or backward scattering; otherwise the ratio is approximately flat. Thus for $\sigma = 1$ we study the analogous quantity
\begin{align}
    \mathcal{F}^{[N](1)}(kr_{c},s) =  \left|\dfrac{\mathcal{M}^{[N](\sigma)}(kr_{c},s,\theta)}{\mathcal{M}(kr_{c},s,\theta)}\right|_{\theta = \pi/2}~.
\end{align}

We also define the overall accuracy of the partial sum over intermediate states using a version of this quantity for which no expansion in powers of energy has been made: 
\begin{align}
    \mathcal{F}^{[N]}(kr_{c},s) \equiv \left|\dfrac{\mathcal{M}^{[N]}(kr_{c},s,\tfrac{\pi}{2})}{\mathcal{M}(kr_{c},s,\tfrac{\pi}{2})}\right|~.
\label{eq:defF2}
\end{align}

\begin{figure*}[t]
\center
\includegraphics[width=\linewidth]{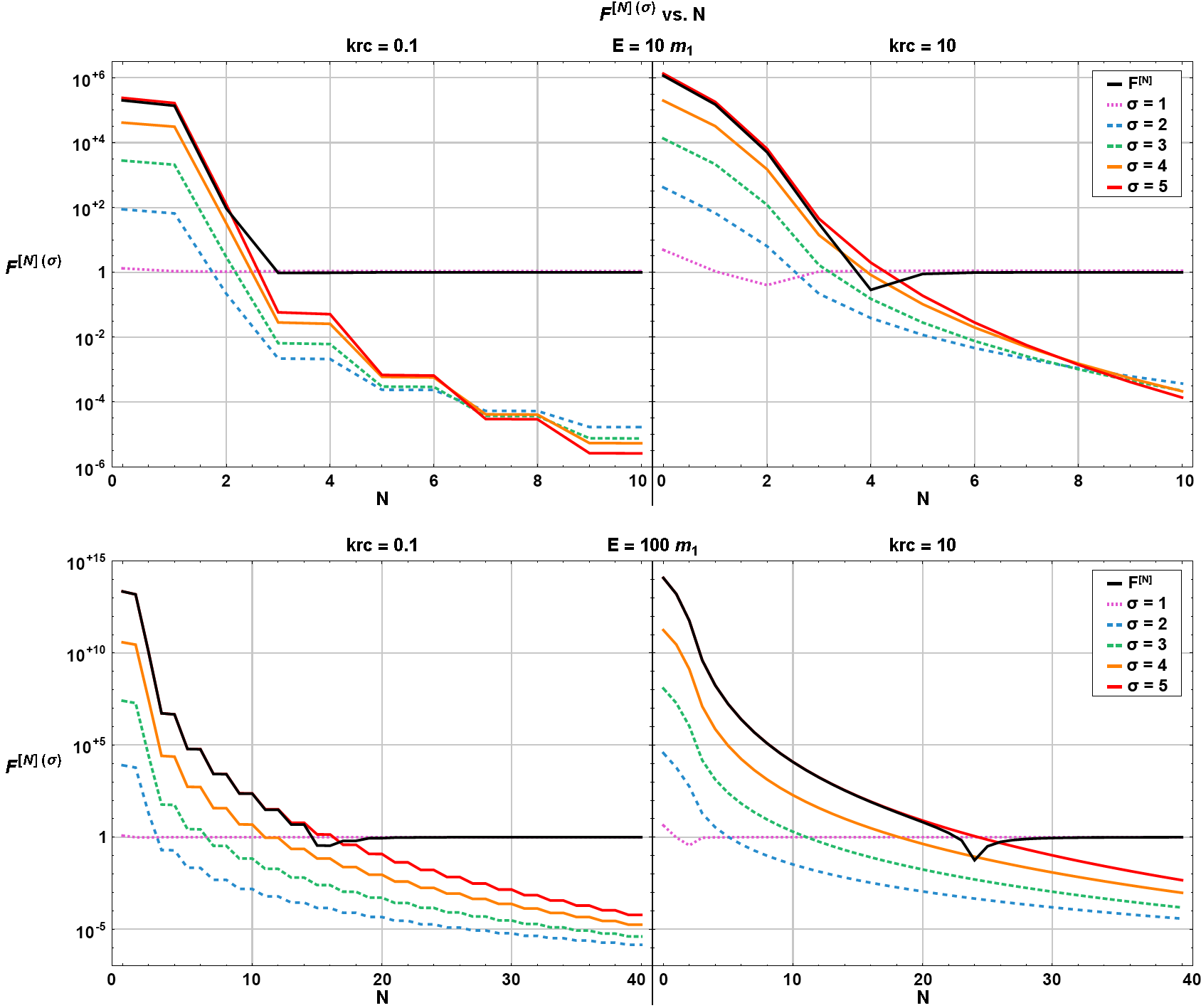}
\caption{These plots show an upper bound on the size of the residual truncation error relative to the size of the full matrix element for the process $(1,1)\to (1,1)$ as a function of the number of included KK modes $N$, for $E=10m_1$ (upper pair) and $E=100m_1$ (lower pair), and $kr_c=0.1$ (left pair) and $kr_c=10$ (right pair). $\mathcal{F}^{[N](\sigma)}(kr_c,s)$ from Eq. (\ref{eq:defF1}) is shown in color, for $\sigma=1$ - $5$, and $\mathcal{F}^{[N]}(kr_c,s)$ from Eq. (\ref{eq:defF2}) is shown in black. We see that the size of the truncation error falls rapidly as the number of included intermediate states $N$ increases. We also see that, for $E \gg m_1$, with a sufficient number of intermediate states $\mathcal{M}^{[N](1)}$ is a good approximation of the full matrix element. Note that if an insufficient number of intermediate KK modes is included, and the truncation error is large, $\mathcal{M}^{[N](5)}$ dominates.}
\label{fig:truncation}
\end{figure*}

Because $\mathcal{F}^{[N](\sigma)}$ ($\mathcal{F}^{[N]}$)  measures the discrepancy between any given contribution $\mathcal{M}^{[N](\sigma)}$ ($\mathcal{M}^{[N]}$) and the full matrix element $\mathcal{M}$, we study these quantities to understand the truncation error. In the upper two panes of Fig. \ref{fig:truncation} we plot the plot these quantities as a function of maximal KK number $N$ for $kr_{c} = 1/10$ and $kr_{c} = 10$ at the representative energy $s=(10 m_{1})^2$, for $m_1=1$ TeV. The lower two panes of Fig. \ref{fig:truncation} plot similar information but at the energy $s=(100 m_{1})^2$. The $kr_{c}=10$ panes contain the more phenomenologically relevant information. In all cases, we find that including sufficiently many modes in the KK tower yields an accurate result for angles away from the forward or backward scattering regime. When including only a small number of modes $N$, the contribution from $\mathcal{M}^{[N](5)}$ (the residual contribution arising from the noncancellation of the ${\cal O}(s^{5})$ contributions) dominates and the truncation yields an inaccurate result. As one increases the number of included modes, this unphysical ${\cal O}(s^{5})$ contribution to the amplitude falls in size until the full amplitude is dominated by $\mathcal{M}^{[N](1)}$, which is itself a good approximation to the complete tree-level amplitude. For $kr_{c}=1/10$, the number of states $N$ required to reach this ``crossover", however, increases from $3$ to $15$ as $\sqrt{s}$ increases from $10 m_{1}$ to $100 m_{1}$. Consistent with our analysis in the previous subsection, however, the truncation error is less dependent on $kr_{c}$; the number of states required to reach crossover increases by less than a factor of $2$ when moving from $kr_{c}=1/10$ to $kr_{c}=10$ at fixed $\sqrt{s}$. 

Finally, we note that the vanishing of $\mathcal{F}^{[N](3)}$ as $N$ increases is a numerical test of the $\mathcal{O}(s^3)$ sum rule in Eq. \eqref{SumRuleOs3}. 

\subsection{\label{sec:level6c}The Strong-Coupling Scale at Large $kr_c$}

In Sec. \ref{sec:level4b} we analyzed the tree-level scattering amplitude $(1,4) \to (2,3)$ and discovered that the 5D gravity compactified on a (flat) orbifolded torus becomes strongly coupled at roughly the Planck scale, $ \Lambda_{\text{strong}}^{(\text{5DOT})} \equiv \sqrt{4\pi} M_{\text{Pl}}$. In the large-$kr_{c}$ limit of the RS1 model, however, we expect that all low-energy mass scales are determined by the emergent scale  \cite{ArkaniHamed:2000ds,Rattazzi:2000hs}
\begin{align}
    \Lambda_\pi = M_{\text{Pl}}\, e^{-\pi k r_{c}}~,
\label{eq:LambdaPi}
\end{align}
which is related to the location $\varphi=\pi$ of the IR brane \cite{Randall:1999ee,Randall:1999vf}. In this section we describe how this emergent scale arises from an analysis of the elastic KK scattering
amplitude in the large-$kr_c$ limit.

Consider the helicity-zero polarized  $(n,n) \to (n,n)$ scattering amplitude. As plotted explicitly for $n=1$ in the previous subsection, at energies $s \gg m^2_n$ the scattering amplitude is dominated by the leading term ${\mathcal M}^{(1)}(kr_c,s,\theta)$ given in Eq.  \eqref{ELM1RS1}. The analogous expression in the 5D Orbifold Torus is given by Eq. (\ref{M15DOTnnnn}). We note that the angular dependence of these two expressions is precisely the same, and therefore we can compare their amplitudes by taking their ratio. This gives
the purely $kr_c$-dependent result\footnote{Formally, as in the case of toroidal compactification, this amplitude has an IR 
divergence due to the exchange of the massless graviton and radion modes. By taking the ratio of the amplitudes in RS1 to that in the 5D Orbifolded Torus, the IR divergences cancel and we can relate the strong-coupling scale in RS to that in the case of toroidal compactification.} 
\begin{align}
    \dfrac{\mathcal{M}^{(1)}(kr_{c})}{\mathcal{M}^{(1)}(0)} = \left[\dfrac{1- e^{-2\pi kr_{c}}}{2 \pi kr_{c}}\right]\cdot \overline{\mathcal{K}}_{nnnn}(kr_{c})~,
\end{align}
where
\begin{align}
    \overline{\mathcal{K}}_{nnnn} &=\dfrac{1}{405}\bigg\{15\sum_{j}\dfrac{m_{j}^{8}}{m_{n}^{8}} a_{nnj}^{2}+28 a_{nnnn}\nonumber\\
    &\hspace{10 pt} - 144\left[ \dfrac{9\s b_{nnr}^{2}}{(m_{n}r_{c})^{4}} - a_{nn0}^{2}\right]\bigg\}~.\label{Kbarnnnn}
\end{align}
From this ratio, we can estimate the strong-coupling scale at nonzero $kr_{c}$:
\begin{align}
    \Lambda^{\rm (RS1)}_{\rm strong} (kr_c) &\equiv \Lambda^{(\text{RS1})}_{\text{strong}}(0)\sqrt{\dfrac{\mathcal{M}^{(1)}(0)}{\mathcal{M}^{(1)}(kr_{c})}}~,\nonumber\\
    &= \dfrac{\Lambda^{\rm (5DOT)}_{\rm strong}}{\sqrt{\overline{\mathcal{K}}_{nnnn}}} \sqrt\frac{2 \pi kr_c}{1 - e^{-2\pi kr_{c}}}~.\label{LambdaStrong}
\end{align}
where we can use our earlier $\Lambda_{\text{strong}}^{(\text{5DOT})} = \sqrt{4\pi} M_{\text{Pl}}$ result.

Now let us consider the $kr_c$ dependence of this expression in the large $kr_c$ limit. To begin with, at large $kr_{c}$, Eq. \eqref{LambdaStrong} becomes
\begin{align}
    \Lambda^{(\text{RS1})}_{\text{strong}}(kr_{c}) \approx \sqrt{4 \pi} M_{\text{Pl}} \sqrt{\dfrac{2\pi kr_{c}}{\overline{\mathcal{K}}_{nnnn}}}~.
    \label{eq:strongscale}
\end{align}
Furthermore, in Appendix \ref{sec:largekrc} we show that at large $kr_{c}$, Eq. \eqref{Kbarnnnn} becomes
\begin{align}
    \frac{\overline{\mathcal{K}}_{nnnn}}{2\pi kr_c} & \approx \dfrac{e^{2\pi kr_{c}}}{810\,\pi\, x_{n}^{8}}\bigg\{ 15\sum_{j=1}^{+\infty}  x_{j}^{8} \, C_{nnj}^{2} + 28 \, x_{n}^{8} \, C_{nnnn} - \nonumber\\
    &\hspace{20 pt}1296\,  x_{n}^{4} \, C_{nnr}^{2} \bigg\}~,
\end{align}
In this expression, the $x_{j,n}$ are the $j$th and $n$th zeros of the Bessel function $J_{1}$, respectively; the constants $C_{nnj}$, $C_{nnnn}$, and $C_{nnr}$ (defined explicitly in Appendix \ref{sec:largekrc}) are integrals depending only on the Bessel functions themselves. Therefore, focusing on the overall $kr_{c}$ dependence, we find that
\begin{align}
    \Lambda_{\text{strong}}^{(\text{RS1})} \propto \sqrt{4 \pi}M_{\text{Pl}} e^{-\pi kr_{c}} = \sqrt{4\pi} \Lambda_{\pi}
\end{align}
at large $kr_{c}$, as anticipated. The precise value of the proportionality constant
depends weakly on the process considered, and in the large-$kr_c$ limit for the processes $(n,n) \to (n,n)$ we find
\begin{align}
    \begin{tabular}{| c | c c c c c |}
    \hline $n$ & $1$ & $2$ & $3$ & $4$ & $5$\\
    \hline $\Lambda^{\rm (RS1)}_{\rm strong}/\sqrt{4\pi} \Lambda_\pi $ & $2.701$ & $2.793$ & $2.812$ & $2.819$ & $2.822$\\
    \hline
    \end{tabular}~.
\end{align}
Since these results for the elastic scattering amplitudes follow from the form of the wavefunctions in Eq. (\ref{eq:asymptotic-wf}), similar results will follow for the inelastic amplitudes as well - and they will also be controlled by $\Lambda_\pi$.

\begin{figure*}
\centering
\includegraphics[scale=0.5]{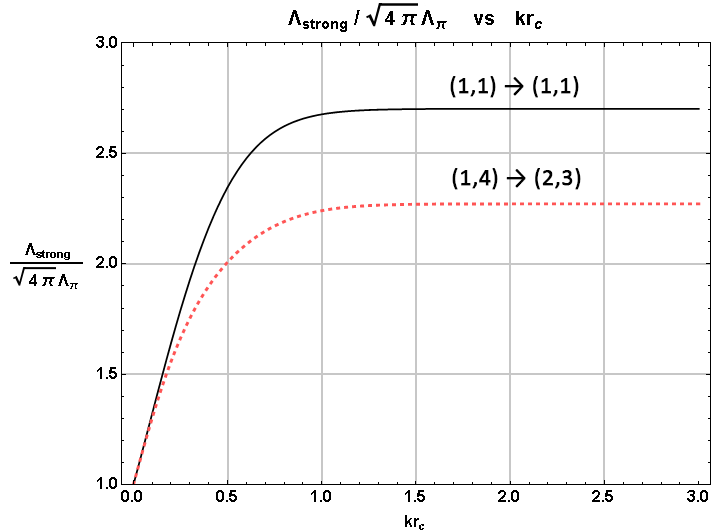}
\caption{The strong-coupling scale $\Lambda^{(\text{RS1})}_{\text{strong}}(kr_{c})$, Eq. (\ref{eq:strongscale}), as a function of $kr_c$ for the processes $(1,1)\to (1,1)$ and $(1,4)\to (2,3)$. We see that this scale is comparable to $\sqrt{4\pi}\Lambda_{\pi}$.}
\label{fig:strongcoupling}
\end{figure*}

We have also examined the dependence for lower values of $kr_c$ via the formula \eqref{LambdaStrong}. We display the dependence of $\Lambda^{\rm (RS1)}_{\rm strong}$ as a function of $kr_{c}$ for the processes $(1,1) \to (1,1)$ and $(1,4) \to (2,3)$ in Fig. \ref{fig:strongcoupling}. In all cases, we find that the strong-coupling scale is roughly $\Lambda_{\pi}$.

Therefore,  in the RS1 model, as conjectured under the AdS/CFT correspondence, all low-energy mass scales are controlled by the single emergent scale $\Lambda_{\pi}$.

\section{\label{sec:level7}Conclusion}

 We have studied the scattering amplitudes of massive spin-2 Kaluza-Klein excitations
in a gravitational theory with a single compact extra dimension, whether flat or warped. Our results have leveraged and expanded upon the work initially reported in \cite{Chivukula:2019rij,Chivukula:2019zkt}.  This paper includes a complete description of the computation of the tree-level two-body scattering amplitudes of the massive spin-2 states in compactified theories of five-dimensional gravity (Secs. \ref{sec:level2} -- \ref{sec:level5}), for all helicities of the incoming and outgoing states.  

These scattering amplitudes are characterized by intricate cancellations between different contributions: although individual contributions may grow as fast as ${\cal O}(s^5)$, the full results grow only as ${\cal O}(s)$ or slower. We have derived sum rules enforcing the cancellations and related them to results obtained by other groups.  We have demonstrated that the cancellations persist for all incoming and outgoing particle helicities and have documented how truncating the computation to only include a finite number of intermediate states impacts the accuracy of the results. 

We have also carefully assessed the range of validity of the low-energy Kaluza-Klein effective field theory (Sec. \ref{sec:level6}).  In particular, for the warped case we  have demonstrated  directly how an emergent low-energy scale controls the size of the scattering amplitude, as conjectured by the AdS/CFT correspondence. 

A number of interesting theoretical and phenomenological questions can now be addressed, including understanding the properties of scattering amplitudes in the presence of brane and/or bulk matter, the effects of radion stabilization, and the application of these results to the phenomenology of these models at colliders and in the early universe. 

This material is based upon work supported by the National Science Foundation under Grant No. PHY-1915147 .

\appendix

\section{\label{sec:levelA}Weak Field Expanded RS1 Lagrangian}\label{AppendixWFE}

This appendix provides details of the weak field expansion in both the RS1 and 5DOT models, including specifying the form of the interactions among up to four 5D fields.

Specifically, we summarize the interaction terms arising in the matter-free RS1 model Lagrangian $\mathcal{L}^{(\text{RS})}_{\text{5D}}$ from (\ref{L5Dy}) according to the expansions given in Eqs. (\ref{eq:weakfieldi}) -- (\ref{GMNRS}) through quartic interactions, as organized by 5D particle content:
\begin{align}
    \mathcal{L}^{(\text{RS})}_{\text{5D}} &= \mathcal{L}^{(\text{RS})}_{hh} + \mathcal{L}^{(\text{RS})}_{rr} + \mathcal{L}^{(\text{RS})}_{hhh} + \cdots +  \mathcal{L}^{(\text{RS})}_{rrr}\\
    &\hspace{15 pt}+ \mathcal{L}^{(\text{RS})}_{hhhh} + \cdots +  \mathcal{L}^{(\text{RS})}_{rrrr} + \cdots ~.\nonumber
\end{align}
We write $\hat{h}\equiv \hat{h}^{\mu}_{\mu}$ for the trace of a single undifferentiated graviton field. Primes indicate derivatives with respect to $y$. The trace of a product of graviton fields is indicated via twice-squared bracket notation, e.g. $\ltr \hat{h} \hat{h}^\prime \hat{h}^\prime\rtr = \hat{h}_{\mu\nu}(\partial_y \hat{h}^{\nu\rho})(\partial_y \hat{h}^{\mu}_{\rho})$. Similarly, $\ltr \hat{h}^\prime \rtr \equiv (\partial_y \hat{h}) = (\partial_y \hat{h}^{\mu}_{\mu})$. 
 
The equivalent weak field expansion of the 5DOT model Lagrangian $\mathcal{L}^{(\text{5DOT})}_{\text{5D}}$ is derived from these results by taking the limit $kr_c \rightarrow 0$ while maintaining finite nonzero $r_c$.

The 4D metric $g$ exactly satisfies
\begin{align}
    g_{\alpha\beta} = \eta_{\alpha\beta} +\kappa \hat{h}_{\alpha\beta}~.
\end{align}
From this, the 4D inverse metric $\tilde{g}$ may be solved for order-by-order by imposing its defining condition, $g_{\alpha\beta}\tilde{g}^{\beta\gamma} = \eta_{\alpha}^{\gamma}$, which implies
\begin{align}
\tilde{g}^{\alpha\beta} = \eta^{\alpha\beta} + \sum_{n=1}^{+\infty} (- \kappa)^{n} \ltr\hat{h}^{n}\rtr^{\alpha\beta}~.
\end{align}
Meanwhile, the 4D determinant equals
\begin{align}
    \sqrt{-\det\tilde{g}} &= \prod_{n=1}^{+\infty} \exp\left[\dfrac{(-1)^{n-1}}{2n}\kappa^n \hspace{2 pt}\ltr \hat{h}^{n}\rtr \right]~.
\end{align}
The first few terms of the determinant equal,
\begin{align}
\sqrt{-\det\tilde{g}}&=1 + \dfrac{\kappa}{2}\hspace{2 pt}\hat{h} + \dfrac{\kappa^2}{8}\left(\hat{h}^{2} - 2 \ltr \hat{h}\hat{h}\rtr\right)\\
&\hspace{10 pt}+ \dfrac{\kappa^3}{48}\left(\hat{h}^3 - 6 \hat{h}\ltr\hat{h}\hat{h}\rtr + 8\ltr\hat{h}\hat{h}\hat{h}\rtr\right) +\mathcal{O}(\kappa^4)~.\nonumber
\end{align}

Finally, separating the (B-Type) interactions that involve $y$ derivatives from the (A-Type) interactions that do not, we define $\overline{\mathcal{L}}_{A}$ and $\overline{\mathcal{L}}_{B}$ according to the following decomposition:
\begin{align}
    \mathcal{L}^{(\text{RS})}_{h^{H}r^{R}} &= \kappa^{H+R-2} \left[e^{-\pi kr_{c}}\varepsilon^{+2}\right]^{R}\nonumber\\
    &\hspace{20 pt}\cdot\bigg[\varepsilon^{-2}\overline{\mathcal{L}}_{A:h^{H}r^{R}}+\varepsilon^{-4}\overline{\mathcal{L}}_{B:h^{H}r^{R}}\bigg]~,
\end{align}
where $\varepsilon \equiv e^{-kr_{c}|\varphi|}$. 

\subsection{\label{sec:levelAa} Quadratic-Level Results}

\begin{align}
    \overline{\mathcal{L}}_{A:hh} &= - \hat{h}_{\mu\nu}(\partial^\mu \partial^\nu \hat{h}) + \hat{h}_{\mu\nu} (\partial^\mu \partial_\rho \hat{h}^{\rho\nu})\nonumber\\
     &\hspace{10 pt}- \dfrac{1}{2} \hat{h}_{\mu\nu}(\square \hat{h}^{\mu\nu}) + \dfrac{1}{2} \hat{h}(\square \hat{h})~,\\
  \overline{\mathcal{L}}_{B:hh} &= - \dfrac{1}{2} \ltr \hat{h}^\prime \hat{h}^\prime \rtr + \dfrac{1}{2} \ltr \hat{h}^\prime \rtr^2~,\\
    &\text{ }\nonumber\\
    \overline{\mathcal{L}}_{A:rr} &= \frac{1}{2} (\partial_\mu \hat{r})(\partial^\mu \hat{r})~.
\end{align}

\subsection{\label{sec:levelAb} Cubic-Level Results}

\begin{align}
    \overline{\mathcal{L}}_{A:hhh} &= \dfrac{1}{2} \hat{h}_{\mu\nu} \hat{h}_{\rho\sigma} (\partial^\mu \partial^\nu \hat{h}^{\rho\sigma})-\dfrac{1}{2} \hat{h} \hat{h}_{\mu\nu}(\partial^\mu \partial^\nu \hat{h})\nonumber\\
    &\hspace{10 pt}-2 \hat{h}_{\mu\nu} (\partial^\mu \hat{h})(\partial_\rho \hat{h}^{\rho\nu})- \hat{h}_{\mu\nu} \hat{h}^{\nu \rho}(\partial^\mu \partial^\sigma \hat{h}_{\sigma\rho})\nonumber\\
    &\hspace{10 pt}+ \hat{h}_{\mu\nu} (\partial^\mu \hat{h}_{\rho\sigma}) (\partial^\sigma \hat{h}^{\nu\rho})-\dfrac{1}{4} \hat{h} \hat{h}_{\mu\nu} (\square \hat{h}^{\mu\nu})\nonumber\\
    &\hspace{10 pt}+\dfrac{3}{4} \hat{h}_{\mu\nu} (\partial_\rho \hat{h}^{\mu\nu})(\partial^\rho \hat{h})+\dfrac{1}{2} \hat{h}_{\mu\nu} \hat{h}^{\nu \rho} (\square \hat{h}^\mu_\rho)\nonumber\\
    &\hspace{10 pt}-\dfrac{1}{2} \hat{h}_{\mu\nu} (\partial^\rho \hat{h}_{\rho\sigma})(\partial^\sigma \hat{h}^{\mu\nu})+\dfrac{1}{2} \hat{h} (\partial_\mu \hat{h}_{\nu\rho})(\partial^\nu \hat{h}^{\mu\rho})\nonumber\\
    &\hspace{10 pt}- \hat{h} (\partial^\mu \hat{h}_{\mu\nu})(\partial_\rho \hat{h}^{\rho \nu})+\dfrac{1}{8} \hat{h}^2 (\square \hat{h})~,\\
    \overline{\mathcal{L}}_{B:hhh} &= - \ltr \hat{h}^\prime \rtr \ltr \hat{h} \hat{h}^\prime \rtr + \ltr \hat{h} \hat{h}^\prime \hat{h}^\prime \rtr-\dfrac{1}{4} \hat{h}\ltr \hat{h}^\prime \hat{h}^\prime \rtr +\dfrac{1}{4} \hat{h} \ltr \hat{h}^\prime \rtr^2~,\\[2 em]
%
    \overline{\mathcal{L}}_{A:hhr} &= 0~,\\
    \overline{\mathcal{L}}_{B:hhr} &= \sqrt{\dfrac{3}{8}}\left[\hat{r}\left(\ltr \hat{h}^\prime \hat{h}^\prime\rtr - \ltr \hat{h}^\prime\rtr^2\right)\right]~,\\[2 em]
%
    \overline{\mathcal{L}}_{A:hrr} &= -\dfrac{1}{24} \left[2 \hat{r}^2 (\partial^\mu\partial^\nu\hat{h}_{\mu\nu}) - 12 \hat{r}(\partial^\mu \partial^\nu \hat{r}) \hat{h}_{\mu\nu}\right.\nonumber\\
    &\hspace{10 pt}\left.+ \hat{r}^2 (\square \hat{h}) + 6 \hat{r}(\square \hat{r}) \hat{h}\right]~,\\
    \overline{\mathcal{L}}_{B:hrr} &= 0~,\\[2 em]
%
    \overline{\mathcal{L}}_{A:rrr} &= \dfrac{1}{2\sqrt{6}} \hat{r}^2 (\square \hat{r})~,\\
    \overline{\mathcal{L}}_{B:rrr} &= 0~.
\end{align}

\subsection{\label{sec:levelAc} Quartic-Level Results}

\begin{align}
    \overline{\mathcal{L}}_{A:hhhh} &= 
    \dfrac{1}{4}\hat{h} \hat{h}_{\mu\nu} \hat{h}_{\rho\sigma} (\partial^\mu \partial^\nu \hat{h}^{\rho\sigma}) - \hat{h}_{\mu\nu} \hat{h}_{\rho\sigma} \hat{h}^{\sigma\tau} (\partial^\mu\partial^\nu \hat{h}^\rho_\tau)
    \nonumber\\
    & -\dfrac{3}{4} \hat{h}_{\mu\nu} \hat{h}_{\rho\sigma} (\partial^\mu \hat{h}^{\rho\sigma})(\partial^\nu \hat{h}) +\hat{h}_{\mu\nu} \hat{h}_{\rho\sigma}\hat{h}^{\sigma\tau} (\partial^\mu \partial^\rho \hat{h}^\nu_\tau) \nonumber\\
    &-\dfrac{1}{4}\ltr \hat{h}\hat{h} \rtr \hat{h}_{\mu\nu}(\partial^\mu \partial_\rho \hat{h}^{\rho\nu}) - \hat{h}_{\mu\nu} \hat{h}_{\rho\sigma} (\partial^\mu \hat{h}^{\rho\tau})(\partial^\nu \hat{h}^\sigma_\tau)\nonumber\\
    &+\hat{h}_{\mu\nu}\hat{h}_{\rho\sigma} (\partial^\mu \hat{h}^{\nu\tau})(\partial^\rho \hat{h}^\sigma_\tau)
    +\dfrac{1}{8} \hat{h}^2 \hat{h}_{\mu\nu} (\partial^\mu \partial_\rho \hat{h}^{\rho \nu})\nonumber\\
    &-\dfrac{1}{8} \hat{h}^2 \hat{h}^{\mu\nu} (\partial^\mu\partial^\nu \hat{h})-\dfrac{1}{2} \hat{h}_{\mu\nu} \hat{h}_{\rho\sigma} (\partial^\mu \hat{h}^{\nu\tau})(\partial_\tau \hat{h}^{\rho\sigma})\nonumber\\
    & +\hat{h}_{\mu\nu} \hat{h}_{\rho\sigma} (\partial_\tau \hat{h}^{\tau \rho}) (\partial^\mu \hat{h}^{\nu\sigma}) +\dfrac{1}{4} \hat{h} \hat{h}_{\mu\nu} (\partial^\mu \hat{h}^{\nu\rho})(\partial_\rho \hat{h})\nonumber\\
    &-\hat{h}_{\mu\nu}\hat{h}_{\rho\sigma} \hat{h}^{\mu\rho}(\partial^\nu \partial^\sigma \hat{h})
    +\hat{h}_{\mu\nu} \hat{h}_{\rho\sigma} \hat{h}^{\mu\rho} (\partial^\nu \partial_\tau \hat{h}^{\tau \sigma})\nonumber\\
    &-\hat{h}_{\mu\nu} \hat{h}_{\rho\sigma} (\partial^\mu \hat{h}^{\rho\tau})(\partial_\tau \hat{h}^{\nu\sigma})+\dfrac{1}{2}\hat{h} \hat{h}_{\mu\nu} \hat{h}^{\nu \rho} (\partial^\mu \partial_\rho \hat{h})\nonumber\\
    &-\dfrac{1}{2}\hat{h} \hat{h}_{\mu\nu} \hat{h}^{\nu\rho} (\partial^\mu \partial^\sigma \hat{h}_{\sigma\rho})
    +\dfrac{1}{2}\hat{h} \hat{h}_{\mu\nu} (\partial^\mu \hat{h}_{\rho\sigma})(\partial^\rho \hat{h}^{\nu\sigma})\nonumber\\
    &-\hat{h}_{\mu\nu}\hat{h}_{\rho\sigma}(\partial^\mu \hat{h}^{\rho\tau})(\partial^\sigma \hat{h}^\nu_\tau)
    -\hat{h}_{\mu\nu} \hat{h}^{\nu\rho} \hat{h}_{\sigma \tau} (\partial^\mu \partial_\rho \hat{h}^{\sigma\tau})\nonumber\\
    &-\dfrac{1}{2} \hat{h}_{\mu\nu} \hat{h}^{\nu\rho}(\partial^\mu \hat{h}_{\sigma\tau})(\partial_\rho \hat{h}^{\sigma\tau})+\hat{h} \hat{h}_{\mu\nu}\hat{h}^{\nu\rho} (\square \hat{h}^\mu_\rho)\nonumber\\
    &+\dfrac{1}{8} \ltr \hat{h} \hat{h} \rtr \hat{h}_{\mu\nu}(\square \hat{h}^{\mu\nu})
    -\dfrac{1}{4} \hat{h}_{\mu\nu} \hat{h}_{\rho\sigma} (\partial^\tau \hat{h}^{\mu\nu})(\partial_\tau \hat{h}^{\rho\sigma})\nonumber\\
    &-\dfrac{1}{16} \hat{h}^2 \hat{h}_{\mu\nu} (\square \hat{h}^{\mu\nu})
    +\dfrac{1}{8}\hat{h} \hat{h}_{\mu\nu} (\partial^\sigma \hat{h}^{\mu\nu})(\partial_\sigma \hat{h})\nonumber\\
    &-\dfrac{1}{2} \hat{h}_{\mu\nu}\hat{h}_{\rho\sigma} \hat{h}^{\mu\rho}(\square \hat{h}^{\nu\sigma}) -\dfrac{1}{2} \hat{h}_{\mu\nu} \hat{h}^{\nu\rho} (\partial^\mu \hat{h}_{\rho\sigma})(\partial^\sigma \hat{h})\nonumber\\
    &+\dfrac{3}{2} \hat{h}\hat{h}_{\mu\nu} (\partial_\rho \hat{h}^{\mu\sigma})(\partial^\rho \hat{h}^\nu_\sigma)+\dfrac{1}{48} \hat{h}^3(\square \hat{h})
    \nonumber\\
    &+\dfrac{1}{4} \hat{h}\hat{h}_{\mu\nu} (\partial^\rho \hat{h}_{\rho\sigma})(\partial^\sigma \hat{h}^{\mu\nu})+\dfrac{1}{4} \hat{h}\ltr \hat{h}\hat{h} \rtr (\partial^\mu\partial^\nu \hat{h}_{\mu\nu})\nonumber\\
    &-\dfrac{1}{8} \hat{h}\ltr \hat{h}\hat{h} \rtr (\square \hat{h})~.\\
    \overline{\mathcal{L}}_{B:hhhh} &= \dfrac{1}{2} \ltr \hat{h} \hat{h}^\prime \rtr^2
    -\dfrac{1}{2} \hat{h} \ltr \hat{h}^\prime \rtr \ltr \hat{h} \hat{h}^\prime \rtr 
    -\dfrac{1}{2} \ltr \hat{h} \hat{h}^\prime \hat{h} \hat{h}^\prime \rtr\nonumber\\
    &+\dfrac{1}{2} \hat{h} \ltr \hat{h} \hat{h}^\prime \hat{h}^\prime \rtr + \ltr \hat{h}^\prime \rtr \ltr \hat{h} \hat{h} \hat{h}^\prime \rtr-\ltr \hat{h} \hat{h} \hat{h}^\prime \hat{h}^\prime \rtr\nonumber\\
    &+\dfrac{1}{8} \ltr \hat{h} \hat{h} \rtr \ltr \hat{h}^\prime \hat{h}^\prime \rtr-\dfrac{1}{8} \ltr \hat{h} \hat{h} \rtr \ltr \hat{h}^\prime \rtr^2
    -\dfrac{1}{16} \hat{h}^2 \ltr \hat{h}^\prime \hat{h}^\prime \rtr\nonumber\\
    &+\dfrac{1}{16} \hat{h}^2 \ltr \hat{h}^\prime \rtr^2~,\\[2 em]
%
    \overline{\mathcal{L}}_{A:hhhr} &=0~,\\
    \overline{\mathcal{L}}_{B:hhhr} &= \sqrt{\dfrac{3}{32}} \left[\hat{r}\hat{h} \left( \ltr \hat{h}^\prime\hat{h}^\prime\rtr - \ltr \hat{h}^\prime \rtr^2 \right)\right.\nonumber\\
    &\hspace{10 pt}\left.+ 4 \hat{r} \left( \ltr \hat{h}^\prime\rtr \ltr \hat{h} \hat{h}^\prime\rtr   - \ltr \hat{h} \hat{h}^\prime \hat{h}^\prime \rtr\right)\right]~,
\end{align}

\begin{align}
    \overline{\mathcal{L}}_{A:hhrr} &= \dfrac{1}{48} \left[2 \hat{r}^2 \hat{h}_{\mu\nu} (\partial^\mu \partial^\nu \hat{h})+ 2\hat{r}^2 \hat{h}^{\mu\nu}(\square \hat{h}_{\mu\nu}) \right.\nonumber\\
    &+8\hat{r}^2 (\partial^\mu \hat{h}_{\mu\rho})(\partial_\nu \hat{h}^{\nu\rho})+ 32 \hat{r} (\partial^\mu \hat{r}) \hat{h}_{\mu\nu} (\partial_\rho \hat{h}^{\nu\rho})\nonumber\\
    &+ 24(\partial^\mu \hat{r})(\partial_\nu \hat{r}) \hat{h}_{\mu\rho}\hat{h}^{\rho\nu}+6 \hat{r} (\square \hat{r})\ltr \hat{h}\hat{h}\rtr \nonumber\\
    &+ 2\hat{r}^2 \hat{h} (\partial^\mu\partial^\nu \hat{h}_{\mu\nu})+8 \hat{r}(\partial^\mu \hat{r})\hat{h} (\partial^\nu \hat{h}_{\mu\nu})\nonumber\\
    &- \hat{r}^2 \hat{h} \square \hat{h}  +12 \hat{r} (\partial^\mu\partial^\nu \hat{r}) \hat{h} \hat{h}_{\mu\nu} - 3 \hat{r}(\square \hat{r})\hat{h}^2\nonumber\\
    &\left. + \hat{r}^2 (\partial^\mu \hat{h})(\partial_\mu \hat{h}) - 4 \hat{r}^2 (\partial_\nu \hat{h}_{\mu \rho})(\partial^\mu \hat{h}^{\nu\rho})\right]~,\\
   \overline{\mathcal{L}}_{B:hhrr} &= - \dfrac{5}{12} \left[\hat{r}^2\left( \ltr \hat{h}^\prime \hat{h}^\prime \rtr - \ltr \hat{h}^\prime \rtr^2 \right)\right]~,\\[2 em]
%
    \overline{\mathcal{L}}_{A:hrrr} &= \dfrac{1}{36 \sqrt{6}}\left[\hat{r}^3 \left(2 (\partial^\mu\partial^\nu \hat{h}_{\mu\nu}) + \square \hat{h}\right)\right.\nonumber\\
    &\hspace{10 pt}\left.- 9 \hat{r}^2 \left(2 (\partial^\mu \partial^\nu \hat{r}) \hat{h}_{\mu\nu} - (\square \hat{r})\hat{h}\right)\right]~,\\
    \overline{\mathcal{L}}_{B:hrrr} &=0~,\\[2 em]
%
    \overline{\mathcal{L}}_{A:rrrr} &= - \dfrac{1}{24}  \hat{r}^3 (\square \hat{r})~,\\
    \overline{\mathcal{L}}_{B:rrrr} &= 0~.
\end{align}

\section{\label{sec:levelB}Massive Bulk Fields and Wavefunctions}\label{AppendixKKGeneral}

This appendix provides a general analysis of the properties of the extra-dimensional wavefunctions. 

Consider a massive 5D field $\Phi_{\vec{\alpha}}(x,y)$ defined over the 5D bulk by a Lagrangian
\begin{align}
    \mathcal{L}_{\text{5D}} &=
    Q_{A}^{\mu\vec{\alpha}\nu\vec{\beta}} e^{-2 k|y|} (\partial_\mu \Phi_{\vec{\alpha}}) (\partial_\nu \Phi_{\vec{\beta}})\label{L5DappB}
    \\&+ Q_{B}^{\vec{\alpha}\vec{\beta}}
        \left\{
            e^{-4k|y|} (\partial_y \Phi_{\vec{\alpha}}) (\partial_y \Phi_{\vec{\beta}})
            + m_{\Phi}^2 e^{-4k|y|}  \Phi_{\vec{\alpha}} \Phi_{\vec{\beta}}
        \right\}~,\nonumber
\end{align}
where the index $\vec{\alpha}$ is a list of Lorentz indices, $k$ is for now a real-valued parameter, and $m_{\Phi}$ is the 5D mass of the field. The tensors $Q_{A}^{\mu\vec{\alpha}\nu\vec{\beta}}$ and $Q_{B}^{\vec{\alpha}\vec{\beta}}$ will have forms (defined below) chosen to yield 4D canonical kinetic terms for the KK modes of different spins. This is a generalization of the quadratic terms in the 5D graviton Lagrangian from the RS1 scenario, and will allow us to consider spin-2 and spin-0 fields simultaneously. By integrating by parts and discarding the surface terms (which vanish when the orbifold symmetry is imposed),
\begin{align}
    \mathcal{L}_{\text{5D}}& =
    Q_{A}^{\mu\vec{\alpha}\nu\vec{\beta}} e^{-2 k|y|} (\partial_\mu \Phi_{\vec{\alpha}}) (\partial_\nu \Phi_{\vec{\beta}})\\
    &\hspace{-10 pt}+ Q_{B}^{\vec{\alpha}\vec{\beta}}
        \left\{
           -\Phi_{\vec{\alpha}}\cdot\partial_y \left[ e^{-4k|y|} (\partial_y \Phi_{\vec{\beta}}) \right]
            + m_{\Phi}^2 e^{-4k|y|} \Phi_{\vec{\alpha}} \Phi_{\vec{\beta}}
        \right\}~.\nonumber
\end{align}
Performing a mode expansion (KK decomposition) according to the ansatz
\begin{align}
    \Phi_{\vec{\alpha}}(x,y) = \dfrac{1}{\sqrt{\pi r_c}}\sum_{n=0}^{+\infty}  \Phi_{\vec{\alpha}}^{(n)}(x)\psi_{n}(y)~,
\end{align}
we obtain
\begin{align}
    \mathcal{L}_{\text{5D}} & =
    \dfrac{1}{\pi r_c} \hspace{-2 pt}\sum_{m,n=0}^{+\infty}\hspace{-5 pt}
    Q_{A}^{\mu\vec{\alpha}\nu\vec{\beta}} (\partial_\mu \Phi^{(m)}_{\vec{\alpha}}) (\partial_\nu \Phi^{(n)}_{\vec{\beta}})\hspace{2 pt}e^{-2 k|y|} \psi^{(m)}\psi^{(n)}\nonumber\\
    & + Q_{B}^{\vec{\alpha}\vec{\beta}} \Phi^{(m)}_{\vec{\alpha}}\Phi^{(n)}_{\vec{\beta}}\hspace{2 pt}
        \psi^{(m)}\bigg\{
            -\partial_y \bigg[ e^{-4k|y|} (\partial_y \psi_{n}) \bigg]\nonumber\\
            &\hspace{50 pt}+ m_{\Phi}^2 e^{-4k|y|} \psi_{n}
        \bigg\}~.
\end{align}
Integrating over the extra dimension then yields the following effective 4D Lagrangian:
\begin{align}
    \mathcal{L}_{\text{4D}}^{(\text{eff})} &= \sum_{m,n=0}^{+\infty} Q_{A}^{\mu\vec{\alpha}\nu\vec{\beta}} (\partial_\mu \Phi^{(m)}_{\vec{\alpha}}) (\partial_\nu \Phi^{(n)}_{\vec{\beta}}) \cdot N_{A}^{(m,n)}\nonumber\\
    &\hspace{50 pt}+ Q_{B}^{\vec{\alpha}\vec{\beta}} \Phi^{(m)}_{\vec{\alpha}}\Phi^{(n)}_{\vec{\beta}}\cdot N_{B}^{(m,n)}~,\label{L4Dgenmass}
\end{align}
where
\begin{align}
    N_{A}^{(m,n)} &= \dfrac{1}{\pi r_c}\int_{-\pi r_c}^{+\pi r_c}dy\hspace{5 pt} e^{-2 k|y|} \psi_{m}\psi_{n}~,\label{NAmn}\\
    N_{B}^{(m,n)} &=  \dfrac{1}{\pi r_c}\int_{-\pi r_c}^{+\pi r_c}dy\hspace{5 pt} \psi_{m}\left\{
            - \partial_y \bigg[ e^{-4k|y|} (\partial_y \psi_{n}) \right]\nonumber\\
            &\hspace{50 pt}+ m_{\Phi}^2 e^{-4k|y|} \psi_{n}\bigg\}~.\label{NBmn}
\end{align}
We desire that this process yield a particle spectrum described by canonical 4D Lagrangians for particles of differing spins and masses. Specifically, a given mode field $\phi(x)$ in the KK spectrum must have canonical kinetic and mass terms in the Lagrangian
\begin{align}
q_{A}^{\mu\vec{\alpha}\nu\vec{\beta}}(\partial_\mu \phi_{\vec{\alpha}})(\partial_\nu \phi_{\vec{\beta}}) + m^2 q_{B}^{\vec{\alpha}\vec{\beta}} \phi_{\vec{\alpha}}\phi_{\vec{\beta}}~,\label{L4DeffappB}
\end{align}
where $m$ is the mass of the KK mode. For a full KK tower, the corresponding canonical quadratic Lagrangian equals (indexing KK number by $n$),
\begin{align}
    \mathcal{L}_{\text{4D}}^{(\text{eff})} = \sum_{n=0}^{+\infty} \hspace{5 pt} q_{A}^{\mu\vec{\alpha}\nu\vec{\beta}}(\partial_\mu \phi^{(n)}_{\vec{\alpha}})(\partial_\nu \phi^{(n)}_{\vec{\beta}}) + m_n^2 q_{B}^{\vec{\alpha}\vec{\beta}} \phi^{(n)}_{\vec{\alpha}}\phi^{(n)}_{\vec{\beta}}~.
\end{align}
Comparing to Eq. \eqref{L4Dgenmass}, one recovers this form for the choices $Q=q$ (i.e. if the 5D quadratic tensor structures mimic the 4D canonical quadratic tensor structures), $\Phi = \phi$, $N_{A}^{(m,n)} = \delta_{m,n}$, and $N_{B}^{(m,n)} = m_n^2 \delta_{m,n}$. Consider the condition on $N_{B}^{(m,n)}$ in more detail:
\begin{align}
    &\dfrac{1}{\pi r_c}\int_{-\pi r_c}^{+\pi r_c}dy\hspace{5 pt} \psi^{(m)}\bigg\{
            -\partial_y \bigg[ e^{-4k|y|} (\partial_y \psi_{n}) \bigg]\nonumber\\
            &\hspace{50 pt}+ m_{\Phi}^2 e^{-4k|y|} \psi_{n}\bigg\} = m_n^2 \delta_{m,n}~.
\end{align}
Using the condition on $N_{A}^{(m,n)}$, this implies
\begin{align}
    &\int_{-\pi r_c}^{+\pi r_c}dy\hspace{5 pt} \psi_{m}\bigg\{
            \partial_y \bigg[ e^{-4k|y|} (\partial_y \psi_{n}) \bigg]\label{DerivingSLEq}\\
            &\hspace{50 pt}+ \bigg( m_n^2 e^{-2k|y|} - m_{\Phi}^2 e^{-4k|y|} \bigg)\psi_{n}\bigg\} = 0~.\nonumber
\end{align}
Anticipating that the $\{\psi_{m}\}$ can be made to form a complete set, Eq. \eqref{DerivingSLEq} would imply that the $\psi_{n}$ are solutions of the following differential equation
\begin{align}
   \partial_y \bigg[ e^{-4k|y|} (\partial_y \psi_{n}) \bigg]
            + \bigg( m_n^2 e^{-2k|y|} - m_{\Phi}^2 e^{-4k|y|} \bigg)\psi_{n} = 0~,
\end{align}
or, when expressed in unitless combinations,
\begin{align}
    &0 = \partial_{\varphi}\bigg[ e^{-4kr_{c}|\varphi|} (\partial_{\varphi} \psi_{n}) \bigg]
             \label{SLeq}\\
    &\hspace{25 pt}+ \bigg( (m_{n}r_{c})^{2} e^{-2kr_{c}|\varphi|}- (m_{\Phi}r_{c})^2 e^{-4kr_{c}|\varphi|} \bigg)\psi_{n}~.\nonumber
\end{align}
In addition, orbifold symmetry requires that the wavefunctions vanish at the orbifold fixed points - providing boundary conditions.
 Finding the solution set $\{\psi_{n}\}$ (and corresponding values of $\{m_nr_{c}\}$) of the above equation is precisely a Sturm-Liouville (SL) problem, for which there is guaranteed a discrete (complete) basis of real wavefunctions satisfying
\begin{align}
    \dfrac{1}{\pi} \int_{-\pi}^{+\pi} d\varphi\hspace{5 pt}e^{-2kr_{c}|\varphi|}\psi_{m}\psi_{n} = N_{A}^{(m,n)} \equiv \delta_{m,n}~, \label{onA}
\end{align}
as required. Hence, by finding wavefunctions $\psi_{n}$ that solve Eqs. \eqref{SLeq} and \eqref{onA}, we can KK decompose the fields in Eq. \eqref{L5DappB} according to the ansatz and (so long as $Q=q$) obtain a tower of canonical quadratic Lagrangians \eqref{L4DeffappB}.

Examining Eq. \eqref{SLeq}, we see that if $m_{\Phi}=0$ there is a massless flat solution, {\it i.e.} with $\partial_y\psi_{0} = 0$.
Hence a massless 5D graviton will give rise to 4D massless graviton and radion modes
in this framework.\footnote{Conversely, to prevent the 4D radion from contributing to long-range gravitational forces we must include interactions which make the physical 4D spin-0 field become massive, as occurs during radion stabilization \cite{Goldberger:1999wh}.}
Normalization fixes  $\psi_{0}$ to equal
\begin{align}
    \psi_{0}=\sqrt{\dfrac{kr_c\pi}{1-e^{-2kr_c\pi}}}~.\label{psi0gen}
\end{align}

By construction, the SL equation combined with $\eqref{onA}$ implies an additional quadratic integral condition:
\begin{align}
    &\dfrac{1}{\pi} \int_{-\pi }^{+\pi} d\varphi\hspace{5 pt} e^{-4kr_{c}|\varphi|}\bigg[ (\partial_{\varphi}\psi_{m})(\partial_{\varphi}\psi_{n})\nonumber\\
    &\hspace{50 pt}+ (m_{\Phi} r_{c})^{2} \psi_{m}\psi_{n}\bigg] = (m_{n}r_{c})^{2} \delta_{m,n} ~.\label{onB}
\end{align}
When $m_\Phi = 0$, this becomes an orthonormality condition on the set $\{\partial_{\varphi} \psi_{n}\}$.

The existence of a discrete solution set of wavefunctions is guaranteed by the SL problem. We now summarize how to find explicit equations for the nonflat wavefunctions in that solution set by following the notation and arguments from \cite{Goldberger:1999wh}. Note that
\begin{align}
\partial_{\varphi}|\varphi| &= \text{sign}(\varphi)~,\\
\partial^{2}_{\varphi}|\varphi| &= 2\big[\delta(\varphi) - \delta(\varphi - \pi)\big]~,
\end{align}
such that $(\partial_{\varphi}|\varphi|)^{2} = 1$ and $\partial_{\varphi}^{2}|\varphi|=0$ when $\varphi\neq 0,\pi$. Thus, away from the orbifold fixed points, Eq. \eqref{SLeq} may be rewritten by defining quantities $z_n = (m_n/k)e^{+k|y|}$ and $f_{n} = (m_n^2/k^2) \psi_{n}/z_n^2$, such that
\begin{align}
    0 = z_n^2 \dfrac{d^2 f_n}{dz_n^2} + z_n \dfrac{df_n}{dz_n} + \left[ z_n^2 - \left( 4 + \dfrac{m_{\Phi}^2}{k^2}\right)\right] f_n~.
\end{align}
When $m_{\Phi} = 0$, this differential equation is solved by $f_n$ equal to Bessel functions $J_2(z_n)$ or $Y_2(z_n)$. When $m_{\Phi} \neq 0$, it is instead solved by Bessel functions $J_\nu(z_n)$ and $Y_\nu(z_n)$ where $\nu^2 \equiv 4 + m_\Phi^2 /k^2$. Taking a superposition of the appropriate Bessel functions yields a generic solution $f_n$, which may then be converted back to $\psi_{n}$. By demanding that the SL boundary condition $\partial_{\varphi}\psi_{n} =0$ is satisfied simultaneously at both orbifold fixed points, the wavefunctions are found to equal
\begin{align}
    \psi_{n} = \dfrac{\varepsilon^{2}}{N_n} \left[J_{\nu} \left(\dfrac{\mu_{n}\varepsilon}{kr_{c}} \right) + b_{n\nu}\s Y_{\nu}\left(\dfrac{\mu_{n}\varepsilon}{kr_{c}}\right)\right]~,
    \label{eq:wavefunction}
\end{align}
where $\varepsilon\equiv e^{+kr_{c}|\varphi|}$ and $\mu_{n}\equiv m_{n}r_{c}$, the normalization $N_{n}$ is determined by Eq. \eqref{onA} (up to a sign that we fix by setting $N_{n}>0$ and which yields $\psi_{n}(0)<0$ for nonzero $n$), and the relative weight $b_{n\nu}$ equals
\begin{align}
    b_{n\nu} = - \dfrac{2J_{\nu}\Big|_{\mu_{n}/kr_{c}} + \dfrac{\mu_{n}}{kr_{c}}(\partial J_{\nu})\Big|_{\mu_{n}/kr_{c}}}{2Y_{\nu}\Big|_{\mu_{n}/kr_{c}} + \dfrac{\mu_{n}}{kr_{c}}(\partial Y_{\nu})\Big|_{\mu_{n}/kr_{c}}}~,\label{bnnu}
\end{align}
where $\partial J_{\nu} \equiv \partial J_{\nu}(z)/\partial z$ and $\partial Y_{\nu} \equiv \partial Y_{\nu}(z)/\partial z$. These wavefunctions satisfy Eq. \eqref{onB} where each $\mu_{n}$ solves
\begin{align}
    0 =& \left[2 J_{\nu} + \dfrac{\mu_{n}\varepsilon}{kr_{c}} (\partial J_{\nu})\right]\bigg|_{\varphi = \pi}\left[2 Y_{\nu} + \dfrac{\mu_{n}\varepsilon}{kr_{c}}  (\partial Y_{\nu})\right]\bigg|_{\varphi = 0}\nonumber\\
    &-\left[2 Y_{\nu} + \dfrac{\mu_{n}\varepsilon}{kr_{c}} (\partial Y_{\nu})\right]\bigg|_{\varphi = \pi}\left[2 J_{\nu} + \dfrac{\mu_{n}\varepsilon}{kr_{c}}  (\partial J_{\nu})\right]\bigg|_{\varphi = 0}~.\label{mn}
\end{align}
Although these wavefunctions were derived by solving Eq. \eqref{SLeq} away from the orbifold fixed points, they solve the equation across the full extra dimension. In particular, they ensure $\partial_{\varphi}^{2}\psi_{n} = [(m_{\Phi}r_{c})^{2} - \varepsilon^{2} \mu_{n}^{2}]\psi_{n}$ at $\varphi = 0,\pi$.

Finally, note that given a 5D Lagrangian consistent with Eq. \eqref{L5DappB}, the wavefunctions $\psi_{n}$ and spectrum $\{\mu_{n}\}$ are entirely determined by the unitless quantities $kr_c$ and $m_{\Phi}r_{c}$. In the RS1 model, the 5D graviton field lacks a bulk mass ($m_{\Phi}=0$ such that $\nu = 2$), so its KK decomposition is dictated by $kr_{c}$ alone.

\section{\label{sec:levelC}4D Effective RS1 Model}\label{AppendixKKResults}

This appendix gives a more detailed description of the 4D interactions.

\subsection{\label{sec:levelCa}General Procedure}

The WFE RS1 Lagrangian equals a sum of terms, wherein each term contains some number of 5D fields and exactly two derivatives. Each derivative in the pair is either a 4D spatial derivative $\partial_\mu$ or an extra-dimension derivative $\partial_y$, and each field is either a radion $\hat{r}$ or a graviton $\hat{h}_{\mu\nu}$. Because the Lagrangian requires an even number of Lorentz indices in order to form a Lorentz scalar, each derivative pair must consist of two copies of the same kind of derivative, i.e. each term in $\mathcal{L}^{(\text{RS})}_{\text{5D}}$ can be classified into one of two categories:
\begin{itemize}
    \item {\bf A-Type:} The term has two spatial derivatives $\partial_\mu \cdot \partial_\nu$; or
    \item {\bf B-Type:} The term has two extra-dimensional derivatives $\partial_y \cdot \partial_y$~.
\end{itemize}
In addition to fields and derivatives, every term in $\mathcal{L}^{(\text{RS})}_{\text{5D}}$ has an exponential prefactor. That exponential's specific form is entirely determined by its type (whether A- or B-Type) and the number of 5D radion fields in the term. Each A-Type term is associated with a factor $\varepsilon^{-2} = e^{-2kr_c|\varphi|}$ whereas each B-Type term is associated with a factor $\varepsilon^{-4} = e^{-4kr_c|\varphi|}$, and every instance of a radion field provides an additional $e^{-\pi kr_{c}}\varepsilon^{+2}$ factor. These assignments correctly reproduce the prefactors of Appendix \ref{AppendixWFE}.

Consider a generic A-Type term with $H$ spin-2 fields and $R$ radion fields. Schematically, it will be of the form,
\begin{align}
    X_A&\equiv \kappa^{(H+R-2)}\left[ \varepsilon^{-2}\right] \left[ e^{-\pi kr_{c}}\varepsilon^{+2} \right]^R (\partial_\mu^2,\hat{h}^{H}, \hat{r}^R) \nonumber\\
    &= \kappa^{(H+R-2)}e^{-R\pi kr_c}\varepsilon^{2(R-1)} \overline{X}_A~,
\end{align}
where the combination $\overline{X}_A \equiv (\partial_\mu^2,\hat{h}^{H}, \hat{r}^R)$ refers to a fully contracted product of two 4D derivatives, $H$ gravitons, and $R$ radions. The $\mu$ label on $\partial_\mu^2$ above is only schematic and not literal. Similarly, an equivalent B-Type term would be of the form,
\begin{align}
    X_B&\equiv \kappa^{(H+R-2)}\left[ \varepsilon^{-4}\right] \left[ e^{-\pi kr_{c}}\varepsilon^{+2}\right]^R(\partial_y^2,\hat{h}^{H},\hat{r}^R)\nonumber\\
    &= \kappa^{(H+R-2)}e^{-R\pi kr_c}\varepsilon^{2(R-2)} \overline{X}_B~,
\end{align}
where the combination $\overline{X}_B \equiv (\partial_y^2,\hat{h}^{H}, \hat{r}^R)$ refers to a fully contracted product of two extra-dimensional derivatives, $H$ gravitons, and $R$ radions. By construction, each B-Type term we consider never has both of its $\partial_y$ derivatives acting on the same field (i.e. any instances of $\partial_y^2\hat{h}$ in our 5D Lagrangian have been removed via integration by parts), and so we assume $\overline{X}_B$ also satisfies this property.

We form a 4D effective Lagrangian by first KK decomposing our 5D fields into states of definite mass (Eq. \eqref{KKreduce}) and then integrating over the extra dimension (Eq. \eqref{L4Deff Def}). For the schematic A-Type term, this procedure yields,
\begin{align}
    X_A^{(\text{eff})} =& \dfrac{r_c}{(\pi r_c)^{(H+R)/2}}\kappa^{(H+R-2)}\\
    &\hspace{-30 pt}\sum_{n_1,\cdots,n_H=0}^{+\infty} \left(\partial_\mu^2,\hspace{6 pt} \hat{h}^{(n_1)}\cdots \hat{h}^{(n_H)},\hspace{6 pt} \left[\hat{r}^{(0)}\right]^R \right)\nonumber\\
    &\hspace{-30 pt}\times e^{-R\pi kr_c}\int_{-\pi}^{+\pi} d\varphi\hspace{5 pt} \varepsilon^{2(R-1)}\psi_{n_1}\cdots\psi_{n_H}\left[\psi_{0}\right]^R~.\nonumber
\end{align}
Define a unitless combination ${a}$ that contains the extra-dimensional overlap integral:
\begin{align}
    {a}_{(R|\vec{n})}&\equiv {a}_{(R|n_1\cdots n_H)}\label{abar}\\
    &\hspace{-30 pt}\equiv \dfrac{1}{\pi} e^{-R\pi kr_c}\int_{-\pi}^{+\pi}d\varphi\hspace{5 pt} \varepsilon^{2(R-1)}\psi_{n_1}\cdots \psi_{n_H} \left[\psi_{0}\right]^R~,\nonumber
\end{align}
so that now we may write
\begin{align}
    X_A^{(\text{eff})} &= \left[\dfrac{\kappa}{\sqrt{\pi r_c}}\right]^{H+R-2}\times\label{KKdecompop}\\ &\hspace{-30 pt}\sum_{n_1,\cdots,n_H=0}^{+\infty} {a}_{(R|n_1\cdots n_H)}\left(\partial_\mu^2,\hspace{6 pt} \hat{h}^{(n_1)}\cdots \hat{h}^{(n_H)},\hspace{6 pt} \left[\hat{r}^{(0)}\right]^R\right)~.\nonumber
\end{align}

To simplify this expression further, we define a  KK decomposition operator $\mathcal{X}_{(\vec{n})}[\bullet]$. The KK decomposition operator maps a product of 5D graviton and radion fields to an analogous product of 4D spin-2 fields labeled by KK numbers $\vec{n}=(n_1,\cdots,n_H)$ and 4D radion fields $\hat{r}^{(0)}$. More specifically, $\mathcal{X}$ maps all $\hat{r}$ in its argument to $\hat{r}^{(0)}$ and applies the specified KK labels to the graviton fields ($\hat{h}_{\mu\nu}\mapsto \hat{h}^{(n_i)}_{\mu\nu}$) per term according to the following prescription: the labels are applied left to right in the order that they occur in $\vec{n}$, and are applied to graviton fields of the form $(\partial_y \hat{h})$ before being applied to all other graviton fields. (This prescription ensures we correctly keep track of KK number relative to the soon-to-be-defined quantity ${b}$.) After KK number assignment, any 4D derivatives $\partial_\mu$ in the argument of $\mathcal{X}$ are kept as is, while each extra-dimensional derivative $\partial_y$ is replaced by $1/r_c$.

Using $\mathcal{X}$, we rewrite the A-Type expression:
\begin{align}
    X_A^{(\text{eff})} =&  \left[\dfrac{\kappa}{\sqrt{\pi r_c}}\right]^{H+R-2}\times \\
    &\sum_{n_1,\cdots,n_H=0}^{+\infty} {a}_{(R|n_1\cdots n_H)}\cdot \mathcal{X}_{(n_1\cdots n_H)}\hspace{-1 pt}\left[\hspace{+2 pt}\overline{X}_A\right]~.\nonumber
\end{align}
This completes the schematic A-Type procedure. B-Type terms admit a similar reorganization. First, we KK decompose and integrate $X_B$ to obtain
\begin{align}
    X_B^{(\text{eff})} = &\dfrac{r_c}{(\pi r_c)^{(H+R)/2}}\kappa^{(H+R-2)}\times \\
    &\hspace{-30 pt}\sum_{n_1,\cdots,n_H=0}^{+\infty} \left(1,\hspace{6 pt}\hat{h}^{(n_1)}\cdots \hat{h}^{(n_H)},\hspace{6 pt} \left[\hat{r}^{(0)}\right]^R \right)\times\nonumber\\
    &\hspace{-30 pt}e^{-R\pi kr_c}\int d\varphi\hspace{5 pt} \varepsilon^{2(R-2)}(\partial_{\varphi}\psi_{n_1})(\partial_{\varphi}\psi_{n_2})\psi_{n_3}\cdots\psi_{n_H}\left[\psi_{0}\right]^R~.\nonumber
\end{align}
We summarize the extra-dimensional overlap integral as a unitless quantity ${b}$:
\begin{align}
    {b}_{(R|\vec{n})} &\equiv {b}_{(R|n_{1}n_{2}|n_{3}\cdots n_{H})} \label{bbar}\\
    &\equiv \dfrac{1}{\pi} e^{-R\pi kr_c}\int_{-\pi}^{+\pi}d\varphi\hspace{5 pt} \varepsilon^{2(R-2)}\times\nonumber\\
    &\hspace{30 pt}(\partial_{\varphi} \psi_{n_1})(\partial_{\varphi} \psi_{n_2}) \psi_{n_3}\cdots \psi_{n_H}\left[\psi_{0}\right]^R~,\nonumber
\end{align}
such that
\begin{align}
    X_B^{(\text{eff})} =& \left[\dfrac{\kappa}{\sqrt{\pi r_c}}\right]^{H+R-2} \sum_{n_1,\cdots,n_H=0}^{+\infty} {b}_{(R|n_1n_2|n_3\cdots n_H)}\nonumber\\
    &\times\dfrac{1}{r_c^2}\left(1,\hspace{6 pt} \hat{h}^{(n_1)}\cdots \hat{h}^{(n_H)},\hspace{6 pt} \left[\hat{r}^{(0)}\right]^R\right)~,
\end{align}
and, via the KK decomposition operator $\mathcal{X}$,
\begin{align}
    X_B^{(\text{eff})} &= \left[\dfrac{\kappa}{\sqrt{\pi r_c}}\right]^{H+R-2}\times\\ &\sum_{n_1,\cdots,n_H=0}^{+\infty} {b}_{(R|n_1n_2|n_3\cdots n_H)}\cdot \mathcal{X}_{(n_1\cdots n_H)}\hspace{-1 pt}\left[\hspace{2 pt}\overline{X}_B\right]~,\nonumber
\end{align}
where we recall that $\mathcal{X}$ maps $\partial_y$ to $1/r_c$ after KK number assignment. This completes the schematic B-Type procedure.

We now connect these procedures to the 4D effective RS1 Lagrangian $\mathcal{L}^{(\text{RS,eff})}_{4D}$, following the arrangement of the 5D Lagrangian described in Sec. \ref{sec:levelA}. Suppose we collect all terms from the WFE RS1 Lagrangian $\mathcal{L}_{5D}^{(\text{RS})}$ that contain $H$ graviton fields and $R$ radion fields. Label this collection $\mathcal{L}^{(\text{RS})}_{h^H r^R}$. In general, we can subdivide those terms into two sets based on their derivative content, i.e. whether they are A-Type or B-Type.
\begin{align}
    \mathcal{L}^{(\text{RS})}_{h^H r^R} = \mathcal{L}^{(\text{RS})}_{A:h^H r^R} + \mathcal{L}_{B:h^H r^R}^{(\text{RS})}~.
\end{align}
We may go a step further by using our existing knowledge to preemptively extract powers of the expansion parameter $\kappa$ and any exponential coefficients:
\begin{align}
    \mathcal{L}^{(\text{RS})}_{h^H r^R} &= \kappa^{(H+R-2)}\bigg[e^{-R\pi kr_c} \varepsilon^{2(R-1)}\hspace{3 pt}\overline{\mathcal{L}}_{A:h^H r^R}\nonumber\\
    &+ e^{-R\pi kr_c} \varepsilon^{2(R-2)}\hspace{3 pt}\overline{\mathcal{L}}_{B:h^H r^R}\bigg]~.
\end{align}
Finally, we can apply the schematic procedures described above to obtain a succinct expression for the effective Lagrangian with $H$ graviton fields and $R$ radion fields:
\begin{align}
    \mathcal{L}^{(\text{RS,eff})}_{h^H r^R} =& \left[\dfrac{\kappa}{\sqrt{\pi r_c}}\right]^{(H+R-2)}\sum_{\vec{n}=\vec{0}}^{+\infty}\bigg\{ {a}_{(R|\vec{n})}\cdot \mathcal{X}_{(\vec{n})}\hspace{-3 pt}\left[\hspace{3 pt}\overline{\mathcal{L}}_{A:h^H r^R}\right]\nonumber\\
    &\hspace{30 pt}+{b}_{(R|\vec{n})}\cdot \mathcal{X}_{(\vec{n})}\hspace{-3 pt}\left[\hspace{3 pt}\overline{\mathcal{L}}_{B:h^H r^R}\right]\bigg\}~.\label{LRSeffPerTerm}
\end{align}

Computationally, a key feature of this Lagrangian is how the dependence on the physical variables arrange themselves. Consider the set $\{M_{\text{Pl}}, kr_{c}, m_{1}\}$. The parameter $kr_{c}$ determines the wavefunctions $\{\psi_{n}\}$ and spectrum $\{\mu_{n}\}\equiv\{m_{n}r_{c}\}$, and thus $\{a_{(R|\vec{n})},b_{(R|\vec{n})}\}$ as well. Additionally fixing the value of $m_{1}$ determines $r_{c} = \mu_{1}/m_{1}$ and $k = (kr_{c})m_{1}/\mu_{1}$. Finally, fixing $M_{\text{Pl}}$ determines $\kappa/\sqrt{\pi r_{c}} = \kappa_{\text{4D}}/\psi_{0}$ $= 2/(M_{\text{Pl}}\psi_{0})$. Therefore, referring back to the specific form of Eq. \eqref{LRSeffPerTerm}, once $kr_{c}$ is fixed, changing $m_{1}$ only affects the relative importance of A-Type vs. B-Type terms via factors of $r_{c}$ introduced by $\mathcal{X}_{(\vec{n})}[\bullet]$ and changing $M_{\text{Pl}}$ only affects the interaction's overall strength via $[\kappa/\sqrt{\pi r_{c}}]^{(H+R-2)}$. Alternatively, by fixing $\kappa$ and $r_{c}$ instead, the couplings $\{a_{(R|\vec{n})},b_{(R|\vec{n})}\}$ encapsulate the effect of varying $k$.

For the specific case of massive spin-2 scattering, we reduce the generality of the preceding notation somewhat by defining
\begin{align}
    \an{\vec{n}} \equiv {a}_{(0|\vec{n})}\hspace{20 pt}\bn{\vec{n}} \equiv {b}_{(0|\vec{n})}\hspace{20 pt}\bhhr{n_1}{n_2} \equiv {b}_{(1|n_1n_2)}
\end{align}
and noting an analogous A-Type radion coupling does not occur in the RS1 model.

\subsection{\label{sec:levelCb}Summary of Results}

Appendix \ref{AppendixWFE} summarized all terms in the WFE RS1 Lagrangian $\mathcal{L}^{(\text{RS})}_{\text{5D}}$ that contain four or fewer fields. In particular, it has listed explicit expressions for all relevant $\overline{\mathcal{L}}_{A}$ and $\overline{\mathcal{L}}_{B}$. Application of \eqref{LRSeffPerTerm} to all of these combinations yields a WFE 4D effective Lagrangian of the following form: 
\begin{align}
    \mathcal{L}^{(\text{RS,eff})}_{4D} =& \mathcal{L}^{(\text{eff})}_{hh} + \mathcal{L}^{(\text{eff})}_{rr} + \mathcal{L}^{(\text{eff})}_{hhh} + \cdots + \mathcal{L}^{(\text{eff})}_{rrr}\nonumber\\
    &+ \mathcal{L}^{(\text{eff})}_{hhhh} + \cdots + \mathcal{L}^{(\text{eff})}_{rrrr} + \mathcal{O}(\kappa^3)~.
\end{align}

Explicitly, we find 
\begin{align}
    \mathcal{L}^{(\text{eff})}_{hh} =& \sum_{n=0}^{+\infty} \left[-\hat{h}^{(n)}_{\mu\nu} (\partial^\mu \partial^\nu \hat{h}^{(n)}) + \hat{h}^{(n)}_{\mu\nu} (\partial^\mu \partial_\rho \hat{h}^{(n)\rho\nu})\right.\nonumber\\
    &\left.- \dfrac{1}{2} \hat{h}^{(n)}_{\mu\nu} (\square \hat{h}^{(n)\mu\nu}) + \dfrac{1}{2} \hat{h}^{(n)} (\square\hat{h}^{(n)})\right]\\
    &\hspace{25 pt}+ m_n^2\left[-\dfrac{1}{2}\ltr \hat{h}^{(n)}\hat{h}^{(n)}\rtr +\dfrac{1}{2} \ltr \hat{h}^{(n)}\rtr \ltr \hat{h}^{(n)} \rtr \right] ~,\nonumber\\
    \mathcal{L}^{(\text{eff})}_{rr} =& \dfrac{1}{2}(\partial_\mu \hat{r}^{(0)})(\partial^\mu \hat{r}^{(0)})~, \\
    \text{ }\nonumber\\
    \mathcal{L}^{(\text{eff})}_{hhh} =& \dfrac{\kappa}{\sqrt{\pi r_c}}\sum_{l,m,n=0}^{+\infty} \bigg\{{a}_{(0|lmn)}\cdot \mathcal{X}_{(lmn)}\hspace{-3 pt}\left[\hspace{1 pt}\overline{\mathcal{L}}_{A:hhh}\right]\\ &+{b}_{(0|lmn)}\cdot\mathcal{X}_{(lmn)}\hspace{-3 pt} \left[\hspace{+1 pt}\overline{\mathcal{L}}_{B:hhh}\right]\bigg\}~,\\
    \mathcal{L}^{(\text{eff})}_{hhr}=& \dfrac{\kappa}{\sqrt{\pi r_c}}\sum_{m,n=0}^{+\infty}\bigg\{ {b}_{(1|mn)}\cdot\mathcal{X}_{(mn)}\hspace{-3 pt} \left[\hspace{1 pt}\overline{\mathcal{L}}_{B:hhr}\right]\bigg\}~,\\
    \mathcal{L}^{(\text{eff})}_{hrr}=& \dfrac{\kappa}{\sqrt{\pi r_c}}\sum_{n=0}^{+\infty}\bigg\{ {a}_{(2|n)}\cdot \mathcal{X}_{(n)}\hspace{-3 pt} \left[\hspace{1 pt}\overline{\mathcal{L}}_{A:hrr}\right]\bigg\}~,\\
    \mathcal{L}^{(\text{eff})}_{rrr}=& \dfrac{\kappa}{\sqrt{\pi r_c}}\bigg\{ {a}_{(3)}\cdot \mathcal{X}\hspace{-3 pt}\left[\hspace{1 pt}\overline{\mathcal{L}}_{A:rrr}\right]\bigg\}~,\\
    \text{ }\nonumber\\
    \mathcal{L}^{(\text{eff})}_{hhhh} =& \left[\dfrac{\kappa}{\sqrt{\pi r_c}}\right]^2\sum_{k,l,m,n=0}^{+\infty} \bigg\{{a}_{(klmn)}\cdot \mathcal{X}_{(klmn)}\hspace{-3 pt} \left[\hspace{1 pt}\overline{\mathcal{L}}_{A:hhhh}\right]\nonumber\\
    &+{b}_{(klmn)}\cdot\mathcal{X}_{(klmn)}\hspace{-3 pt} \left[\hspace{1 pt}\overline{\mathcal{L}}_{B:hhhh}\right]\bigg\}~,\\
    \mathcal{L}^{(\text{eff})}_{hhhr} =& \left[\dfrac{\kappa}{\sqrt{\pi r_c}}\right]^2\sum_{l,m,n=0}^{+\infty}\bigg\{ {b}_{(1|lmn)}\cdot\mathcal{X}_{(lmn)}\hspace{-3 pt} \left[\hspace{1 pt}\overline{\mathcal{L}}_{B:hhhr}\right]\bigg\}~,\\
    \mathcal{L}^{\text{(eff)}}_{hhrr} =& \left[\dfrac{\kappa}{\sqrt{\pi r_c}}\right]^2\sum_{m,n=0}^{+\infty} \bigg\{{a}_{(2|mn)}\cdot\mathcal{X}_{(mn)}\hspace{-3 pt} \left[\hspace{1 pt}\overline{\mathcal{L}}_{A:hhrr}\right]\nonumber\\
    &+{b}_{(2|mn)}\cdot\mathcal{X}_{(mn)}\hspace{-3 pt} \left[\hspace{1 pt}\overline{\mathcal{L}}_{B:hhrr}\right]\bigg\}~,\\
    \mathcal{L}^{(\text{eff})}_{hrrr} =& \left[\dfrac{\kappa}{\sqrt{\pi r_c}}\right]^2\sum_{n=0}^{+\infty} \bigg\{{a}_{(3|n)} \cdot\mathcal{X}_{(n)}\hspace{-3 pt} \left[\hspace{1 pt}\overline{\mathcal{L}}_{A:hrrr}\right]\bigg\}~,\\
    \mathcal{L}^{(\text{eff})}_{rrrr} =& \left[\dfrac{\kappa}{\sqrt{\pi r_c}}\right]^2 \bigg\{ {a}_{(4)}\cdot\mathcal{X}\hspace{-3 pt} \left[\hspace{1 pt}\overline{\mathcal{L}}_{A:rrrr}\right]\bigg\}~.
\end{align}
The quantity $a_{(R|\vec{n})}$ is defined in Eq. \eqref{abar}; $b_{(R|\vec{n})}$ is shown in Eq. \eqref{bbar}, and the KK decomposition operator $\mathcal{X}$ is introduced below Eq. \eqref{KKdecompop}.

\section{\label{sec:levelD}Elastic Sum Rules}\label{AppendixSumRules}

 This appendix provides a new analytic proof for a relation arising from the $s^3$ and $s^2$ sum rules. Specifically, we prove Eq. \eqref{SumRuleOsAx} from the main text. Combining that equation with a proof of either the $\mathcal{O}(s^{3})$ or $\mathcal{O}(s^{2})$ sum rule automatically implies a proof of the other sum rule. 
 
 In the main text, we calculated $\sum_{j} a_{nnj}^{2}$ and then $\sum_{j} \mu_{j}^{2}a_{nnj}^{2}$, and thereby proved the $\mathcal{O}(s^{5})$ and $\mathcal{O}(s^{4})$ sum rules respectively. Consider the next sum in that sequence: $\sum_{j} \mu_{j}^{4}a_{nnj}^{2}$. Two results are possible depending on how many factors of $\mu_{j}^{2}$ are absorbed into A-Type couplings (via the sequence of relations $\mu_{j}^{2} a_{nnj}= 2b_{jnn} = 2(\mu_{n}^{2}a_{nnj}-b_{nnj})$). If only one factor of $\mu_{j}^{2}$ is absorbed into an A-Type coupling, the sum equals
\begin{align}
    &\sum_{j}\mu_{j}^{4}a^{2}_{nnj} = 2\sum_{j} \mu_{j}^{2}\left[\mu_{n}^{2}a_{nnj} - b_{nnj}\right] a_{nnj}\nonumber\\
    &\hspace{25 pt}= \dfrac{8}{3}\mu_{n}^{4} a_{nnnn} - 2\sum_{j} \mu_{j}^{2} b_{nnj}a_{nnj}~.
\end{align}
If instead both factors of $\mu_{j}^{2}$ are absorbed, the sum equals
\begin{align}
    &\sum_{j} \mu_{j}^{4} a_{nnj}^{2} = 4\sum_{j}\left[\mu_{n}^{4}a_{nnj}^{2} - 2\mu_{n}^{2} b_{nnj} a_{nnj}+ b_{nnj}^{2}\right]\nonumber\\
    &\hspace{25 pt}=\dfrac{4}{3}\mu_{n}^{4}a_{nnnn} + 4\sum_{j} b_{nnj}^{2}~.\label{muj4annj2x}
\end{align}
Because these results must be equal, together these imply
\begin{align}
    &\sum_{j} \mu_{j}^{2} b_{nnj} a_{nnj} = \dfrac{2}{3} \mu_{n}^{4}a_{nnnn} - 2\sum_{j}b^{2}_{nnj}~.\label{mu2bnnjannj}
\end{align}

Continuing along in the sequence, the next sum to consider is $\sum_{j} \mu_{j}^{6} a_{nnj}^{2}$. If as many factors of $\mu_{j}^{2}$ are absorbed into A-Type couplings as possible, we find
\begin{align}
    &\sum_{j}\mu_{j}^{6}a_{nnj}^{2} = 4\sum_{j} \mu_{j}^{2}\left[\mu_{n}^{4}a_{nnj}^{2} - 2\mu_{n}^{2}b_{nnj}a_{nnj} + b_{nnj}^{2}\right]\nonumber\\
    &\hspace{25 pt}= \dfrac{16}{3} \mu_{n}^{6}a_{nnnn} - 8\mu_{n}^{2}\sum_{j}\mu_{j}^{2} b_{nnj} a_{nnj} + 4\sum_{j} \mu_{j}^{2}b_{nnj}^{2}\nonumber\\
    &\hspace{25 pt}=4 \sum_{j} \left[\mu_{j}^{2} + 4 \mu_{n}^{2} \right] b_{nnj}^{2}~,\label{muj2bnnj2x}
\end{align}
where Eq. \eqref{mu2bnnjannj} was utilized. 

To ultimately obtain our desired result, we require additional details about the nature of the exponential $\varepsilon$. Namely, because
\begin{align}
\partial_{\varphi}|\varphi| &= \text{sign}(\varphi)~,\\
\partial^{2}_{\varphi}|\varphi| &= 2\big[\delta(\varphi) - \delta(\varphi - \pi)\big]~,
\end{align}
the exponential $\varepsilon$ satisfies
\begin{align}
    (\partial_{\varphi} \varepsilon)^{2} &= (kr_{c})^{2} \varepsilon^{2}~,\\
    \partial^{2}_{\varphi} \varepsilon &= (kr_{c})^{2} \varepsilon + 2 (kr_{c})\big[\delta(\varphi) - \delta(\varphi - \pi)\big] \varepsilon~.
\end{align}
Furthermore, because $(\partial_{\varphi}\psi_{n})=0$ for $\varphi\in\{0,\pi\}$,
\begin{align}
    (\partial^{2}_{\varphi}\varepsilon) \varepsilon (\partial_{\varphi}\psi_{n}) = (\partial_{\varphi}\varepsilon)^{2} (\partial_{\varphi}\psi_{n}) = (kr_{c})^{2}\varepsilon^{2}(\partial_{\varphi}\psi_{n})~.\label{eeDpsi}
\end{align}
This will allow us to simplify $\mu_{j}^{2} b_{nnj}$ in Eq. \eqref{muj2bnnj2x} and thereby derive Eq. \eqref{SumRuleOsAx}. 

Define the commonly occurring combination $\mathcal{D} \equiv \varepsilon^{-4}\partial_{\varphi}$ for convenience. Because
\begin{align}
    \mathcal{D}\left[\varepsilon^{+6} (\mathcal{D}\psi_{n})^{2}\right] &= 6(\partial_{\varphi}\varepsilon)\varepsilon(\mathcal{D}\psi_{n})^{2}-2\mu_{n}^{2}(\mathcal{D}\psi_{n})\psi_{n}~,
\end{align}
it is the case that
\begin{align}
    \partial_{\varphi}\mathcal{D}\left[\varepsilon^{+6} (\mathcal{D}\psi_{n})^{2}\right] =& 12(kr_{c})^{2}\varepsilon^{2}(\mathcal{D}\psi_{n})^{2}\\
    &-12\mu_{n}^{2}(\partial_{\varphi}\varepsilon)\varepsilon^{-1} (\mathcal{D}\psi_{n})\psi_{n}\nonumber\\ 
    &+2\mu_{n}^{4}\varepsilon^{-2}\psi_{n}^{2} - 2\mu_{n}^{2}\varepsilon^{4}(\mathcal{D}\psi_{n})^{2}~,\nonumber
\end{align}
where Eq. \eqref{eeDpsi} was used to eliminate factors of $(\partial^{2}_{\varphi}\varepsilon)$ and $(\partial_{\varphi}\varepsilon)^{2}$. Thus,
\begin{align}
    \mu_{j}^{2} b_{nnj} =& -\dfrac{1}{\pi}\int d\varphi\hspace{5 pt}\varepsilon^{6}(\mathcal{D}\psi_{n})^{2}\left[-\mu_{j}^{2}\varepsilon^{-2}\psi_{j}\right]\nonumber\\
    =& -\dfrac{1}{\pi}\int d\varphi\hspace{5 pt}\varepsilon^{6}(\mathcal{D}\psi_{n})^{2}\left[\partial_{\varphi}\mathcal{D}_{\varphi}\psi_{j}\right]\nonumber\\
    =& -\dfrac{1}{\pi}\int d\varphi\hspace{5 pt}\psi_{j}\s\partial_{\varphi}\mathcal{D}\left[\varepsilon^{+6} (\mathcal{D}\psi_{n})^{2}\right]\nonumber\\
    =& -\dfrac{12}{\pi}(kr_{c})^{2} \int d\varphi \hspace{5 pt} \varepsilon^{2}(\mathcal{D}\psi_{n})^{2}\psi_{j}\nonumber\\
    & +\dfrac{12}{\pi}\mu_{n}^{2}\int d\varphi\hspace{5 pt}(\partial_{\varphi}\varepsilon)\varepsilon^{-1}(\mathcal{D}\psi_{n})\psi_{n}\nonumber\\
    &  -2 \mu_{n}^{4} a_{nnj} + 2 \mu_{n}^{2} b_{nnj}~,
\end{align}
such that
\begin{align}
    \sum_{j} \mu_{j}^{2}b_{nnj}^{2} =& -\dfrac{12}{\pi}(kr_{c})^{2}\int d\varphi\hspace{5 pt}\varepsilon^{8}(\mathcal{D}\psi_{n})^{4}\nonumber\\
    & +\dfrac{12}{\pi}\mu_{n}^{2} \int d\varphi\hspace{5 pt}(\partial_{\varphi}\varepsilon)\varepsilon^{5}(\mathcal{D}\psi_{n})^{3}\psi_{n}\nonumber\\
    & - 2\mu_{n}^{4} b_{nnnn} + 2\mu_{n}^{2}\sum_{j} b_{nnj}^{2}~.\label{muj2bnnj2y}
\end{align}
The second term can be rewritten in terms of B-Type couplings
\begin{align}
\dfrac{6}{\pi}\int d\varphi\hspace{5 pt}(\partial_{\varphi}\varepsilon)\varepsilon^{5}(\mathcal{D}\psi_{n})^{3}\psi_{n} &= \dfrac{1}{\pi} \int d\varphi\hspace{5 pt}(\partial_{\varphi}\varepsilon^{6})(\mathcal{D}\psi_{n})^{3}\psi_{n}\nonumber\\
&\hspace{-100 pt}=-\dfrac{1}{\pi}\int d\varphi\hspace{5 pt}\varepsilon^{6} \partial_{\varphi}\left[(\mathcal{D}\psi_{n})^{3}\psi_{n}\right]\nonumber\\
&\hspace{-100 pt}=3\mu_{n}^{2}b_{nnnn} - \sum_{j} b_{nnj}^{2}~,
\end{align}
so that Eq. \eqref{muj2bnnj2y} becomes
\begin{align}
    \sum_{j}\mu_{j}^{2}b_{nnj}^{2} &= -\dfrac{12}{\pi}(kr_{c})^{2}\int d\varphi\hspace{5 pt}\varepsilon^{8}(\mathcal{D}\psi_{n})^{4}+ 4\mu_{n}^{4} b_{nnnn}~.\label{muj2bnnj2z}
\end{align}

The only noncoupling integral $\mathcal{I}$ that remains may also be rewritten in terms of B-Type couplings by carefully reorganizing terms and applying Eq. \eqref{eeDpsi}:
\begin{align}
    \mathcal{I}\equiv& -\dfrac{12}{\pi}(kr_{c})^2 \int d\varphi\hspace{5 pt}\varepsilon^{8}(\mathcal{D}\psi_{n})^{4}\nonumber\\
    =&-\dfrac{3}{2\pi} \int d\varphi\hspace{5 pt}\left[(\partial^{2}_{\varphi}\varepsilon)\varepsilon^{7}+7(\partial_{\varphi}\varepsilon)^{2}\varepsilon^{6}\right] (\mathcal{D}\psi_{n})^{4}\nonumber\\
    =&-\dfrac{3}{2\pi} \int d\varphi\hspace{5 pt}\partial_{\varphi}\left[(\partial_{\varphi}\varepsilon)\varepsilon^{7}\right] (\mathcal{D}\psi_{n})^{4}\nonumber\\
    =&\dfrac{3}{2\pi} \int d\varphi\hspace{5 pt}(\partial_{\varphi}\varepsilon)\varepsilon^{7}\s \partial_{\varphi}\left[(\mathcal{D}\psi_{n})^{4}\right]\nonumber\\
    =&-\dfrac{6}{\pi}\mu_{n}^{2}\int d\varphi\hspace{5 pt}(\partial_{\varphi}\varepsilon)\varepsilon^{5}\s (\mathcal{D}\psi_{n})^{3}\psi_{n}\nonumber\\
    =&-\dfrac{1}{\pi}\mu_{n}^{2}\int d\varphi\hspace{5 pt}(\partial_{\varphi}\varepsilon^{6})\s (\mathcal{D}\psi_{n})^{3}\psi_{n}\nonumber\\
    =&\dfrac{1}{\pi}\mu_{n}^{2}\int d\varphi\hspace{5 pt}\varepsilon^{6}\s\partial_{\varphi}\left[ (\mathcal{D}\psi_{n})^{3}\psi_{n}\right]\nonumber\\
    =&-3\mu_{n}^{4}b_{nnnn} +\mu_{n}^{2}\sum_{j}b_{nnj}^{2}~.
\end{align}
Thus, Eq. \eqref{muj2bnnj2z} becomes
\begin{align}
    \sum_{j} \mu_{j}^{2}b_{nnj}^{2} = \mu_{n}^{4}b_{nnnn} + \mu_{n}^{2}\sum_{j}b_{nnj}^{2}~,
\end{align}
which when applied to Eq. \eqref{muj2bnnj2x} yields
\begin{align}
    \sum_{j}\mu_{j}^{6}a^{2}_{nnj} = 28\mu_{n}^{4}b_{nnnn} + 12\mu_{n}^{2}\sum_{j} b_{nnj}^{2}~.
\end{align}

Finally, we can eliminate $\sum_{j} b_{nnj}^{2}$ from this expression in favor of $\sum_{j} \mu_{j}^{4} a_{nnj}^{2}$ by utilizing Eq. \eqref{muj4annj2x}, such that
\begin{align}
    \sum_{j=0}^{+\infty}\left[\mu_{j}^{2} - 5 \mu_{n}^{2} \right] \mu_{j}^{4} a_{nnj}^{2} &= -\dfrac{16}{3}\mu_{n}^{6} a_{nnnn}~.
\end{align}
This is the desired result.

\section{\label{sec:levelE}Connecting to the Literature}

As previously noted, following the appearance of \cite{Chivukula:2019rij}, and as \cite{Chivukula:2019zkt} was being completed, the authors of \cite{Bonifacio:2019ioc} independently proved that the scattering amplitudes of massive spin-2 KK modes in extra-dimensional theories grow only like ${\cal O}(s)$ for compactifications on arbitrary {\it Ricci-flat} manifolds (note that RS1 is not Ricci-flat). In addition, in Sec. 3.5 of that work, the authors of \cite{Bonifacio:2019ioc} consider the on-shell tree-level elastic scattering amplitude of an arbitrary massive spin-2 state (in four dimensions) which is coupled to one massless graviton, a tower of massive spin-2 states, and possible additional scalar and vector states. They consider interaction vertices involving the most general two-derivative parity-even on-shell cubic vertices, and quartic contact interactions which involve the contractions of polarization tensors and momenta, and containing up to six derivatives.  {\it By assuming} that the elastic scattering amplitude of spin-0 modes of the massive spin-2 state grows no faster than $\mathcal{O}(s)$, they derive a set of consistency conditions (sum rules) that must be satisfied by these couplings. In this appendix, we present the relationships of our sum rules to those presented in  \cite{Bonifacio:2019ioc}, demonstrating that our RS1 sum rules in Eqs. (\ref{SumRuleO5}), (\ref{SumRuleO4}), (\ref{SumRuleOs3}), and (\ref{SumRuleOs2}), obey the consistency conditions given there.

Our weak field expanded RS1 Lagrangian (as written and prior to applying any coupling relations) matches their parameterization (their Eqs. 3.78-82 as written) when
\begin{align}
    \tilde{a}_{1} &= -3\s a_{nnn} + \dfrac{6}{\mu_{n}^{2}} b_{nnn}~,\\
    \tilde{a}_{2} &= \tfrac{1}{2} \tilde{a}_{3} = -a_{nnn}~,\\
\tilde{b}_{1} &= \tfrac{1}{2} \tilde{b}_{2} = \tilde{b}_{3} = -a_{nn0}~,\label{b1b2b3ann0}\\
    \tilde{c}_{1} &=\dfrac{1}{\mu_{n}^{2}} \sqrt{\dfrac{3}{2}} b_{nnr}~,\hspace{60 pt}\tilde{c}_{2} = 0~,\\
    \tilde{e}_{1} &=  \dfrac{1}{\mu_{n}^2}\bigg[ 2\bigg(2\s b_{jnn} + b_{nnj}\bigg) - \bigg(2\s \mu_{n}^{2} + \mu_{j}^{2}\bigg) a_{nnj}\bigg]~,\\
    \tilde{e}_{2} &= \tfrac{1}{2}\tilde{e}_{3} = \tilde{e}_{4} = \tfrac{1}{2} \tilde{e}_{5} = -a_{nnj}~,
\end{align}
where $\mu_{n} \equiv m_{n} r_{c}$ and $\tilde{x} \equiv \sqrt{\pi r_{c}} x/\kappa$. The minus signs here reflect a difference of metric convention: we use the mostly-minus convention for $\eta_{\mu\nu}$, whereas Ref. \cite{Bonifacio:2019ioc} uses mostly-plus. To ensure gauge invariance (i.e. to not generate terms that violate 4D diffeomorphism invariance), it must be the case that
\begin{align}
    b_{nn0} = \mu_{n}^{2} \s a_{nn0}~,
\end{align}
and (Eq. 3.83 of \cite{Bonifacio:2019ioc})
\begin{align}
    2b_{1} = b_{2} = \dfrac{4}{M_{\text{Pl}}}~,
\end{align}
which implies, via Eq. \eqref{b1b2b3ann0} above,
\begin{align}
    \dfrac{\kappa}{\sqrt{\pi r_{c}}}a_{nn0} = -\dfrac{2}{M_{\text{Pl}}} = -\kappa_{\text{4D}}~.
\end{align}
Diffeomorphism invariance therefore implies the relationship between 4D and 5D Planck scales which is necessary to ensure our graviton reproduces the 4D Einstein-Hilbert Lagrangian at cubic order.

Next, we substitute these explicit RS1 parameters into the sum rules of \cite{Bonifacio:2019ioc} (Eqs. 3.85a-c). Let $j$ denote summation over all KK numbers and let $a$ denote summation over all KK numbers {\it except} $0$ and $n$. Their first sum rule, Eq. 3.85a, becomes
\begin{align}
    0 &= a_{2}^{2} + 4\s  b_{1}^{2} + \dfrac{1}{4}\sum_{a} \bigg[ 4- 3 \dfrac{m_{a}^{2}}{m_{n}^{2}}\bigg] e_{5,a}^{2}\nonumber\\
    &= a_{nnn}^{2} + 4\s  a_{nn0}^{2} + \dfrac{1}{4} \sum_{a} \bigg[4 - 3 \dfrac{m_{a}^{2}}{m_{n}^{2}}\bigg] (2\s a_{nna})^{2}\nonumber\\
    &= \dfrac{3}{\mu_{n}^{2}} \sum_{j} \bigg[ \dfrac{4}{3} \mu_{n}^{2}-  \mu_{j}^{2}\bigg] a_{nnj}^{2}\nonumber\\
    &= \dfrac{3}{m_{n}^{2}} \bigg[ \dfrac{4}{3} \mu_{n}^{2} a_{nnnn} - \sum_{j} \mu_{j}^{2} a_{nnj}^{2} \bigg]~,
\end{align}
which is equivalent to our $\mathcal{O}(s^{4})$ rule, Eq. \eqref{SumRuleO4}. Next, their Eq. 3.85b reduces as follows:
\begin{align}
    0 &= a_{2}^{2} + 4\s b_{1}^{2} - 24\s c_{1}^{2}+ \dfrac{1}{4} \sum_{a} \bigg[5\dfrac{m_{a}^{2}}{m_{n}^{2}} - 4\bigg] \dfrac{m_{a}^{2}}{m_{n}^{2}} e_{5,a}^{2}\nonumber\\
    &= a_{nnn}^{2} + 4\s  a_{nn0}^{2} - 24\bigg(\dfrac{1}{(m_{n} r_{c})^{2}} \sqrt{\dfrac{3}{2}} b_{nnr}\bigg)^{2}\nonumber\\
    &\hspace{10 pt}+ \dfrac{1}{4} \sum_{a} \bigg[ 5 \dfrac{m_{a}^{2}}{m_{n}^{2}} -4\bigg] \dfrac{m_{a}^{2}}{m_{n}^{2}} (2 \s a_{nna})^{2}\nonumber\\
    &= a_{nnn}^{2} + 4\s a_{nn0}^{2} - \dfrac{36}{\mu_{n}^{2}} b_{nnr}^{2} + \sum_{a} \bigg[5 \dfrac{\mu_{a}^{2}}{\mu_{n}^{2}} - 4\bigg]\dfrac{\mu_{a}^{2}}{\mu_{n}^{2}} a_{nna}^{2}\nonumber\\
    &= -\dfrac{5}{\mu_{n}^{4}}\bigg[-\dfrac{1}{5} \mu_{n}^{4} a_{nnn}^{2} - \dfrac{4}{5} \mu_{n}^{4} a^{2}_{nn0} + \dfrac{36}{5} b_{nnr}^{2}\bigg.\nonumber\\
    &\hspace{50 pt}\bigg.- \sum_{a} \mu_{a}^{4} a_{nna}^{2} + \dfrac{4}{5} \mu_{n}^{2} \sum_{a} \mu_{a}^{2} a_{nna}^{2}\bigg]\nonumber\\
    &= -\dfrac{5}{\mu_{n}^{4}}\bigg[\dfrac{4}{5}\bigg(9 b_{nnr}^{2} -\mu_{n}^{4} a_{nn0}^{2} \bigg) - \sum_{j} \mu_{j}^{4} a_{nnj}^{2}\nonumber\\
    &\hspace{50 pt}+ \dfrac{4}{5}\mu_{n}^{2} \sum_{j} \mu_{j}^{2} a_{nnj}^{2}\bigg]~.
\end{align}
The last term may be simplified using our $\mathcal{O}(s^{4})$ rule, such that
\begin{align}
0 &= - \dfrac{5}{\mu_{n}^{4}}\bigg[\dfrac{4}{5}\bigg(9 b_{nnr}^{2} -\mu_{n}^{4} a_{nn0}^{2} \bigg)  - \sum_{j} \mu_{j}^{4} a_{nnj}^{2}\bigg.\nonumber\\
&\hspace{50 pt}\bigg.+ \dfrac{16}{15}\mu_{n}^{4}a_{nnnn}\bigg]\nonumber\\
&=- \dfrac{5}{\mu_{n}^{4}}\bigg[ \dfrac{16}{15}\mu_{n}^{4}a_{nnnn} + \dfrac{4}{5}\bigg(9 b_{nnr}^{2} -\mu_{n}^{4} a_{nn0}^{2} \bigg)  \bigg.\nonumber\\
&\hspace{50 pt}\bigg.- \sum_{j} \mu_{j}^{4} a_{nnj}^{2}\bigg]~,
\end{align}
which is equivalent to our $\mathcal{O}(s^{3})$ relation, Eq. \eqref{SumRuleOs3}. Lastly, consider their Eq. 3.85c (and recall that our only scalar is the massless radion):
\begin{align}
    0 &= \dfrac{1}{4} \sum_{a}\bigg(\dfrac{m_{a}^{2}}{m_{n}^{2}} -4\bigg) \bigg( \dfrac{m_{a}^{2}}{m_{n}^{2}} - 1\bigg) m_{a}^{2} (2\s  a_{nna})^{2}\nonumber\\
    &= \dfrac{1}{4} \sum_{j}\bigg(\dfrac{\mu_{j}^{2}}{\mu_{n}^{2}} -4\bigg) \bigg( \dfrac{\mu_{j}^{2}}{\mu_{n}^{2}} - 1\bigg) \mu_{j}^{2} (2\s  a_{nnj})^{2}\nonumber\\
    &= -\dfrac{1}{\mu_{n}^{4}}\bigg[\sum_{j} \bigg(-\mu_{j}^{6} + 5 \mu_{n}^{2} \mu_{j}^{4} -4\s \mu_{n}^{4} \mu_{j}^{2}\bigg)a_{nnj}^{2}\bigg]\nonumber\\
    &= - \dfrac{1}{\mu^{4}_{n}} \bigg[5 \mu_{n}^{2}\bigg(\sum_{j} \mu_{j}^{4} a_{nnj}^{2}\bigg) - 4\mu_{n}^{4}\bigg( \sum_{j} \mu_{j}^{2} a_{nnj}^{2} \bigg)\bigg.\nonumber\\
    &\hspace{50 pt}\bigg.- \sum_{j} \mu_{j}^{6} a_{nnj}^{2}\bigg]~.
\end{align}
Using the $\mathcal{O}(s^{4})$ relation again, this becomes
\begin{align}
    0 = \dfrac{1}{\mu_{n}^{4}}\bigg[\sum_{j}\bigg(\mu_{j}^{2}-5\mu_{n}^{2}\bigg) \mu_{j}^{4} a_{nnj}^{2} + \dfrac{16}{3} \mu_{n}^{6} a_{nnnn}\bigg]~,
\end{align}
which implies Eq. \eqref{SumRuleOsA} and is equivalent to our $\mathcal{O}(s^{2})$ sum rule Eq. \eqref{SumRuleOs2} once our $\mathcal{O}(s^{3})$ sum rule is invoked.

\section{KK Mode Couplings at Large $kr_{c}$}
\label{sec:largekrc}

In this appendix, we consider the behavior of the KK mode couplings in the large-$kr_c$ limit, to support the discussion in Sec. \ref{sec:level6c}.

\subsection{General Considerations}

At large values of $kr_{c}$, for nonzero $n$, the behavior of the irregular Bessel functions $Y_\nu$ implies that the coefficients $b_{n\nu}$ in Eq. (\ref{bnnu}) are small. The wavefunctions of Eq. (\ref{eq:wavefunction}) can then be approximated
\begin{align}
    \psi_{n}(\varphi) \approx \dfrac{1}{N_{n}}e^{+2 kr_{c} |\phi|} J_{2}\left[x_{n} e^{kr_{c}(|\phi|-\pi)}\right]~,
    \label{eq:asymptotic-wf}
\end{align}
where $x_{n}$ is the $n$th root of $J_{1}$ and
\begin{align}
    N_{n} \approx \dfrac{e^{\pi kr_{c}}}{\sqrt{\pi kr_{c}}} J_{0}(x_{n})~,
\end{align}
corresponding to a state with mass
\begin{align}
    m_{n} \approx x_{n} k\, e^{-\pi kr_{c}}~.
\end{align}
In these expressions, we neglect terms suppressed by higher powers of $kr_c$.

Adopting the expressions derived under the above approximations allows us to consider the $(n,n)\rightarrow(n,n)$ coupling integrals analytically. Specifically, we convert $\varphi$ integrals
\begin{align}
    \int_{-\pi}^{+\pi} d\varphi \hspace{5 pt}e^{-A kr_{c}|\varphi|} \, f(|\varphi|) = 2 \int_{0}^{+\pi} d\varphi \hspace{5 pt}e^{-A kr_{c}|\varphi|} \,  f(\varphi)~,
\end{align}
to $u \equiv x_{n} e^{kr_{c}(\varphi-\pi)}$ integrals, noting $d\varphi = du/(kr_{c} u)$,
\begin{align}
    &\int_{-\pi}^{+\pi} d\varphi \hspace{5 pt}e^{-A kr_{c}|\varphi|} \, f(|\varphi|) \nonumber\\
    &= \dfrac{2x_{n}^{A}e^{-A kr_{c}\pi}}{kr_{c}}\,\left[ \int_{u_{0}}^{u_{\pi}} \dfrac{du}{u^{A+1}}\hspace{5 pt}f\left(\varphi(u)\right)\right]~, 
\end{align}
for any $n$. Note the limits of integration become independent of $kr_{c}$ in this limit:
\begin{align}
    u_{0} = e^{-kr_{c} \pi} x_{n} \rightarrow 0\hspace{50 pt}u_{\pi} = x_{n}~.
\end{align}
Furthermore, in terms of $u$ the $n\neq 0$ wavefunction factorizes into separate $u$ and $kr_{c}$-dependent pieces.
\begin{align}
    \psi_{n}(u) \approx  \dfrac{\sqrt{\pi}}{x_{n}^{2}\,|J_{0}(x_{n})|}\, \left[ u^{2}\, J_{2}(u) \right]\,\sqrt{kr_{c}}\,e^{\pi kr_{c}}~.
\end{align}
More generally, for generic $j\neq 0$,
\begin{align}
    \psi_{j}(u) \approx  \dfrac{\sqrt{\pi}}{x_{n}^{2}\,|J_{0}(x_{j})|}\, \left[ u^{2}\, J_{2}\left(\tfrac{x_{j}}{x_{n}}u\right) \right]\,\sqrt{kr_{c}}\,e^{\pi kr_{c}}~,
\end{align}
and
\begin{align}
    (\partial_{\varphi} \psi_{j})(u) \approx  \dfrac{\sqrt{\pi}\, x_{j}}{x_{n}^{3}\,|J_{0}(x_{j})|}\, \left[ u^{3}\, J_{1}\left(\tfrac{x_{j}}{x_{n}}u\right) \right]\,(kr_{c})^{3/2}\,e^{\pi kr_{c}}~.
\end{align}
For the zero mode we have
\begin{align}
    \psi_{0} \sim \sqrt{\pi kr_{c}}~.
\end{align}
By combining all of the preceding elements, we factor $kr_{c}$ dependence out of the coupling in the large-$kr_c$ limit.

\subsection{The Integrals}

Using these results, in the large-$kr_c$ limit we find
\begin{align}
    a_{nnnn} &\approx C_{nnnn}\, (kr_{c})\, e^{2\pi kr_{c}}~,\\
    a_{nn0} &\approx C_{nn0}\, \sqrt{kr_{c}}~,\\
    b_{nnr} &\approx C_{nnr}\, (kr_{c})^{5/2}\, e^{-\pi kr_{c}}~,\\
    a_{nnj} &\approx C_{nnj}\, \sqrt{kr_{c}}\, e^{\pi kr_{c}}~,
\end{align}
where the coefficients $C$ are given by the $kr_c$-independent integrals
\begin{align}
    C_{nnnn} &\equiv \left[\dfrac{2\pi }{x_{n}^{6}\, J_{0}(x_{n})^{4}}\, \int_0^{x_n} du\hspace{5 pt}u^{5}\, J_{2}(u)^{4}\right]~,\\
    C_{nn0} &\equiv \left[\dfrac{2\sqrt{\pi }}{x_{n}^{2}\,J_{0}(x_{n})^{2}}\, \int_0^{x_n} du\hspace{5 pt}u\, J_{2}(u)^{2}\right]~,\\
    C_{nnr} &\equiv \left[\dfrac{2 \sqrt{\pi} }{x_{n}^{2}\,J_{0}(x_{n})^{2} }\, \int_0^{x_n} du\hspace{5 pt}u^{3}\, J_{1}(u)^{2}\right]~,\\
    C_{nnj} &\equiv \left[\dfrac{2\sqrt{\pi}}{x_{n}^{4}\,|J_{0}(x_{j})|\, J_{0}(x_{n})^{2}}\right.\nonumber\\
    &\times \left.\, \int_0^{x_n} du\hspace{5 pt}u^{3}\, J_{2}(u)^{2}\, J_{2}\left(\tfrac{x_{j}}{x_{n}}u\right)\right]
\end{align}

\subsection{Scattering Amplitudes}
The following combination of couplings occurs in the helicity-zero $(n,n)\rightarrow(n,n)$ matrix element:
\begin{align}
    \overline{\mathcal{K}}_{nnnn} &=\dfrac{1}{405}\bigg\{15\sum_{j}\dfrac{m_{j}^{8}}{m_{n}^{8}} a_{nnj}^{2}+28 a_{nnnn}\nonumber\\
    &\hspace{10 pt} - 144\left[ \dfrac{9\s b_{nnr}^{2}}{(m_{n}r_{c})^{4}} - a_{nn0}^{2}\right]\bigg\}~.
\end{align}
Based on our previous argument,
\begin{align}
    \dfrac{m_{j}^{8}}{m_{n}^{8}} a_{nnj}^{2} &\approx \dfrac{x_{j}^{8}}{x_{n}^{8}} C_{nnj}^{2}\, (kr_{c})\, e^{2\pi kr_{c}}~,\\\
    a_{nnnn} &\approx C_{nnnn}\, (kr_{c})\, e^{2\pi kr_{c}}~,\\
    \dfrac{b_{nnr}^{2}}{(m_{n}r_{c})^{4}} &\approx \dfrac{1}{x_{n}^{4}} C_{nnr}^{2}\, (kr_{c})\, e^{2 \pi kr_{c}}~,\\
    a_{nn0}^{2} &\approx C_{nn0}\, (kr_{c})~.
\end{align}
Therefore, at large $kr_{c}$,
\begin{align}
    \dfrac{\overline{\mathcal{K}}_{nnnn}}{2\pi kr_c} &= \dfrac{ e^{2\pi kr_{c}}}{810\pi\, x_{n}^{8}}\bigg\{ 15\sum_{j=1}^{+\infty}  x_{j}^{8} \, C_{nnj}^{2} + 28 \, x_{n}^{8} \, C_{nnnn}\nonumber \\
    & \hspace{25 pt} - 1296\,  x_{n}^{4} \, C_{nnr}^{2} \bigg\}~,
\end{align}
as quoted in Sec. \ref{sec:level6c}.


%

\end{document}